\documentclass[12pt,titlepage,letterpaper]{utarticle}

\usepackage{amsmath}
\usepackage{amsfonts}
\usepackage{amssymb}
\usepackage{amsthm}
\usepackage{xparse}
\usepackage{mathtools}
\usepackage{graphicx}
\usepackage{color}
\usepackage[utf8x]{inputenc}
\usepackage{tocloft}
\usepackage[titletoc,title]{appendix}
\usepackage{hyperref}
\usepackage{longtable}

\definecolor{aqua}{rgb}{0, 1.0, 1.0}
\definecolor{fuschia}{rgb}{1.0, 0, 1.0}
\definecolor{gray}{rgb}{0.502, 0.502, 0.502}
\definecolor{lime}{rgb}{0, 1.0, 0}
\definecolor{maroon}{rgb}{0.502, 0, 0}
\definecolor{navy}{rgb}{0, 0, 0.502}
\definecolor{olive}{rgb}{0.502, 0.502, 0}
\definecolor{purple}{rgb}{0.502, 0, 0.502}
\definecolor{silver}{rgb}{0.753, 0.753, 0.753}
\definecolor{teal}{rgb}{0, 0.502, 0.502}

\makeatletter
\newdimen\itex@wd%
\newdimen\itex@dp%
\newdimen\itex@thd%
\def\itexspace#1#2#3{\itex@wd=#3em%
\itex@wd=0.1\itex@wd%
\itex@dp=#2ex%
\itex@dp=0.1\itex@dp%
\itex@thd=#1ex%
\itex@thd=0.1\itex@thd%
\advance\itex@thd\the\itex@dp%
\makebox[\the\itex@wd]{\rule[-\the\itex@dp]{0cm}{\the\itex@thd}}}
\makeatother

\makeatletter
\newif\if@sup
\newtoks\@sups
\def\append@sup#1{\edef\act{\noexpand\@sups={\the\@sups #1}}\act}%
\def\reset@sup{\@supfalse\@sups={}}%
\def\mk@scripts#1#2{\if #2/ \if@sup ^{\the\@sups}\fi \else%
  \ifx #1_ \if@sup ^{\the\@sups}\reset@sup \fi {}_{#2}%
  \else \append@sup#2 \@suptrue \fi%
  \expandafter\mk@scripts\fi}
\def\tensor#1#2{\reset@sup#1\mk@scripts#2_/}
\def\multiscripts#1#2#3{\reset@sup{}\mk@scripts#1_/#2%
  \reset@sup\mk@scripts#3_/}
\makeatother

\makeatletter
\newbox\slashbox \setbox\slashbox=\hbox{$/$}
\def\itex@pslash#1{\setbox\@tempboxa=\hbox{$#1$}
  \@tempdima=0.5\wd\slashbox \advance\@tempdima 0.5\wd\@tempboxa
  \copy\slashbox \kern-\@tempdima \box\@tempboxa}
\def\slash{\protect\itex@pslash}
\makeatother

\def\clap#1{\hbox to 0pt{\hss#1\hss}}

\let\oldroot\root
\def\root#1#2{\oldroot #1 \of{#2}}
\renewcommand{\sqrt}[2][]{\oldroot #1 \of{#2}}

\DeclareSymbolFont{symbolsC}{U}{txsyc}{m}{n}
\SetSymbolFont{symbolsC}{bold}{U}{txsyc}{bx}{n}
\DeclareFontSubstitution{U}{txsyc}{m}{n}

\DeclareSymbolFont{stmry}{U}{stmry}{m}{n}
\SetSymbolFont{stmry}{bold}{U}{stmry}{b}{n}

\DeclareFontFamily{OMX}{MnSymbolE}{}
\DeclareSymbolFont{mnomx}{OMX}{MnSymbolE}{m}{n}
\SetSymbolFont{mnomx}{bold}{OMX}{MnSymbolE}{b}{n}
\DeclareFontShape{OMX}{MnSymbolE}{m}{n}{
    <-6>  MnSymbolE5
   <6-7>  MnSymbolE6
   <7-8>  MnSymbolE7
   <8-9>  MnSymbolE8
   <9-10> MnSymbolE9
  <10-12> MnSymbolE10
  <12->   MnSymbolE12}{}

\makeatletter
\def\re@DeclareMathSymbol#1#2#3#4{%
    \let#1=\undefined
    \DeclareMathSymbol{#1}{#2}{#3}{#4}}
\re@DeclareMathSymbol{\neArrow}{\mathrel}{symbolsC}{116}
\re@DeclareMathSymbol{\neArr}{\mathrel}{symbolsC}{116}
\re@DeclareMathSymbol{\seArrow}{\mathrel}{symbolsC}{117}
\re@DeclareMathSymbol{\seArr}{\mathrel}{symbolsC}{117}
\re@DeclareMathSymbol{\nwArrow}{\mathrel}{symbolsC}{118}
\re@DeclareMathSymbol{\nwArr}{\mathrel}{symbolsC}{118}
\re@DeclareMathSymbol{\swArrow}{\mathrel}{symbolsC}{119}
\re@DeclareMathSymbol{\swArr}{\mathrel}{symbolsC}{119}
\re@DeclareMathSymbol{\nequiv}{\mathrel}{symbolsC}{46}
\re@DeclareMathSymbol{\Perp}{\mathrel}{symbolsC}{121}
\re@DeclareMathSymbol{\Vbar}{\mathrel}{symbolsC}{121}
\re@DeclareMathSymbol{\sslash}{\mathrel}{stmry}{12}
\re@DeclareMathSymbol{\bigsqcap}{\mathop}{stmry}{"64}
\re@DeclareMathSymbol{\biginterleave}{\mathop}{stmry}{"6}
\re@DeclareMathSymbol{\invamp}{\mathrel}{symbolsC}{77}
\re@DeclareMathSymbol{\parr}{\mathrel}{symbolsC}{77}
\makeatother

\makeatletter
\def\Decl@Mn@Delim#1#2#3#4{%
  \if\relax\noexpand#1%
    \let#1\undefined
  \fi
  \DeclareMathDelimiter{#1}{#2}{#3}{#4}{#3}{#4}}
\def\Decl@Mn@Open#1#2#3{\Decl@Mn@Delim{#1}{\mathopen}{#2}{#3}}
\def\Decl@Mn@Close#1#2#3{\Decl@Mn@Delim{#1}{\mathclose}{#2}{#3}}
\Decl@Mn@Open{\llangle}{mnomx}{'164}
\Decl@Mn@Close{\rrangle}{mnomx}{'171}
\Decl@Mn@Open{\lmoustache}{mnomx}{'245}
\Decl@Mn@Close{\rmoustache}{mnomx}{'244}
\makeatother

\makeatletter
\DeclareRobustCommand\widecheck[1]{{\mathpalette\@widecheck{#1}}}
\def\@widecheck#1#2{%
    \setbox\z@\hbox{\m@th$#1#2$}%
    \setbox\tw@\hbox{\m@th$#1%
       \widehat{%
          \vrule\@width\z@\@height\ht\z@
          \vrule\@height\z@\@width\wd\z@}$}%
    \dp\tw@-\ht\z@
    \@tempdima\ht\z@ \advance\@tempdima2\ht\tw@ \divide\@tempdima\thr@@
    \setbox\tw@\hbox{%
       \raise\@tempdima\hbox{\scalebox{1}[-1]{\lower\@tempdima\box
\tw@}}}%
    {\ooalign{\box\tw@ \cr \box\z@}}}
\makeatother

\makeatletter
\NewDocumentCommand\mathraisebox{moom}{%
\IfNoValueTF{#2}{\def\@temp##1##2{\raisebox{#1}{$\m@th##1##2$}}}{%
\IfNoValueTF{#3}{\def\@temp##1##2{\raisebox{#1}[#2]{$\m@th##1##2$}}%
}{\def\@temp##1##2{\raisebox{#1}[#2][#3]{$\m@th##1##2$}}}}%
\mathpalette\@temp{#4}}
\makeatletter

\makeatletter
\def\udots{\mathinner{\mkern2mu\raise\p@\hbox{.}
\mkern2mu\raise4\p@\hbox{.}\mkern1mu
\raise7\p@\vbox{\kern7\p@\hbox{.}}\mkern1mu}}
\makeatother

\newcommand{\gt}{>}

\theoremstyle{plain}

\theoremstyle{definition}

\theoremstyle{remark}

\begin{document}

%-------------------------------------------------------------------
\preprint{
UTTG--XX--15\\
TCC--XXX--15\\
ICTP--SAIFR/2015--001\\
}

\title{Tinkertoys for the Twisted $E_6$ Theory}

\author{Oscar Chacaltana
    \address{
    ICTP South American Institute for\\ Fundamental Research,\\
    Instituto de F\'isica Te\'orica,\\Universidade Estadual Paulista,\\
    01140-070 S\~{a}o Paulo, SP, Brazil\\
    {~}\\
    \email{chacaltana@ift.unesp.br}\\
    },
    Jacques Distler ${}^\mathrm{b}$ and Anderson Trimm
     \address{
      Theory Group and\\
      Texas Cosmology Center\\
      Department of Physics,\\
      University of Texas at Austin,\\
      Austin, TX 78712, USA \\
      {~}\\
      \email{distler@golem.ph.utexas.edu}\\
      \email{atrimm@physics.utexas.edu}
      }
}
%\date{\today}
\date{January 1, 2015}

\Abstract{
We study $4D$ $\mathcal{N}=2$ superconformal field theories that arise as the compactification of the six-dimensional $(2,0)$ theory of type $E_6$ on a punctured Riemann surface in the presence of $\mathbb{Z}_2$ outer-automorphism twists. We explicitly carry out the classification of these theories in terms of three-punctured spheres and cylinders, and provide tables of properties of the $\mathbb{Z}_2$-twisted punctures. An expression is given for the superconformal index of a fixture with twisted punctures of type $E_6$, which we use to check our identifications. Several of our fixtures have Higgs branches which are isomorphic to instanton moduli spaces, and we find that S-dualities involving these fixtures imply interesting isomorphisms between hyperK\"ahler quotients of these spaces. Additionally, we find families of fixtures for which the Sommers-Achar group, which was previously a Coulomb branch concept, acts non-trivially on the Higgs branch operators.
}

\maketitle

\tocloftpagestyle{empty}
\tableofcontents
\vfill
\newpage
\setcounter{page}{1}

\section{Introduction}\label{introduction}

In recent years, remarkable progress has been made in the study of $4D$ $\mathcal{N}=2$ superconformal field theories by realizing them as partially-twisted compactifications of $6D$ $(2,0)$ theories of type $\mathfrak{j}=A,D,E$ on a punctured Riemann surface, $C$ \cite{Gaiotto:2009hg,Gaiotto:2009we,Alday:2009aq,Gaiotto:2009gz,Gadde:2010te,Gadde:2011ik,Beem:2014rza}. In addition to ordinary $\mathcal{N}=2$ gauge theories, this class of theories (sometimes called ``class $\mathcal{S}$'') contains many strongly-interacting SCFTs, with no known Lagrangian description \cite{Minahan:1996fg,Minahan:1996cj}.

An even larger class of theories can be constructed by including punctures which carry a non-trivial action of the outer-automorphism group of $\mathfrak{j}$ \cite{Tachikawa:2010vg,Chacaltana:2012zy}. In a series of papers \cite{Chacaltana:2010ks,Chacaltana:2011ze,Chacaltana:2012zy,Chacaltana:2012ch,Chacaltana:2013oka,Chacaltana:2014jba,Chacaltana:2014ica}, we have presented a method of classification of these theories. By listing the allowed three-punctured spheres (``fixtures''), and the cylinders connecting them, which can occur in a pants decomposition of $C$, and giving the rules for gluing these together, one can build up an arbitrary theory in this class. Different pants decompositions of the same surface $C$ give different weakly-coupled presentations of the same theory, related by S-duality.

In this paper, we turn our attention to the theories obtained by compactifying the $(2,0)$ theory of type $E_6$ in the presence of punctures twisted by a $\mathbb{Z}_2$ outer-automorphism. (The analysis of the untwisted theories of type $E_6$ can be found in \cite{Chacaltana:2014jba}.) The twisted punctures are in 1-1 correspondence with embeddings $\rho: \mathfrak{su}(2) \hookrightarrow \mathfrak{f}_4$, and we label them by the Bala-Carter label of the corresponding nilpotent orbit. For a given puncture, we compute all the local properties which contribute to determining the $4D$ $\mathcal{N}=2$ SCFT and record them in Table \ref{Table1}. We also determine a projection matrix implementing the branching rule under each embedding, which we use to compute the expansion of the superconformal index. These can be found in Appendix \ref{projection_matrices}.

\section{The twisted $E_6$ theory}\label{the_twisted__theory}
\subsection{The Hitchin system}\label{the_hitchin_system}

For a choice of Riemann surface $C$, the compactification of the 6D (2,0) theory of type $E_6$ on $\mathbb{R}^{3,1}\times C$ yields a 4D $\mathcal{N}=2$ theory on $\mathbb{R}^{3,1}$. The $(2,0)$ theory of type $E_6$ has an outer-automorphism group which is isomorphic to $\mathbb{Z}_2$. This allows us to introduce a class of ``twisted" punctures, around which the fields on $C$ undergo a monodromy by a non-trivial element of the outer-automorphism group\footnote{We also allow for the fields on $C$ to undergo a monodromy upon traversing a homologically non-trivial cycle.}. The properties of these punctures are listed in table \ref{Table1}.

In \cite{Chacaltana:2014jba} we studied the theories that arise from compactifying the $E_6$ $(2,0)$ on a Riemann surface with untwisted punctures. These punctures are classified by nilpotent orbits in the complexified Lie algebra $\mathfrak{e}_6$, and obey a Hitchin boundary condition of the form

\begin{displaymath}
\Phi(z) = \frac{A}{z} + \mathfrak{e}_6
\end{displaymath}
where $\Phi$ is the Higgs field, $z$ is a local coordinate on $C$ such that the puncture is at $z=0$, $A$ is a nilpotent element in $\mathfrak{e}_6$, and $\mathfrak{e}_6$ in the boundary condition above denotes a generic element of $\mathfrak{e}_6$ (or a regular function of $z$ taking values in $\mathfrak{e}_6$).

By contrast, twisted punctures are classified by nilpotent orbits in the complexified Lie algebra $\mathfrak{f}_4$, and obey a twisted boundary condition,

\begin{displaymath}
\Phi(z) = \frac{A}{z} + \frac{o_{-1}}{z^{1/2}} + \mathfrak{f}_4
\end{displaymath}
Here, we have split $\mathfrak{e}_6$ into eigenspaces under the action of the $\mathbb{Z}_2$ outer-automorphism, as $\mathfrak{e}_6=\mathfrak{f}_4 \oplus o_{-1}$, where $\mathfrak{f}_4$ ($o_{-1}$) is the even (odd) eigenspace. Also, $A$ is a nilpotent element in $\mathfrak{f}_4$, and $o_{-1}$ and $\mathfrak{f}_4$ above represent generic elements in the respective spaces.

\subsection{$k$-differentials}\label{differentials}

We use the basis of $E_6$ Casimir $k$-differentials $\{\phi_2,\phi_5,\phi_6,\phi_8,\phi_9,\phi_{12}\}$ of our previous paper \cite{Chacaltana:2014jba}. For the reader's convenience, we repeat here how to construct this basis in terms of the trace invariants $P_k=Tr(\Phi^k)$ for $\Phi$ in the adjoint representation of $E_6$.

\begin{displaymath}
\begin{aligned}
\phi_2 =& \tfrac{1}{48}P_2\\
\phi_6 =& \tfrac{1}{24}\left(P_6 -\tfrac{7}{4608}(P_2)^3\right)\\
\phi_8 =& \tfrac{1}{30}\left(P_8 -\tfrac{2}{9}P_6P_2 +\tfrac{155}{663552}(P_2)^4\right)\\
\phi_{10} =& -\tfrac{1}{105}\left(P_{10}-\tfrac{17}{96} P_8P_2+\tfrac{77}{6912}P_6(P_2)^2-\tfrac{427}{63700992}(P_2)^5\right)\\
\phi_{12} =& \tfrac{1}{155}\left(P_{12} - \tfrac{107}{504} P_{10}P_2+ \tfrac{515}{32256}P_8(P_2)^2- \tfrac{41}{108} (P_6)^2+ \tfrac{295}{497664}P_6(P_2)^3-\tfrac{5669}{9172942848} (P_2)^6\right)\\
\phi_{14} =& \tfrac{1}{4389}\left(P_{14}-\tfrac{3479}{14880}P_{12}P_2 + \tfrac{61391}{3214080}P_{10}(P_2)^2- \tfrac{539}{2160}P_8P_6- \tfrac{139733}{617103360}P_8(P_2)^3\right.\\
& \left.+ \tfrac{165781}{4821120}(P_6)^2P_2- \tfrac{3488947}{44431441920}P_6(P_2)^4+\tfrac{19596907}{409480168734720}(P_2)^7 \right)
\end{aligned}
\end{displaymath}
These relations define all the Casimirs except $\phi_5$ and $\phi_9$. These can be computed from $\phi_{10}$ and $\phi_{14}$, which factorize as

\begin{displaymath}
\begin{aligned}
\phi_{10}&=\phi_5^2,\\
\phi_{14}&=\phi_5\phi_9.
\end{aligned}
\end{displaymath}
Notice that the choice of sign of $\phi_5$ determines also the sign of $\phi_9$. This is precisely the action of the $\mathbb{Z}_2$ outer-automorphism of $E_6$ on the Casimir $k$-differentials,

\begin{displaymath}
\begin{aligned}
\phi_5 &\mapsto -\phi_5\\
\phi_9 &\mapsto -\phi_9\\
\phi_k &\mapsto \phi_k,\qquad k=2,6,8,12
\end{aligned}
\end{displaymath}
So, we can expect that the leading pole orders of the $\phi_k$ for twisted punctures will be half-integer for $k=5,9$, and integer for $k=2,6,8,12$, corresponding to the orders of the Casimirs of $F_4$.

\section{Tinkertoys}\label{tinkertoys}

We find 2078 fixtures with 3 regular punctures, two twisted and one untwisted, which correspond to either an interacting SCFT, a mix of an interacting SCFT and free hypers, or a gauge theory. Of these, we find 1757 SCFTs without global symmetry enhancement, 122 SCFTs with enhanced global symmetry, 32 mixed fixtures, and 167 gauge theory fixtures.

Additionally, there are 23 fixtures with one irregular puncture: 15 free-field fixtures, 6 interacting fixtures, 1 mixed fixture, and 1 gauge theory fixture.

Below, we give tables of the twisted punctures and their properties, as well as tables of the twisted fixtures. For the mixed fixtures, we list $\{d_k\}$ and $(n_h,n_v)$ of the interacting SCFT, and the representation of the free hypermultiplets. We do not list the fixtures without global symmetry enhancement, as their properties can be readily computed from the tables of punctures. Tables of untwisted punctures and fixtures can be found in our previous paper \cite{Chacaltana:2014jba}. Following the conventions of that paper, in the tables we denote the Bala-Carter labels of twisted punctures by underlining them; in the figures, twisted punctures are denoted in gray.

\subsection{Twisted punctures}\label{twisted_punctures}

Twisted punctures in the $E_6$ theory are labeled by nilpotent orbits in $\mathfrak{f}_4$, which we denote by the corresponding Bala-Carter label. As discussed in \cite{Chacaltana:2014jba}, the Bala-Carter notation provides a systematic way to label nilpotent orbits in any exceptional semisimple Lie algebra, and a concise review can be found in appendix A of \cite{Chacaltana:2014jba}. Here we merely add that for the $\mathfrak{f}_4$ nilpotent orbits, components of the Levi subalgebra in the Bala-Carter label with (without) a tilde are constructed from the short (long) roots of $\mathfrak{f}_4$. (So, e.g., $A_2 + \widetilde{A}_1$ and $\widetilde{A}_2 + A_1$ represent different orbits.)

The pole structure of the $k$-differentials is denoted by $\{p_2,p_5,p_6,p_8,p_9,p_{12}\}$, and, for twisted punctures, $p_5$ are $p_9$ are half-integer. The contributions to the graded Coulomb branch dimensions are denoted by $\{d_2,d_3,d_4,d_5,d_6,d_8,d_9,d_{12}\}$, allowing for new Coulomb branch parameters (introduced by a-constraints) of dimensions 3 and 4, which are not degrees of $E_6$ Casimirs. The constraints are shown separately in Appendix \ref{appendix_constraints}.

{\footnotesize
\renewcommand{\arraystretch}{2.25}

\begin{longtable}{|c|c|c|c|c|c|}
\hline
\mbox{\shortstack{\\Nahm\\orbit}}&\mbox{\shortstack{\\Hitchin\\orbit}}&Pole structure&\mbox{\shortstack{\\Coulomb branch\\contributions}}&Flavour group&$(\delta n_h,\delta n_v)$\\
\hline 
\endhead
$\underline{0}$&$F_4$&$\{1,\tfrac{9}{2},5,7,\tfrac{17}{2},11\}$&$\{1,0,0,\tfrac{9}{2},5,7,\tfrac{17}{2},11\}$&$(F_4)_{18}$&$(624,601)$\\
\hline
$\underline{A_1}$&$(F_4(a_1),\mathbb{Z}_2)$&$\{1,\tfrac{9}{2},5,7,\tfrac{15}{2},11\}$&$\{1,0,0,\tfrac{9}{2},5,7,\tfrac{15}{2},11\}$&$Sp(3)_{13}$&$(599,584)$\\
\hline
$\underline{\widetilde{A}_1}$&$F_4(a_1)$&$\{1,\tfrac{9}{2},5,7,\tfrac{15}{2},11\}$&$\{1,0,0,\tfrac{9}{2},6,7,\tfrac{15}{2},10\}$&$SU(4)_{12}$&$(584,572)$\\
\hline
$\underline{A_1 + \widetilde{A}_1}$&$F_4(a_2)$&$\{1,\tfrac{9}{2},5,7,\tfrac{15}{2},10\}$&$\{1,0,0,\tfrac{9}{2},5,7,\tfrac{15}{2},10\}$&$SU(2)_{64}\times SU(2)_{10}$&$(570,561)$\\
\hline
$\underline{A_2}$&$B_3$&$\{1,\tfrac{7}{2},5,7,\tfrac{15}{2},10\}$&$\{1,0,0,\tfrac{7}{2},5,7,\tfrac{15}{2},10\}$&$SU(3)_{16}$&$(560,552)$\\
\hline
$\underline{\widetilde{A}_2}$&$C_3$&$\{1,\tfrac{9}{2},5,7,\tfrac{15}{2},10\}$&$\{1,1,0,\tfrac{9}{2},5,6,\tfrac{15}{2},9\}$&$(G_2)_{10}$&$(536,528)$\\
\hline
$\underline{A_2 + \widetilde{A}_1}$&$(F_4(a_3),S_4)$&$\{1,\tfrac{7}{2},5,6,\tfrac{15}{2},10\}$&$\{1,0,0,\tfrac{7}{2},5,6,\tfrac{15}{2},10\}$&$SU(2)_{39}$&$(543,537)$\\
\hline
$\underline{B_2}$&$(F_4(a_3),\mathbb{Z}_2\times\mathbb{Z}_2)$&$\{1,\tfrac{7}{2},5,6,\tfrac{15}{2},10\}$&$\{1,1,0,\tfrac{7}{2},6,6,\tfrac{13}{2},9\}$&${SU(2)}_7^2$&$(518,513)$\\
\hline
$\underline{\widetilde{A}_2 + A_1}$&$(F_4(a_3),S_3)$&$\{1,\tfrac{7}{2},5,6,\tfrac{15}{2},10\}$&$\{1,1,0,\tfrac{7}{2},5,6,\tfrac{15}{2},9\}$&$SU(2)_{20}$&$(524,519)$\\
\hline
$\underline{C_3(a_1)}$&$(F_4(a_3),\mathbb{Z}_2)$&$\{1,\tfrac{7}{2},5,6,\tfrac{15}{2},10\}$&$\{1,2,0,\tfrac{7}{2},5,6,\tfrac{13}{2},9\}$&$SU(2)_7$&$(511,507)$\\
\hline
$\underline{F_4(a_3)}$&$F_4(a_3)$&$\{1,\tfrac{7}{2},5,6,\tfrac{15}{2},10\}$&$\{1,3,0,\tfrac{7}{2},4,6,\tfrac{13}{2},9\}$&$-$&$(504,501)$\\
\hline
$\underline{B_3}$&$A_2$&$\{1,\tfrac{7}{2},4,6,\tfrac{13}{2},9\}$&$\{1,0,1,\tfrac{7}{2},4,5,\tfrac{11}{2},8\}$&$SU(2)_{24}$&$(440,438)$\\
\hline
$\underline{C_3}$&$\widetilde{A}_2$&$\{1,\tfrac{7}{2},5,6,\tfrac{15}{2},10\}$&$\{1,1,1,\tfrac{7}{2},4,5,\tfrac{11}{2},7\}$&$SU(2)_6$&$(422,420)$\\
\hline
$\underline{F_4(a_2)}$&$A_1+\widetilde{A}_1$&$\{1,\tfrac{7}{2},4,6,\tfrac{13}{2},9\}$&$\{1,0,1,\tfrac{7}{2},4,5,\tfrac{11}{2},7\}$&$-$&$(416,415)$\\
\hline
$\underline{F_4(a_1)}$&$\widetilde{A}_1$&$\{1,\tfrac{7}{2},4,6,\tfrac{13}{2},9\}$&$\{2,1,0,\tfrac{5}{2},4,4,\tfrac{9}{2},6\}$&$-$&$(352,352)$\\
\hline
$\underline{F_4}$&$0$&$\{1,\tfrac{5}{2},3,4,\tfrac{9}{2},6\}$&$\{1,1,0,\tfrac{3}{2},2,2,\tfrac{5}{2},3\}$&$-$&$(184,185)$\\
\hline
\end{longtable}\label{Table1}

}

There is a special piece consisting of \emph{five} nilpotent orbits,

\begin{displaymath}
A_2+\tilde{A}_1,\quad \tilde{A}_2+A_1,\quad B_2,\quad C_3(a_1),\quad F_4(a_3).
\end{displaymath}
The corresponding Hitchin boundary conditions are $(F_4(a_3),\Gamma)$, where the Sommers-Achar group, $\Gamma$, is a subgroup of $S_4$. The leading pole coefficients,

\begin{equation}
\begin{split}
c^{(6)}_5      &= - \left(6a^2+3{a'}^2+{a''}^2\right)\\
c^{(9)}_{15/2} &= \tfrac{1}{3}(a+a') \left({(2a-a')}^2-{a''}^2\right)\\
c^{(12)}_{10}  &= 3{a'}^2 {(4a+a')}^2+2(8a^2 - 12 a a' +{a'}^2){a''}^2
     +\tfrac{1}{3}{a''}^4-\tfrac{4}{3} {\left(c^{(6)}_5\right)}^2
\end{split}
\label{piecepoles}\end{equation}
are invariant under the $S_4$ action, $\left(\begin{smallmatrix}a\\ a'\\a''\end{smallmatrix}\right)\mapsto \gamma \left(\begin{smallmatrix}a\\ a'\\a''\end{smallmatrix}\right)$, generated by

\begin{displaymath}
\sigma_{1 2} = \frac{1}{3}\begin{pmatrix}-1&2&0\\ 4& 1&0\\ 0&0& 3\end{pmatrix},\quad
  \sigma_{2 3} = \frac{1}{2}\begin{pmatrix} 2&0&0\\ 0&-1&1\\ 0&3& 1\end{pmatrix},\quad
  \sigma_{3 4} =            \begin{pmatrix} 1&0&0\\ 0& 1&0\\ 0&0&-1\end{pmatrix},\quad
\end{displaymath}
\begin{itemize}%
\item For the special orbit, $F_4(a_3)$, the Sommers-Achar group is trivial, and $a,\,a',\, a''$ are invariants.
\item For $C_3(a_1)$, the Sommers-Achar group is the $\mathbb{Z}_2$ generated by $\sigma_{3 4}$ and the invariants are $a,\, a',\, {a''}^2$.
\item For $B_2$, the Sommers-Achar group is the $\mathbb{Z}_2\times \mathbb{Z}_2$ generated by $\sigma_{1 2},\, \sigma_{3 4}$ and the invariants are $a+2a',\, {a''}^2,\, 2a^2+{a'}^2$.
\item For $\tilde{A}_2+A_1$, the Sommers-Achar group is the $S_3$ generated by $\sigma_{2 3},\, \sigma_{3 4}$ and the invariants are $a,\, 3a'^2+{a''}^2,\, c^{(9)}_{15/2}$.
\item Finally, for $A_2+\tilde{A}_1$, the Sommers-Achar group is the full $S_4$, and the invariants are $c^{(6)}_5,\, c^{(9)}_{15/2},\, c^{(12)}_{10}$.
\end{itemize}

In \S\ref{enhanced_global_symmetries_the_sommersachar_group_on_the_higgs_branch},
we will discover an action of this $S_4$ group on the Higgs branch  of certain fixtures obtained by varying one of the punctures over this special piece. 

\subsection{Free-field fixtures}\label{freefield_fixtures}
{
\renewcommand{\arraystretch}{1.375}
\begin{longtable}{|c|l|c|l|}
\hline
\#&Fixture&$n_h$&Representation\\
\hline 
\endhead
1&$\begin{matrix}\underline{F_4}\\ \underline{F_4}\end{matrix}\quad (D_4,SU(3)_0)$&0&empty\\
\hline
2&$\begin{matrix}\underline{F_4}\\ \underline{F_4(a_1)}\end{matrix}\quad (A_2,{SU(3)}^2_0)$&0&empty\\
\hline
3&$\begin{matrix}\underline{F_4}\\ \underline{F_4(a_2)}\end{matrix}\quad (A_1,SU(6)_6)$&10&$\tfrac{1}{2}(20)$\\
\hline
4&$\begin{matrix}\underline{F_4}\\ \underline{B_3}\end{matrix}\quad (0,SU(6)_0)$&0&empty\\
\hline
5&$\begin{matrix}\underline{F_4}\\ E_6(a_1)\end{matrix}\quad (\underline{F_4(a_1)},\empty)$&0&empty\\
\hline
6&$\begin{matrix}\underline{F_4}\\ E_6(a_3)\end{matrix}\quad (\underline{C_3(a_1)},SU(2)_1)$&1&$\tfrac{1}{2}(2)$\\
\hline
7&$\begin{matrix}\underline{F_4}\\ A_5\end{matrix}\quad (\underline{B_2},SU(2)_1)$&1&$\tfrac{1}{2}(2)$\\
\hline
8&$\begin{matrix}\underline{F_4}\\ D_5(a_1)\end{matrix}\quad (\underline{\widetilde{A}_2},SU(3)_2)$&1&$1(3)$\\
\hline
9&$\begin{matrix}\underline{F_4}\\ D_5\end{matrix}\quad (\underline{C_3},SU(2)_2)$&2&$1(2)$\\
\hline
10&$\begin{matrix}\underline{F_4}\\ A_4+A_1\end{matrix}\quad (\underline{\widetilde{A}_1},SU(3)_0)$&0&empty\\
\hline
11&$\begin{matrix}\underline{F_4}\\ A_4\end{matrix}\quad (\underline{\widetilde{A}_1},SU(4)_4)$&8&$(2,4)$\\
\hline
12&$\begin{matrix}\underline{F_4(a_1)}\\ E_6(a_1)\end{matrix}\quad (\underline{B_2},{SU(2)}_1^2)$&2&$\tfrac{1}{2}(2,1)+\tfrac{1}{2}(1,2)$\\
\hline
13&$\begin{matrix}\underline{F_4(a_2)}\\ E_6(a_1)\end{matrix}\quad (\underline{\widetilde{A}_1},Sp(2)_0)$&0&empty\\
\hline
14&$\begin{matrix}\underline{C_3}\\ E_6(a_1)\end{matrix}\quad (\underline{\widetilde{A}_1},SU(4)_4)$&6&$\tfrac{1}{2}(2,6)$\\
\hline
15&$\begin{matrix}\underline{B_3}\\ E_6(a_1)\end{matrix}\quad (\underline{A_1},Sp(3)_3)$&9&$\tfrac{1}{2}(3,6)$\\
\hline
\end{longtable}
}

\subsection{Interacting fixtures with one irregular puncture}\label{interacting_fixtures_with_one_irregular_puncture}
{
\renewcommand{\arraystretch}{1.375}
\begin{tabular}{|l|c|c|c|l|}
\hline
$\#$&Fixture&$(n_2,n_3,n_4,n_5,n_6,n_8,n_9,n_{12})$&$(n_h,n_v)$&Theory\\
\hline
1&$\begin{matrix}\underline{F_4}\\ D_4\end{matrix}\quad (\underline{\widetilde{A}_2},G_2)$&$(0,1,0,0,0,0,0,0)$&$(16,5)$&${(E_6)}_{6}$ SCFT\\
\hline
2&$\begin{matrix}\underline{F_4}\\ D_4(a_1)\end{matrix}\quad (\underline{0},Spin(8))$&$(0,1,0,0,0,0,0,0)$&$(16,5)$&${(E_6)}_{6}$ SCFT\\
\hline
3&$\begin{matrix}\underline{F_4}\\ \underline{C_3}\end{matrix}\quad (A_1,SU(6)_6)$&$(0,1,0,0,0,0,0,0)$&$(16,5)$&${(E_6)}_{6}$ SCFT\\
\hline
4&$\begin{matrix}\underline{F_4}\\ A_3\end{matrix}\quad (\underline{0},Spin(9))$&$(0,1,0,1,0,0,0,0)$&$(36,14)$&$Spin(14)_{10}\times U(1)$ SCFT\\
\hline
5&$\begin{matrix}\underline{C_3}\\ D_5\end{matrix}\quad (\underline{0},Spin(9))$&$(0,1,1,1,0,0,0,0)$&$(38,21)$&$Spin(9)_{10} \times SU(2)_6 \times U(1)$\\
\hline
6&$\begin{matrix}\underline{F_4(a_2)}\\ D_5\end{matrix}\quad (\underline{0},Spin(9))$&$(0,0,1,1,0,0,0,0)$&$(32,16)$&$Spin(9)_{10} \times U(1)$\\
\hline
\end{tabular}
}

\subsection{Interacting fixtures with enhanced global symmetry}\label{interacting_fixtures_with_enhanced_global_symmetry}
{\footnotesize
\renewcommand{\arraystretch}{1.375}
\begin{longtable}{|c|l|c|c|c|}
\hline
\#&Fixture&$(n_2,n_3,n_4,n_5,n_6,n_8,n_9,n_{12})$&$(n_h,n_v)$&$G_k$\\
\hline
\endhead
1&$\begin{matrix}\underline{F_4}\\0\end{matrix}\quad \underline{F_4(a_3)}$&$(0,4,0,0,0,0,0,0)$&$(64, 20)$&$[(E_6)_6 \text{ SCFT}]^4$\\
\hline
2&$\begin{matrix}\underline{F_4}\\0\end{matrix}\quad \underline{C_3(a_1)}$&$(0,3,0,0,1,0,0,0)$&$(71, 26)$&$[(E_6)_6 \text{ SCFT}]^2\times [(E_6)_{12} \times SU(2)_7 \text{ SCFT}]$\\
\hline
3&$\begin{matrix}\underline{F_4}\\0\end{matrix}\quad \underline{\widetilde{A}_2+A_1}$&$(0,2,0,0,1,0,1,0)$&$(84, 38)$&$[(E_6)_6\text{ SCFT}]\times [(E_6)_{18} \times SU(2)_{20} \text{ SCFT}]$\\
\hline
4&$\begin{matrix}\underline{F_4}\\0\end{matrix}\quad \underline{B_2}$&$(0,2,0,0,2,0,0,0)$&$(78, 32)$&$[(E_6)_{12} \times SU(2)_7 \text{ SCFT}]^2$\\
\hline
5&$\begin{matrix}\underline{F_4}\\0\end{matrix}\quad \underline{\widetilde{A}_2}$&$(0,2,0,1,1,0,1,0)$&$(96, 47)$&$[(E_6)_6\text{ SCFT}]\times [(E_6)_{18}\times (G_2)_{10}\text{ SCFT}]$\\
\hline
6&$\begin{matrix}\underline{F_4}\\2A_1\end{matrix}\quad \underline{A_2}$&$(0,1,0,0,1,1,0,0)$&$(64, 31)$&$Spin(13)_{16} \times U(1)$\\
\hline
7&$\begin{matrix}\underline{F_4}\\A_1\end{matrix}\quad \underline{A_2}$&$(0,1,0,0,1,1,1,0)$&$(86, 48)$&$(G_2)_{16}\times SU(6)_{18}$\\
\hline
8&$\begin{matrix}\underline{F_4}\\2A_1\end{matrix}\quad \underline{A_1+\widetilde{A}_1}$&$(0,1,0,1,1,1,0,0)$&$(74, 40)$&$Spin(10)_{16} \times SU(2)_{10} \times SU(2)_{32} \times U(1)$\\
\hline
9&$\begin{matrix}\underline{F_4}\\A_1\end{matrix}\quad \underline{A_1+\widetilde{A}_1}$&$(0,1,0,1,1,1,1,0)$&$(96, 57)$&$SU(6)_{18} \times SU(2)_{64-k}\times SU(2)_k \times SU(2)_{10}$\\
\hline
10&$\begin{matrix}\underline{F_4}\\2A_1\end{matrix}\quad \underline{\widetilde{A}_1}$&$(0,1,0,1,2,1,0,0)$&$(88, 51)$&$Spin(8)_{16}\times SU(4)_{12} \times {U(1)}^2$\\
\hline
11&$\begin{matrix}\underline{F_4}\\A_1\end{matrix}\quad \underline{\widetilde{A}_1}$&$(0,1,0,1,2,1,1,0)$&$(110, 68)$&$SU(6)_{18} \times SU(4)_{12} \times {U(1)}$\\
\hline
12&$\begin{matrix}\underline{F_4}\\3A_1\end{matrix}\quad \underline{A_1}$&$(0,1,0,1,1,0,0,1)$&$(84, 48)$&$Sp(4)_{13}\times SU(3)_{24}$\\
\hline
13&$\begin{matrix}\underline{F_4}\\A_2+2A_1\end{matrix}\quad \underline{0}$&$(0,1,0,1,0,0,1,0)$&$(70, 31)$&$(E_7)_{18}\times U(1)$\\
\hline
14&$\begin{matrix}\underline{F_4}\\2A_2\end{matrix}\quad \underline{0}$&$(0,1,0,0,1,0,0,0)$&$(56, 16)$&$[(E_8)_{12}\text{ SCFT}]\times[(E_6)_6 \text{ SCFT}]$\\
\hline
15&$\begin{matrix}\underline{F_4}\\A_2+A_1\end{matrix}\quad \underline{0}$&$(0,1,0,1,1,0,1,0)$&$(83, 42)$&$(E_6)_{18} \times SU(3)_{12} \times U(1)$\\
\hline
16&$\begin{matrix}\underline{F_4}\\A_2\end{matrix}\quad \underline{0}$&$(0,1,0,1,2,0,1,0)$&$(96, 53)$&$(E_6)_{18} \times {SU(3)}^2_{12}$\\
\hline
17&$\begin{matrix}\underline{F_4(a_2)}\\A_2+2A_1\end{matrix}\quad \underline{F_4(a_2)}$&$(0,0,2,2,1,1,1,0)$&$(94, 75)$&$SU(2)_{54-k}\times SU(2)_{k} \times U(1)$\\
\hline
18&$\begin{matrix}\underline{F_4(a_2)}\\2A_2\end{matrix}\quad \underline{F_4(a_2)}$&$(0,0,2,1,2,1,0,0)$&$(80, 60)$&$Spin(7)_{12}\times U(1)$\\
\hline
19&$\begin{matrix}\underline{F_4(a_2)}\\A_2+A_1\end{matrix}\quad \underline{F_4(a_2)}$&$(0,0,2,2,2,1,1,0)$&$(107, 86)$&$SU(3)_{12}\times {U(1)}^2$\\
\hline
20&$\begin{matrix}\underline{F_4(a_2)}\\A_2\end{matrix}\quad \underline{F_4(a_2)}$&$(0,0,2,2,3,1,1,0)$&$(120, 97)$&${SU(3)}_{12}^2\times U(1)$\\
\hline
21&$\begin{matrix}\underline{F_4(a_2)}\\A_2+2A_1\end{matrix}\quad \underline{C_3}$&$(0,1,2,2,1,1,1,0)$&$(100, 80)$&$SU(2)_{36}\times SU(2)_{18}\times SU(2)_6 \times U(1)$\\
\hline
22&$\begin{matrix}\underline{F_4(a_2)}\\2A_2\end{matrix}\quad \underline{C_3}$&$(0,1,2,1,2,1,0,0)$&$(86, 65)$&$Spin(7)_{12}\times SU(2)_6 \times U(1)$\\
\hline
23&$\begin{matrix}\underline{F_4(a_2)}\\A_2+A_1\end{matrix}\quad \underline{C_3}$&$(0,1,2,2,2,1,1,0)$&$(113, 91)$&$SU(3)_{12}\times SU(2)_6 \times {U(1)}^2$\\
\hline
24&$\begin{matrix}\underline{F_4(a_2)}\\A_2\end{matrix}\quad \underline{C_3}$&$(0,1,2,2,3,1,1,0)$&$(126, 102)$&${SU(3)}_{12}^2\times SU(2)_6 \times U(1)$\\
\hline
25&$\begin{matrix}\underline{F_4(a_2)}\\D_4(a_1)\end{matrix}\quad \underline{B_3}$&$(0,0,4,1,1,0,0,0)$&$(64, 48)$&${SU(2)}^3_{8}\times {U(1)}^2$\\
\hline
26&$\begin{matrix}\underline{F_4(a_2)}\\A_3+A_1\end{matrix}\quad \underline{B_3}$&$(0,0,3,1,1,1,0,0)$&$(73, 56)$&$SU(2)_{16}\times SU(2)_8 \times SU(2)_9 \times U(1)$\\
\hline
27&$\begin{matrix}\underline{F_4(a_2)}\\A_3\end{matrix}\quad \underline{B_3}$&$(0,0,3,2,1,1,0,0)$&$(84, 65)$&$SU(2)_{16}\times SU(2)_8 \times Sp(2)_{10} \times U(1)$\\
\hline
28&$\begin{matrix}\underline{F_4(a_2)}\\A_4+A_1\end{matrix}\quad \underline{C_3(a_1)}$&$(0,2,1,1,2,1,0,0)$&$(79, 63)$&$SU(2)_7\times {U(1)}^3$\\
\hline
29&$\begin{matrix}\underline{F_4(a_2)}\\A_4\end{matrix}\quad \underline{C_3(a_1)}$&$(0,2,2,1,2,1,0,0)$&$(87, 70)$&$SU(2)_8\times SU(2)_7 \times {U(1)}^3$\\
\hline
30&$\begin{matrix}\underline{F_4(a_2)}\\A_4+A_1\end{matrix}\quad \underline{\widetilde{A}_2+A_1}$&$(0,1,1,1,2,1,1,0)$&$(92, 75)$&$SU(2)_{20}\times {U(1)}^2$\\
\hline
31&$\begin{matrix}\underline{F_4(a_2)}\\A_4\end{matrix}\quad \underline{\widetilde{A}_2+A_1}$&$(0,1,2,1,2,1,1,0)$&$(100, 82)$&$SU(2)_{20}\times SU(2)_8 \times {U(1)}^2$\\
\hline
32&$\begin{matrix}\underline{F_4(a_2)}\\A_4+A_1\end{matrix}\quad \underline{B_2}$&$(0,1,1,1,3,1,0,0)$&$(86, 69)$&${SU(2)}_7^2 \times {U(1)}^2$\\
\hline
33&$\begin{matrix}\underline{F_4(a_2)}\\A_4\end{matrix}\quad \underline{B_2}$&$(0,1,2,1,3,1,0,0)$&$(94, 76)$&$SU(2)_8\times {SU(2)}_7^2 \times {U(1)}^2$\\
\hline
34&$\begin{matrix}\underline{F_4(a_2)}\\A_4+A_1\end{matrix}\quad \underline{\widetilde{A}_2}$&$(0,1,1,2,2,1,1,0)$&$(104, 84)$&$(G_2)_{10}\times {U(1)}^2$\\
\hline
35&$\begin{matrix}\underline{F_4(a_2)}\\A_4\end{matrix}\quad \underline{\widetilde{A}_2}$&$(0,1,2,2,2,1,1,0)$&$(112, 91)$&$(G_2)_{10}\times SU(2)_8 \times {U(1)}^2$\\
\hline
36&$\begin{matrix}\underline{F_4(a_2)}\\E_6(a_3)\end{matrix}\quad \underline{A_2}$&$(0,1,1,0,1,1,0,0)$&$(56, 38)$&$Spin(7)_{16}\times U(1)$\\
\hline
37&$\begin{matrix}\underline{F_4(a_2)}\\A_5\end{matrix}\quad \underline{A_2}$&$(0,0,1,0,2,1,0,0)$&$(63, 44)$&$(G_2)_{16} \times SU(2)_7 \times {U(1)}^2$\\
\hline
38&$\begin{matrix}\underline{F_4(a_2)}\\D_5(a_1)\end{matrix}\quad \underline{A_2}$&$(0,1,1,1,1,1,1,0)$&$(83, 64)$&$(G_2)_{16} \times U(1)$\\
\hline
39&$\begin{matrix}\underline{F_4(a_2)}\\D_4\end{matrix}\quad \underline{A_2}$&$(0,1,1,1,2,1,1,0)$&$(96, 75)$&$(G_2)_{16}\times SU(3)_{12}$\\
\hline
40&$\begin{matrix}\underline{F_4(a_2)}\\E_6(a_3)\end{matrix}\quad \underline{A_1+\widetilde{A}_1}$&$(0,1,1,1,1,1,0,0)$&$(66, 47)$&$\begin{gathered}SU(2)_{64-k_1-k_2}\times {SU(2)}_{k_1}\times {SU(2)}_{k_2}\\ \times SU(2)_{10}\end{gathered}$\\
\hline
41&$\begin{matrix}\underline{F_4(a_2)}\\A_5\end{matrix}\quad \underline{A_1+\widetilde{A}_1}$&$(0,0,1,1,2,1,0,0)$&$(73, 53)$&${SU(2)}^2_{32}\times SU(2)_{10}\times SU(2)_7$\\
\hline
42&$\begin{matrix}\underline{F_4(a_2)}\\D_5(a_1)\end{matrix}\quad \underline{A_1+\widetilde{A}_1}$&$(0,1,1,2,1,1,1,0)$&$(93, 73)$&$SU(2)_{64-k} \times SU(2)_k \times SU(2)_{10} \times U(1)$\\
\hline
43&$\begin{matrix}\underline{F_4(a_2)}\\D_4\end{matrix}\quad \underline{A_1+\widetilde{A}_1}$&$(0,1,1,2,2,1,1,0)$&$(106, 84)$&$SU(3)_{12}\times SU(2)_{64-k}\times SU(2)_k \times SU(2)_{10}$\\
\hline
44&$\begin{matrix}\underline{F_4(a_2)}\\E_6(a_3)\end{matrix}\quad \underline{\widetilde{A}_1}$&$(0,1,1,1,2,1,0,0)$&$(80, 58)$&$SU(4)_{12} \times {U(1)}^2$\\
\hline
45&$\begin{matrix}\underline{F_4(a_2)}\\A_5\end{matrix}\quad \underline{\widetilde{A}_1}$&$(0,0,1,1,3,1,0,0)$&$(87, 64)$&$SU(4)_{12} \times SU(2)_7 \times U(1)$\\
\hline
46&$\begin{matrix}\underline{F_4(a_2)}\\D_5(a_1)\end{matrix}\quad \underline{\widetilde{A}_1}$&$(0,1,1,2,2,1,1,0)$&$(107, 84)$&$SU(4)_{12}\times {U(1)}^2$\\
\hline
47&$\begin{matrix}\underline{F_4(a_2)}\\D_4\end{matrix}\quad \underline{\widetilde{A}_1}$&$(0,1,1,2,3,1,1,0)$&$(120, 95)$&$SU(4)_{12}\times SU(3)_{12} \times U(1)$\\
\hline
48&$\begin{matrix}\underline{C_3}\\A_2+2A_1\end{matrix}\quad \underline{C_3}$&$(0,2,2,2,1,1,1,0)$&$(106, 85)$&$SU(2)_{36} \times SU(2)_{18} \times {SU(2)}_6^2\times U(1)$\\
\hline
49&$\begin{matrix}\underline{C_3}\\2A_2\end{matrix}\quad \underline{C_3}$&$(0,2,2,1,2,1,0,0)$&$(92, 70)$&$Spin(7)_{12}\times {SU(2)}_6^2\times U(1)$\\
\hline
50&$\begin{matrix}\underline{C_3}\\A_2+A_1\end{matrix}\quad \underline{C_3}$&$(0,2,2,2,2,1,1,0)$&$(119, 96)$&$SU(3)_{12}\times {SU(2)}_6^2\times {U(1)}^2$\\
\hline
51&$\begin{matrix}\underline{C_3}\\A_2\end{matrix}\quad \underline{C_3}$&$(0,2,2,2,3,1,1,0)$&$(132, 107)$&${SU(3)}_{12}^2\times {SU(2)}_6^2\times U(1)$\\
\hline
52&$\begin{matrix}\underline{C_3}\\D_4(a_1)\end{matrix}\quad \underline{B_3}$&$(0,1,4,1,1,0,0,0)$&$(70, 53)$&${SU(2)}^3_8\times SU(2)_6 \times {U(1)}^2$\\
\hline
53&$\begin{matrix}\underline{C_3}\\A_3+A_1\end{matrix}\quad \underline{B_3}$&$(0,1,3,1,1,1,0,0)$&$(79, 61)$&$\begin{gathered}SU(2)_9\times SU(2)_{16} \times\\ SU(2)_8 \times SU(2)_6 \times U(1)\end{gathered}$\\
\hline
54&$\begin{matrix}\underline{C_3}\\A_3\end{matrix}\quad \underline{B_3}$&$(0,1,3,2,1,1,0,0)$&$(90, 70)$&$\begin{gathered}Sp(2)_{10} \times SU(2)_{16}\times\\ SU(2)_8 \times SU(2)_6 \times U(1)\end{gathered}$\\
\hline
55&$\begin{matrix}\underline{C_3}\\A_4+A_1\end{matrix}\quad \underline{C_3(a_1)}$&$(0,3,1,1,2,1,0,0)$&$(85, 68)$&$SU(2)_7\times SU(2)_6 \times {U(1)}^3$\\
\hline
56&$\begin{matrix}\underline{C_3}\\A_4\end{matrix}\quad \underline{C_3(a_1)}$&$(0,3,2,1,2,1,0,0)$&$(93, 75)$&$SU(2)_8\times SU(2)_7 \times SU(2)_6 \times {U(1)}^3$\\
\hline
57&$\begin{matrix}\underline{C_3}\\A_4+A_1\end{matrix}\quad \underline{\widetilde{A}_2+A_1}$&$(0,2,1,1,2,1,1,0)$&$(98, 80)$&$SU(2)_{20}\times SU(2)_6 \times {U(1)}^2$\\
\hline
58&$\begin{matrix}\underline{C_3}\\A_4\end{matrix}\quad \underline{\widetilde{A}_2+A_1}$&$(0,2,2,1,2,1,1,0)$&$(106, 87)$&$SU(2)_{20}\times SU(2)_8 \times SU(2)_6 \times {U(1)}^2$\\
\hline
59&$\begin{matrix}\underline{C_3}\\A_4+A_1\end{matrix}\quad \underline{B_2}$&$(0,2,1,1,3,1,0,0)$&$(92, 74)$&${SU(2)}_7^2\times SU(2)_6 \times {U(1)}^2$\\
\hline
60&$\begin{matrix}\underline{C_3}\\A_4\end{matrix}\quad \underline{B_2}$&$(0,2,2,1,3,1,0,0)$&$(100, 81)$&$SU(2)_8\times {SU(2)}_7^2\times SU(2)_6 \times {U(1)}^2$\\
\hline
61&$\begin{matrix}\underline{C_3}\\A_4+A_1\end{matrix}\quad \underline{\widetilde{A}_2}$&$(0,2,1,2,2,1,1,0)$&$(110, 89)$&$(G_2)_{10}\times SU(2)_6 \times {U(1)}^2$\\
\hline
62&$\begin{matrix}\underline{C_3}\\A_4\end{matrix}\quad \underline{\widetilde{A}_2}$&$(0,2,2,2,2,1,1,0)$&$(118, 96)$&$(G_2)_{10}\times SU(2)_8 \times SU(2)_6 \times {U(1)}^2$\\
\hline
63&$\begin{matrix}\underline{C_3}\\E_6(a_3)\end{matrix}\quad \underline{A_2}$&$(0,2,1,0,1,1,0,0)$&$(62, 43)$&$Spin(7)_{16} \times SU(2)_6$\\
\hline
64&$\begin{matrix}\underline{C_3}\\A_5\end{matrix}\quad \underline{A_2}$&$(0,1,1,0,2,1,0,0)$&$(69, 49)$&$(G_2)_{16}\times SU(2)_7 \times SU(2)_6 \times U(1)$\\
\hline
65&$\begin{matrix}\underline{C_3}\\D_5(a_1)\end{matrix}\quad \underline{A_2}$&$(0,2,1,1,1,1,1,0)$&$(89, 69)$&$(G_2)_{16} \times SU(2)_6 \times U(1)$\\
\hline
66&$\begin{matrix}\underline{C_3}\\D_4\end{matrix}\quad \underline{A_2}$&$(0,2,1,1,2,1,1,0)$&$(102, 80)$&$(G_2)_{16}\times SU(3)_{12} \times SU(2)_6$\\
\hline
67&$\begin{matrix}\underline{C_3}\\E_6(a_3)\end{matrix}\quad \underline{A_1+\widetilde{A}_1}$&$(0,2,1,1,1,1,0,0)$&$(72, 52)$&$SU(2)_{32}\times {SU(2)}^2_{16}\times SU(2)_{10} \times SU(2)_6$\\
\hline
68&$\begin{matrix}\underline{C_3}\\A_5\end{matrix}\quad \underline{A_1+\widetilde{A}_1}$&$(0,1,1,1,2,1,0,0)$&$(79, 58)$&${SU(2)}_{32}^2 \times SU(2)_{10} \times SU(2)_7 \times SU(2)_6$\\
\hline
69&$\begin{matrix}\underline{C_3}\\D_5(a_1)\end{matrix}\quad \underline{A_1+\widetilde{A}_1}$&$(0,2,1,2,1,1,1,0)$&$(99, 78)$&$\begin{gathered}SU(2)_{64-k}\times SU(2)_{k}\times\\ SU(2)_{10} \times SU(2)_6 \times U(1)\end{gathered}$\\
\hline
70&$\begin{matrix}\underline{C_3}\\D_4\end{matrix}\quad \underline{A_1+\widetilde{A}_1}$&$(0,2,1,2,2,1,1,0)$&$(112, 89)$&$\begin{gathered}SU(3)_{12}\times SU(2)_{64-k}\times SU(2)_k\\ \times SU(2)_{10} \times SU(2)_6\end{gathered}$\\
\hline
71&$\begin{matrix}\underline{C_3}\\E_6(a_3)\end{matrix}\quad \underline{\widetilde{A}_1}$&$(0,2,1,1,2,1,0,0)$&$(86, 63)$&$SU(4)_{12}\times SU(2)_6 \times {U(1)}^2$\\
\hline
72&$\begin{matrix}\underline{C_3}\\A_5\end{matrix}\quad \underline{\widetilde{A}_1}$&$(0,1,1,1,3,1,0,0)$&$(93, 69)$&$SU(4)_{12}\times SU(2)_7 \times SU(2)_6 \times U(1)$\\
\hline
73&$\begin{matrix}\underline{C_3}\\D_5(a_1)\end{matrix}\quad \underline{\widetilde{A}_1}$&$(0,2,1,2,2,1,1,0)$&$(113, 89)$&$SU(4)_{12} \times SU(2)_6 \times {U(1)}^2$\\
\hline
74&$\begin{matrix}\underline{C_3}\\D_4\end{matrix}\quad \underline{\widetilde{A}_1}$&$(0,2,1,2,3,1,1,0)$&$(126, 100)$&$SU(4)_{12}\times SU(3)_{12} \times SU(2)_6 \times U(1)$\\
\hline
75&$\begin{matrix}\underline{B_3}\\D_5(a_1)\end{matrix}\quad \underline{F_4(a_3)}$&$(0,4,1,1,0,0,0,0)$&$(51, 36)$&${SU(2)}_6^4 \times {U(1)}^3$\\
\hline
76&$\begin{matrix}\underline{B_3}\\D_4\end{matrix}\quad \underline{F_4(a_3)}$&$(0,4,1,1,1,0,0,0)$&$(64, 47)$&$SU(3)_{12}\times {SU(2)}^4_{6}$\\
\hline
77&$\begin{matrix}\underline{B_3}\\D_5(a_1)\end{matrix}\quad \underline{C_3(a_1)}$&$(0,3,1,1,1,0,0,0)$&$(58, 42)$&$SU(2)_{12}\times {SU(2)}^2_{6}\times SU(2)_7\times {U(1)}^2$\\
\hline
78&$\begin{matrix}\underline{B_3}\\D_4\end{matrix}\quad \underline{C_3(a_1)}$&$(0,3,1,1,2,0,0,0)$&$(71, 53)$&$SU(3)_{12}\times SU(2)_{12}\times {SU(2)}^2_{6}\times SU(2)_7$\\
\hline
79&$\begin{matrix}\underline{B_3}\\D_5(a_1)\end{matrix}\quad \underline{\widetilde{A}_2+A_1}$&$(0,2,1,1,1,0,1,0)$&$(71, 54)$&$SU(2)_{20}\times SU(2)_{18}\times SU(2)_6\times U(1)$\\
\hline
80&$\begin{matrix}\underline{B_3}\\D_4\end{matrix}\quad \underline{\widetilde{A}_2+A_1}$&$(0,2,1,1,2,0,1,0)$&$(84, 65)$&$SU(3)_{12}\times SU(2)_{20} \times SU(2)_{18}\times SU(2)_6$\\
\hline
81&$\begin{matrix}\underline{B_3}\\D_5(a_1)\end{matrix}\quad \underline{B_2}$&$(0,2,1,1,2,0,0,0)$&$(65, 48)$&${SU(2)}_7^2\times {SU(2)}^2_{12} \times {U(1)}^2$\\
\hline
82&$\begin{matrix}\underline{B_3}\\D_4\end{matrix}\quad \underline{B_2}$&$(0,2,1,1,3,0,0,0)$&$(78, 59)$&$SU(3)_{12}\times {SU(2)}_7^2\times {SU(2)}^2_{12}$\\
\hline
83&$\begin{matrix}\underline{B_3}\\E_6(a_3)\end{matrix}\quad \underline{A_2+\widetilde{A}_1}$&$(0,1,1,0,1,0,0,1)$&$(63, 46)$&$SU(2)_k \times SU(2)_{39-k}\times SU(2)_{24}$\\
\hline
84&$\begin{matrix}\underline{B_3}\\A_5\end{matrix}\quad \underline{A_2+\widetilde{A}_1}$&$(0,0,1,0,2,0,0,1)$&$(70, 52)$&$SU(2)_7\times SU(2)_{26} \times SU(2)_{13}\times SU(2)_{24}$\\
\hline
85&$\begin{matrix}\underline{B_3}\\E_6(a_3)\end{matrix}\quad \underline{\widetilde{A}_2}$&$(0,2,1,1,1,0,0,0)$&$(56, 37)$&$Spin(7)_{10}\times SU(2)_{12} \times {SU(2)}^2_{6}$\\
\hline
86&$\begin{matrix}\underline{B_3}\\A_5\end{matrix}\quad \underline{\widetilde{A}_2}$&$(0,1,1,1,2,0,0,0)$&$(63, 43)$&$Spin(7)_{10}\times SU(2)_7 \times {SU(2)}^2_{12}$\\
\hline
87&$\begin{matrix}\underline{B_3}\\D_5(a_1)\end{matrix}\quad \underline{\widetilde{A}_2}$&$(0,2,1,2,1,0,1,0)$&$(83, 63)$&$(G_2)_{10}\times SU(2)_{18}\times SU(2)_6\times U(1)$\\
\hline
88&$\begin{matrix}\underline{B_3}\\D_4\end{matrix}\quad \underline{\widetilde{A}_2}$&$(0,2,1,2,2,0,1,0)$&$(96, 74)$&$(G_2)_{10}\times SU(3)_{12}\times SU(2)_{18}\times SU(2)_6$\\
\hline
89&$\begin{matrix}\underline{B_3}\\E_6(a_3)\end{matrix}\quad \underline{A_2}$&$(0,1,1,0,1,1,0,1)$&$(80, 61)$&$SU(3)_{16}\times SU(2)_{24} \times U(1)$\\
\hline
90&$\begin{matrix}\underline{B_3}\\A_5\end{matrix}\quad \underline{A_2}$&$(0,0,1,0,2,1,0,1)$&$(87, 67)$&$SU(3)_{16}\times SU(2)_7 \times SU(2)_{24} \times U(1)$\\
\hline
91&$\begin{matrix}\underline{B_3}\\D_5\end{matrix}\quad \underline{0}$&$(0,0,2,1,0,0,0,0)$&$(56, 23)$&$(E_7)_8\times (F_4)_{10} \times U(1)$\\
\hline
92&$\begin{matrix}\underline{F_4(a_3)}\\A_4+A_1\end{matrix}\quad \underline{F_4(a_2)}$&$(0,3,1,1,1,1,0,0)$&$(72, 57)$&${U(1)}^4$\\
\hline
93&$\begin{matrix}\underline{F_4(a_3)}\\A_4\end{matrix}\quad \underline{F_4(a_2)}$&$(0,3,2,1,1,1,0,0)$&$(80, 64)$&$SU(2)_8\times {U(1)}^4$\\
\hline
94&$\begin{matrix}\underline{F_4(a_3)}\\A_4+A_1\end{matrix}\quad \underline{C_3}$&$(0,4,1,1,1,1,0,0)$&$(78, 62)$&$SU(2)_6\times {U(1)}^4$\\
\hline
95&$\begin{matrix}\underline{F_4(a_3)}\\A_4\end{matrix}\quad \underline{C_3}$&$(0,4,2,1,1,1,0,0)$&$(86, 69)$&$SU(2)_8\times SU(2)_6 \times {U(1)}^4$\\
\hline
96&$\begin{matrix}\underline{F_4(a_3)}\\D_5\end{matrix}\quad \underline{A_2}$&$(0,3,1,0,0,1,0,0)$&$(56, 37)$&$Spin(8)_{16}\times U(1)$\\
\hline
97&$\begin{matrix}\underline{F_4(a_3)}\\D_5\end{matrix}\quad \underline{A_1+\widetilde{A}_1}$&$(0,3,1,1,0,1,0,0)$&$(66, 46)$&${SU(2)}^4_{16} \times SU(2)_{10} \times U(1)$\\
\hline
98&$\begin{matrix}\underline{F_4(a_3)}\\D_5\end{matrix}\quad \underline{\widetilde{A}_1}$&$(0,3,1,1,1,1,0,0)$&$(80, 57)$&$SU(4)_{12}\times {U(1)}^4$\\
\hline
99&$\begin{matrix}\underline{F_4(a_3)}\\E_6(a_1)\end{matrix}\quad \underline{0}$&$(0,3,0,0,0,0,0,0)$&$(48, 15)$&$[{(E_6)_6}\text{ SCFT}]^3$\\
\hline
100&$\begin{matrix}\underline{C_3(a_1)}\\D_5\end{matrix}\quad \underline{A_2}$&$(0,2,1,0,1,1,0,0)$&$(63, 43)$&$Spin(7)_{16}\times SU(2)_7 \times U(1)$\\
\hline
101&$\begin{matrix}\underline{C_3(a_1)}\\D_5\end{matrix}\quad \underline{A_1+\widetilde{A}_1}$&$(0,2,1,1,1,1,0,0)$&$(73, 52)$&$\begin{gathered}SU(2)_{32}\times {SU(2)}^2_{16}\times\\ SU(2)_{10} \times SU(2)_7 \times U(1)\end{gathered}$\\
\hline
102&$\begin{matrix}\underline{C_3(a_1)}\\D_5\end{matrix}\quad \underline{\widetilde{A}_1}$&$(0,2,1,1,2,1,0,0)$&$(87, 63)$&$SU(4)_{12}\times SU(2)_7 \times {U(1)}^3$\\
\hline
103&$\begin{matrix}\underline{C_3(a_1)}\\E_6(a_1)\end{matrix}\quad \underline{0}$&$(0,2,0,0,1,0,0,0)$&$(55, 21)$&$[(E_6)_{6} \text{ SCFT}]\times [(E_6)_{12} \times SU(2)_7 \text{ SCFT}]$\\
\hline
104&$\begin{matrix}\underline{\widetilde{A}_2+A_1}\\D_5\end{matrix}\quad \underline{A_2}$&$(0,1,1,0,1,1,1,0)$&$(76, 55)$&$(G_2)_{16}\times SU(2)_{20} \times U(1)$\\
\hline
105&$\begin{matrix}\underline{\widetilde{A}_2+A_1}\\D_5\end{matrix}\quad \underline{A_1+\widetilde{A}_1}$&$(0,1,1,1,1,1,1,0)$&$(86, 64)$&$\begin{gathered}SU(2)_{48}\times SU(2)_{16} \times\\ SU(2)_{10} \times SU(2)_{20} \times U(1)\end{gathered}$\\
\hline
106&$\begin{matrix}\underline{\widetilde{A}_2+A_1}\\D_5\end{matrix}\quad \underline{\widetilde{A}_1}$&$(0,1,1,1,2,1,1,0)$&$(100, 75)$&$SU(4)_{12}\times SU(2)_{20} \times {U(1)}^2$\\
\hline
107&$\begin{matrix}\underline{\widetilde{A}_2+A_1}\\E_6(a_1)\end{matrix}\quad \underline{0}$&$(0,1,0,0,1,0,1,0)$&$(68, 33)$&$(E_6)_{18} \times SU(2)_{20} \text{ SCFT}$\\
\hline
108&$\begin{matrix}\underline{B_2}\\D_5\end{matrix}\quad \underline{A_2}$&$(0,1,1,0,2,1,0,0)$&$(70, 49)$&$(G_2)_{16}\times {SU(2)}_7^2\times {U(1)}^2$\\
\hline
109&$\begin{matrix}\underline{B_2}\\D_5\end{matrix}\quad \underline{A_1+\widetilde{A}_1}$&$(0,1,1,1,2,1,0,0)$&$(80, 58)$&${SU(2)}_{32}^2\times SU(2)_{10}\times{SU(2)}_7^2\times U(1)$\\
\hline
110&$\begin{matrix}\underline{B_2}\\D_5\end{matrix}\quad \underline{\widetilde{A}_1}$&$(0,1,1,1,3,1,0,0)$&$(94, 69)$&$SU(4)_{12}\times {SU(2)}^2_7 \times {U(1)}^2$\\
\hline
111&$\begin{matrix}\underline{B_2}\\E_6(a_1)\end{matrix}\quad \underline{0}$&$(0,1,0,0,2,0,0,0)$&$(62, 27)$&$[(E_6)_6 \text{ SCFT}] \times [(F_4)_{12} \times {SU(2)}_{7}^2 \text{ SCFT}]$\\
\hline
112&$\begin{matrix}\underline{A_2+\widetilde{A}_1}\\D_5\end{matrix}\quad \underline{A_2+\widetilde{A}_1}$&$(0,0,1,0,1,0,1,1)$&$(78, 58)$&$Sp(2)_{39} \times U(1)$\\
\hline
113&$\begin{matrix}\underline{A_2+\widetilde{A}_1}\\E_6(a_1)\end{matrix}\quad \underline{A_1}$&$(0,0,0,0,1,0,0,1)$&$(62, 34)$&$Sp(4)_{13}\times SU(2)_{26}$\\
\hline
114&$\begin{matrix}\underline{\widetilde{A}_2}\\D_5\end{matrix}\quad \underline{A_2}$&$(0,1,1,1,1,1,1,0)$&$(88, 64)$&$(G_2)_{16}\times(G_2)_{10}\times U(1)$\\
\hline
115&$\begin{matrix}\underline{\widetilde{A}_2}\\D_5\end{matrix}\quad \underline{A_1+\widetilde{A}_1}$&$(0,1,1,2,1,1,1,0)$&$(98, 73)$&$\begin{gathered}(G_2)_{10} \times SU(2)_{64-k}\times SU(2)_k\\ \times SU(2)_{10} \times U(1)\end{gathered}$\\
\hline
116&$\begin{matrix}\underline{\widetilde{A}_2}\\D_5\end{matrix}\quad \underline{\widetilde{A}_1}$&$(0,1,1,2,2,1,1,0)$&$(112, 84)$&$SU(4)_{12}\times (G_2)_{10} \times {U(1)}^2$\\
\hline
117&$\begin{matrix}\underline{\widetilde{A}_2}\\E_6(a_1)\end{matrix}\quad \underline{0}$&$(0,1,0,1,1,0,1,0)$&$(80, 42)$&$(E_6)_{18} \times (G_2)_{10} \text{ SCFT}$\\
\hline
118&$\begin{matrix}\underline{A_2}\\E_6(a_1)\end{matrix}\quad \underline{\widetilde{A}_1}$&$(0,0,0,0,2,1,0,0)$&$(64, 37)$&$Spin(7)_{12}\times (G_2)_{16}\times U(1)$\\
\hline
119&$\begin{matrix}\underline{A_2}\\E_6(a_1)\end{matrix}\quad \underline{A_1}$&$(0,0,0,0,1,1,0,1)$&$(79, 49)$&$Sp(3)_{13} \times SU(3)_{16} \times U(1)$\\
\hline
120&$\begin{matrix}\underline{A_1+\widetilde{A}_1}\\E_6(a_1)\end{matrix}\quad \underline{A_1+\widetilde{A}_1}$&$(0,0,0,1,1,1,0,0)$&$(60, 35)$&$SU(4)_{32}\times Sp(2)_{10}$\\
\hline
121&$\begin{matrix}\underline{A_1+\widetilde{A}_1}\\E_6(a_1)\end{matrix}\quad \underline{\widetilde{A}_1}$&$(0,0,0,1,2,1,0,0)$&$(74, 46)$&$SU(4)_{12}\times {SU(2)}^2_{32}\times SU(2)_{10}$\\
\hline
122&$\begin{matrix}\underline{\widetilde{A}_1}\\E_6(a_1)\end{matrix}\quad \underline{\widetilde{A}_1}$&$(0,0,0,1,3,1,0,0)$&$(88, 57)$&${SU(4)}_{12}^2\times U(1)$\\
\hline
\end{longtable}
}

\subsection{Mixed fixtures}\label{mixed_fixtures}
{\footnotesize
\renewcommand{\arraystretch}{1.75}
\begin{longtable}{|c|l|c|c|c|}
\hline
\#&Fixture&$(n_2,n_3,n_4,n_5,n_6,n_8,n_9,n_{12})$&$(n_h,n_v)$&Theory\\
\hline 
\endhead
1&$\begin{matrix}\underline{F_4}\\A_3+A_1\end{matrix}\quad (\underline{0},Spin(9))$&$(0,1,0,0,0,0,0,0)$&$(16,5)$&$(E_6)_{6}\,\text{SCFT} + 1(9)$\\
\hline
2&$\begin{matrix}\underline{F_4}\\2A_1\end{matrix}\quad \underline{A_2+\widetilde{A}_1}$&$(0,1,0,0,1,0,0,0)$&$(38, 16)$&$(E_6)_{12} \times SU(2)_7 + \frac{1}{2}(1,2)+\frac{1}{2}(7,2)$\\
\hline
3&$\begin{matrix}\underline{F_4}\\A_1\end{matrix}\quad \underline{A_2+\widetilde{A}_1}$&$(0,1,0,0,1,0,1,0)$&$(68, 33)$&$(E_6)_{18}\times SU(2)_{20}+\frac{1}{2}(1,2)$\\
\hline
4&$\begin{matrix}\underline{F_4}\\3A_1\end{matrix}\quad \underline{A_2}$&$(0,1,0,0,1,0,0,0)$&$(39, 16)$&$(E_6)_{12} \times SU(2)_7 + (1,2,3)$\\
\hline
5&$\begin{matrix}\underline{F_4}\\3A_1\end{matrix}\quad \underline{A_1+\widetilde{A}_1}$&$(0,1,0,1,1,0,0,0)$&$(52, 25)$&$SU(6)_{12}\times Spin(7)_{10} + \frac{1}{2}(1,2,3,1)$\\
\hline
6&$\begin{matrix}\underline{F_4}\\3A_1\end{matrix}\quad \underline{\widetilde{A}_1}$&$(0,1,0,1,2,0,0,0)$&$(68, 36)$&$SU(6)_{12} \times {SU(3)}_{12}^2 + \frac{1}{2}(1,2,1)$\\
\hline
7&$\begin{matrix}\underline{F_4}\\A_2+2A_1\end{matrix}\quad \underline{A_1}$&$(0,1,0,1,0,0,0,0)$&$(36, 14)$&$Spin(14)_{10}\times U(1) +\frac{1}{2}(3,6)$\\
\hline
8&$\begin{matrix}\underline{F_4}\\A_2+A_1\end{matrix}\quad \underline{A_1}$&$(0,1,0,1,1,0,0,0)$&$(55, 25)$&$SU(9)_{12}\times U(1)+\frac{1}{2}(1,6)$\\
\hline
9&$\begin{matrix}\underline{F_4}\\A_2\end{matrix}\quad \underline{A_1}$&$(0,1,0,1,2,0,0,0)$&$(68, 36)$&$SU(6)_{12} \times {SU(3)}_{12}^2 + \frac{1}{2}(1,1,6)$\\
\hline
10&$\begin{matrix}\underline{F_4}\\2A_2+A_1\end{matrix}\quad \underline{0}$&$(0,1,0,0,0,0,0,0)$&$(16, 5)$&$(E_6)_{6}\,\text{SCFT} + \frac{1}{2}(26,2)$\\
\hline
11&$\begin{matrix}\underline{F_4(a_2)}\\2A_2+A_1\end{matrix}\quad \underline{F_4(a_2)}$&$(0,0,2,1,1,1,0,0)$&$(65, 49)$&$SU(2)_{25-k} \times SU(2)_k \times U(1) + \frac{1}{2}(2)$\\
\hline
12&$\begin{matrix}\underline{F_4(a_2)}\\2A_2+A_1\end{matrix}\quad \underline{C_3}$&$(0,1,2,1,1,1,0,0)$&$(71, 54)$&$\begin{gathered}SU(2)_{16} \times SU(2)_9 \times SU(2)_6 \times U(1)\\ + \frac{1}{2}(2,1)\end{gathered}$\\
\hline
13&$\begin{matrix}\underline{F_4(a_2)}\\E_6(a_3)\end{matrix}\quad \underline{A_2+\widetilde{A}_1}$&$(0,1,1,0,1,0,0,0)$&$(37, 23)$&$Sp(2)_{12}\times SU(2)_7 \times SU(2)_6 + 2$\\
\hline
14&$\begin{matrix}\underline{F_4(a_2)}\\A_5\end{matrix}\quad \underline{A_2+\widetilde{A}_1}$&$(0,0,1,0,2,0,0,0)$&$(45, 29)$&$Sp(2)_7 \times SU(2)_7 \times {SU(2)}_{12}^2+\frac{1}{2}(1,2)$\\
\hline
15&$\begin{matrix}\underline{F_4(a_2)}\\D_5(a_1)\end{matrix}\quad \underline{A_2+\widetilde{A}_1}$&$(0,1,1,1,1,0,1,0)$&$(65, 49)$&$SU(2)_{38-k}\times SU(2)_k \times U(1) + \frac{1}{2}(2)$\\
\hline
16&$\begin{matrix}\underline{F_4(a_2)}\\D_4\end{matrix}\quad \underline{A_2+\widetilde{A}_1}$&$(0,1,1,1,2,0,1,0)$&$(78, 60)$&$SU(3)_{12}\times SU(2)_{20} \times SU(2)_{18} + \frac{1}{2}(1,2)$\\
\hline
17&$\begin{matrix}\underline{C_3}\\2A_2+A_1\end{matrix}\quad \underline{C_3}$&$(0,2,2,1,1,1,0,0)$&$(77, 59)$&$\begin{gathered}SU(2)_{16}\times SU(2)_9 \times {SU(2)}_6^2 \times U(1)\\ + \frac{1}{2}(2,1,1)\end{gathered}$\\
\hline
18&$\begin{matrix}\underline{C_3}\\E_6(a_3)\end{matrix}\quad \underline{A_2+\widetilde{A}_1}$&$(0,2,1,0,1,0,0,0)$&$(43, 28)$&${SU(2)}_{12}^2 \times {SU(2)}_6^2 \times SU(2)_7 + (2,1)$\\
\hline
19&$\begin{matrix}\underline{C_3}\\A_5\end{matrix}\quad \underline{A_2+\widetilde{A}_1}$&$(0,1,1,0,2,0,0,0)$&$(51, 34)$&$\begin{gathered}Sp(2)_7 \times SU(2)_{24} \times SU(2)_7 \times SU(2)_6\\ + \frac{1}{2}(1,2,1)\end{gathered}$\\
\hline
20&$\begin{matrix}\underline{C_3}\\D_5(a_1)\end{matrix}\quad \underline{A_2+\widetilde{A}_1}$&$(0,2,1,1,1,0,1,0)$&$(71, 54)$&$\begin{gathered}SU(2)_{20}\times SU(2)_{18} \times SU(2)_6 \times U(1)\\ + \frac{1}{2}(2,1)\end{gathered}$\\
\hline
21&$\begin{matrix}\underline{C_3}\\D_4\end{matrix}\quad \underline{A_2+\widetilde{A}_1}$&$(0,2,1,1,2,0,1,0)$&$(84, 65)$&$\begin{gathered}SU(3)_{12}\times SU(2)_{20}\times SU(2)_{18} \times SU(2)_6\\ +\frac{1}{2}(1,2,1)\end{gathered}$\\
\hline
22&$\begin{matrix}\underline{B_3}\\E_6(a_3)\end{matrix}\quad \underline{\widetilde{A}_2+A_1}$&$(0,2,1,0,1,0,0,0)$&$(43, 28)$&${SU(2)}_{12}^2 \times {SU(2)}_6^2 \times SU(2)_7+\frac{1}{2}(2,1)$\\
\hline
23&$\begin{matrix}\underline{B_3}\\A_5\end{matrix}\quad \underline{\widetilde{A}_2+A_1}$&$(0,1,1,0,2,0,0,0)$&$(50, 34)$&$SU(2)_7\times {SU(2)}^3_{12} \times SU(2)_7 + \frac{1}{2}(1,2,1)$\\
\hline
24&$\begin{matrix}\underline{F_4(a_3)}\\D_5\end{matrix}\quad \underline{A_2+\widetilde{A}_1}$&$(0,3,1,0,0,0,0,0)$&$(36, 22)$&${SU(2)}_6^6\times U(1) + \frac{3}{2}(2)$\\
\hline
25&$\begin{matrix}\underline{C_3(a_1)}\\D_5\end{matrix}\quad \underline{A_2+\widetilde{A}_1}$&$(0,2,1,0,1,0,0,0)$&$(44, 28)$&$\begin{gathered}Sp(2)_7 \times {SU(2)}_{12}^2 \times SU(2)_6 \times U(1)\\ + (2,1)\end{gathered}$\\
\hline
26&$\begin{matrix}\underline{\widetilde{A}_2+A_1}\\D_5\end{matrix}\quad \underline{A_2+\widetilde{A}_1}$&$(0,1,1,0,1,0,1,0)$&$(58, 40)$&$Sp(2)_{20}\times SU(2)_{18} \times U(1) + \frac{1}{2}(2,1)$\\
\hline
27&$\begin{matrix}\underline{\widetilde{A}_2+A_1}\\E_6(a_1)\end{matrix}\quad \underline{A_1}$&$(0,1,0,0,1,0,0,0)$&$(39, 16)$&$(E_6)_{12}\times SU(2)_7 + \frac{1}{2}(6,1)+\frac{1}{2}(1,2)$\\
\hline
28&$\begin{matrix}\underline{B_2}\\D_5\end{matrix}\quad \underline{A_2+\widetilde{A}_1}$&$(0,1,1,0,2,0,0,0)$&$(52, 34)$&${Sp(2)}^2_7\times SU(2)_{24} \times U(1) + \frac{1}{2}(2,1,1)$\\
\hline
29&$\begin{matrix}\underline{A_2+\widetilde{A}_1}\\E_6(a_1)\end{matrix}\quad \underline{A_1+\widetilde{A}_1}$&$(0,0,0,0,1,0,0,0)$&$(27, 11)$&$Sp(5)_7 + \frac{1}{2}(3,1,2)+\frac{1}{2}(1,2,3)$\\
\hline
30&$\begin{matrix}\underline{A_2+\widetilde{A}_1}\\E_6(a_1)\end{matrix}\quad \underline{\widetilde{A}_1}$&$(0,0,0,0,2,0,0,0)$&$(46, 22)$&$(F_4)_{12}\times {SU(2)}_7^2 + \frac{1}{2}(1,2)$\\
\hline
31&$\begin{matrix}\underline{\widetilde{A}_2}\\D_5\end{matrix}\quad \underline{A_2+\widetilde{A}_1}$&$(0,1,1,1,1,0,1,0)$&$(70, 49)$&$\begin{gathered}SU(2)_{20}\times SU(2)_{18} \times (G_2)_{10} \times U(1)\\ + \frac{1}{2}(2,1)\end{gathered}$\\
\hline
32&$\begin{matrix}\underline{\widetilde{A}_2}\\E_6(a_1)\end{matrix}\quad \underline{A_1}$&$(0,1,0,1,1,0,0,0)$&$(52, 25)$&$SU(6)_{12}\times Spin(7)_{10}+\frac{1}{2}(6,1)$\\
\hline
33&$\begin{matrix}\underline{A_2}\\E_6(a_1)\end{matrix}\quad \underline{A_1+\widetilde{A}_1}$&$(0,0,0,0,1,1,0,0)$&$(49, 26)$&$SU(6)_{16}\times SU(2)_9 + \frac{1}{2}(1,2,1)$\\
\hline
\end{longtable}
}

We note that for mixed fixture 22 on our list, the order $q$ (equivalently, $\tau^2$) term in the expansion of its superconformal index implies that the global symmetry is enhanced to $SU(2)_{19-k} \times SU(2)_{k} \times SU(2)_{24-k_1-k_2} \times SU(2)_{k_1}\times SU(2)_{k_2}$. Since we are not able to gauge any of the punctures, we cannot determine the levels $k,k_1,k_2$ using an S-duality.

However, by setting $k=7,k_1=k_2=6$, its properties agree with that of mixed fixture 18, up to the addition of a half-hypermultiplet. As further evidence, we have checked that the next non-trivial term in the expansion of the superconformal index is the same for each theory:

\begin{displaymath}
\begin{split}
\mathcal{I}_{\# 18}&=\mathcal{I}_{\# 22} \times \mathcal{I}_{\text{free}}\\
&=(1+2q^{\frac{1}{2}}+18q+66q^{\frac{3}{2}}+\dots)(1+2q^{\frac{1}{2}}+3q+6q^{\frac{3}{2}}+\dots)\\
&=1+4q^{\frac{1}{2}}+25q+114q^\frac{3}{2}+\dots 
\end{split}
\end{displaymath}
Thus we conjecture that the SCFT realized by fixture 22 is the same as that of 18, and fill in the levels in the table above.

\subsection{Gauge theory fixtures}\label{gauge_theory_fixtures}

Gauge theory fixtures are 3-punctured spheres with 1 or 2 insertions of $F_4(a_1)$. There are 167 such 3-punctured spheres involving three regular punctures and 1 involving an irregular puncture. Of the 167, all but 1 are resolved by replacing the $F_4(a_1)$ by the pair $F_4$, $E_6(a_1)$; that is, they can be thought-of as 4-punctured spheres in disguise. The remaining case involves \emph{two} $F_4(a_1)$ punctures and is really a 5-punctured sphere in disguise. The two exceptional cases are listed in the table below. Note that the latter involves two decoupled copies of a theory to be discussed at greater length in \S\ref{ME81SU3}.

{
\renewcommand{\arraystretch}{1.5}

\begin{longtable}{|c|c|c|c|}
\hline
\#&Fixture&$(n_2,n_3,n_4,n_5,n_6,n_8,n_9,n_{12})$&Theory\\
\hline
\endhead
1&$\begin{matrix}\underline{F_4(a_1)}\\D_5\end{matrix}\quad (\underline{\widetilde{A}_1},SU(4)_4)$&$(1,0,0,0,0,0,0,0)$&$SU(2)+4(2)$\\
\hline
2&$\begin{matrix}\underline{F_4(a_1)}\\0\end{matrix}\quad \underline{F_4(a_1)}$&$(2,2,0,0,2,0,0,0)$&$\begin{gathered}{\bigl[SU(3) + {(E_8)}_{12}\bigr]}^2\\ \simeq {\bigl[SU(2) +\tfrac{1}{2}(2) + {(E_6)}_{12}\times {SU(2)}_7\bigr]}^2\end{gathered}$\\
\hline
\end{longtable}
}

\section{Global symmetries and the superconformal index}\label{global_symmetries_and_the_superconformal_index}

To determine the global symmetry of each SCFT and the number of free hypermultiplets for each fixture, we use the superconformal index \cite{Gadde:2011ik,Kinney:2005ej,Gadde:2009kb,Gadde:2011uv,Lemos:2012ph}. This analysis was carried out for the untwisted $E_6$ fixtures in \cite{Chacaltana:2014jba}. We leave many of the details to that paper, and the references therein.

\subsection{Superconformal index for twisted fixtures}\label{superconformal_index_for_twisted_fixtures}

Following \cite{Chacaltana:2013oka,Lemos:2012ph,Mekareeya:2012tn}, we assume that the superconformal index for an $E_6$ fixture with a two twisted punctures and one untwisted puncture takes the usual form:

\begin{equation}
\mathcal{I}(\mathbf{a}_i;\tau)=\mathcal{A}(\tau)\sum_{(a_1,a_2,a_3,a_4)}\frac{\mathcal{K}(\mathbf{a}_1;\tau)P_{E_6}^{(a_1,a_2,a_3,a_2,a_1,a_4)}(\mathbf{a}_1;\tau)\prod_{i=2}^3\mathcal{K}(\mathbf{a}_i;\tau)P_{F_4}^{(a_1,a_2,a_3,a_4)}(\mathbf{a}_i;\tau)}{P_{E_6}^{(a_1,a_2,a_3,a_2,a_1,a_4)}(\mathbf{a}_\text{triv};\tau)}
\label{SCI}\end{equation}
where the sum runs over finite-dimensional irreducible representations of $F_4$, and the Dynkin labels of each $E_6$ representation are determined by those of the corresponding $F_4$ representation, as indicated.

To obtain this formula, one can use the fact that, when $C$ has genus zero, the Hall-Littlewood limit of the superconformal index coincides with the Coulomb branch Hilbert series of the $3d$ mirror of the $(2,0)$ theory on $C \times S^1$. For a fixture of type $E_6$ with twisted punctures, the 3d mirror is obtained by assigning the 3d $\mathcal{N}=4$ SCFT $T_{\tilde{\rho}}(F_4)$ to each twisted puncture $\tilde{\rho}$ and the SCFT $T_\rho(E_6)$ to the untwisted puncture $\rho$, and gauging the common centerless flavor symmetry $F_4/Z(F_4)$ \cite{Gaiotto:2008ak,Benini:2010uu}. The Coulomb branch Hilbert series can then be computed following \cite{Cremonesi:2014kwa, Cremonesi:2014vla}, giving \eqref{SCI}.

The Taylor expansion of the superconformal index is given by \cite{Gaiotto:2012uq}

\begin{displaymath}
\mathcal{I}(\mathbf{a_i};\tau)=1+ \tau \chi^R_{free}(\mathbf{a_i}) + \tau^2 (\chi^{adj}_{free}(\mathbf{a_i})+\chi^{adj}_{SCFT}(\mathbf{a_i}))+\dots
\end{displaymath}
allowing us to read off the number of free hypermultiplets and the global symmetry group of the interacting SCFT for a given fixture. Examples of this type of calculation can be found in \cite{Chacaltana:2013oka,Chacaltana:2014jba,Gaiotto:2012uq}.

\subsection{Higher-order expansion of the index}\label{higherorder_expansion_of_the_index}

Computing the expansion of \eqref{SCI} to higher-order becomes very tedious due to the sum over the Weyl group in the definition of the Hall-Littlewood polynomials. We will therefore also be interested in the Schur limit of the superconformal index, where the Hall-Littlewood polynomials are replaced by characters of the corresponding representations\footnote{This limit corresponds to the $(0,q,t=q)$ slice in the space of superconformal fugacities \cite{Gadde:2011uv}.}. For a twisted fixture, this is given by

\begin{equation}
\mathcal{I}(\mathbf{a}_i;q)=\prod_{j=2,5,6,8,9,12}(q^j;q)\sum_{(a_1,a_2,a_3,a_4)}\frac{\prod_{i=1}^2\mathcal{K}(\mathbf{a}_i)\chi^{(a_1,a_2,a_3,a_4)}_{F_4}(\mathbf{a}_i)\mathcal{K}(\mathbf{a}_3)\chi^{(a_1,a_2,a_3,a_2,a_1,a_4)}_{E_6}(\mathbf{a}_3)}{\chi^{(a_1,a_2,a_3,a_2,a_1,a_4)}_{E_6}(\mathbf{a}_\text{triv})}
\label{Schur}\end{equation}
where $(a;q)\equiv \prod_{j=0}^\infty (1-aq^j)$ is the $q$-Pochhammer symbol. We expand each character $\chi^\lambda$ in \eqref{Schur} in terms of $\mathfrak{su}(2) \times \mathfrak{f}$ characters as determined by the $\mathfrak{su}(2)$ embedding which defines the puncture, where the $\mathfrak{su}(2)$ fugacity is set equal to $q^\frac{1}{2}$. Decomposing the adjoint representation as

\begin{displaymath}
\mathfrak{g}=\bigoplus_n V_n \otimes R_n
\end{displaymath}
where $V_n$ is the $n$-dimensional irrep of $\mathfrak{su}(2)$ and $R_n$ is the corresponding representation of $\mathfrak{f}$, $\mathcal{K}(\mathbf{a})$ is defined by 

\begin{displaymath}
\mathcal{K}(\mathbf{a})=PE[\sum_n q^{\frac{n+1}{2}}\chi^{R_n}_{\mathfrak{f}}(\mathbf{a})],
\end{displaymath}
where $PE$ denotes the plethystic exponential. By their definitions, one can easily see that the coefficient of $\tau$ $(\tau^2)$ in the Hall-Littlewood index is the same as that of $q^\frac{1}{2}$ $(q)$ in the Schur index (though the higher-order terms are different). However, while the Schur index removes the difficulty of explicitly summing over the Weyl group, we find that the number of terms in the sum in \eqref{Schur} grows very quickly at each order in $q^\frac{1}{2}$ and begins to involve large-dimensional representations of $E_6$, also making the calculation very tedious. Therefore, in most of the calculations that follow, we compute only the next 1-2 terms in the expansion of the Schur index. It would be very useful to find a more efficient way to explicitly calculate \eqref{SCI}, \eqref{Schur}.

\section{Enhanced global symmetries: the Sommers-Achar group on the Higgs branch}\label{enhanced_global_symmetries_the_sommersachar_group_on_the_higgs_branch}

Consider a family of fixtures where we keep two of the punctures fixed, and vary the third puncture over a special piece, $\{O\}$. Let $O_s$ be the special puncture in this special piece. The Sommers-Achar group $C(O)$, for each of the punctures $O$ in the special piece, is a subgroup of Lusztig's canonical quotient group, $\overline{A}(d(O))\simeq S_n$. Let $O_m$ be the puncture with the maximal Sommers-Achar group, i.e, the one whose Hitchin pole is $(d(O),S_n)$.

It frequently happens that, when $O=O_s$, a simple factor (associated to one of the \emph{other} punctures, which we are holding fixed) in the manifest global symmetry of the fixture is enhanced as

\begin{displaymath}
F_{k n} \to {(F_k)}^n
\end{displaymath}
(There may, in addition, be further enhancements of the global symmetry but, by examining the fugacity-dependence of the superconformal index, we know unambiguously which ones are associated to the enhancement of $F_{k n}$.) When this enhancement takes place, for $O=O_s$, then, for $O=O_m$, the $F_{k n}$ is unenhanced and, as $O$ varies over the special piece, the enhancement is the subgroup of ${(F_k)}^n$ which is invariant under $C(O)$ acting by permutations of the $n$ copies of $F_k$.

In particular, this gives an explicit action of the Sommers-Achar group on the holomorphic moment map operators, which are generators of the Higgs branch chiral ring. Heretofore, the Sommers-Achar group was purely a Coulomb-branch concept \cite{Trimm-Yan-in-progress}.

We found numerous examples of this in \cite{Chacaltana:2014jba} and were able to verify, using various S-dualities (see, e.g., Section 4 of \cite{Chacaltana:2014jba}) that the levels of the factors of $F$ in the global symmetry behave as predicted by this permutation action.

One example eluded us there. We were unable to verify, using S-duality, the levels of the $SU(3)$s in the first two fixtures in

\begin{displaymath}
 \includegraphics[width=264pt]{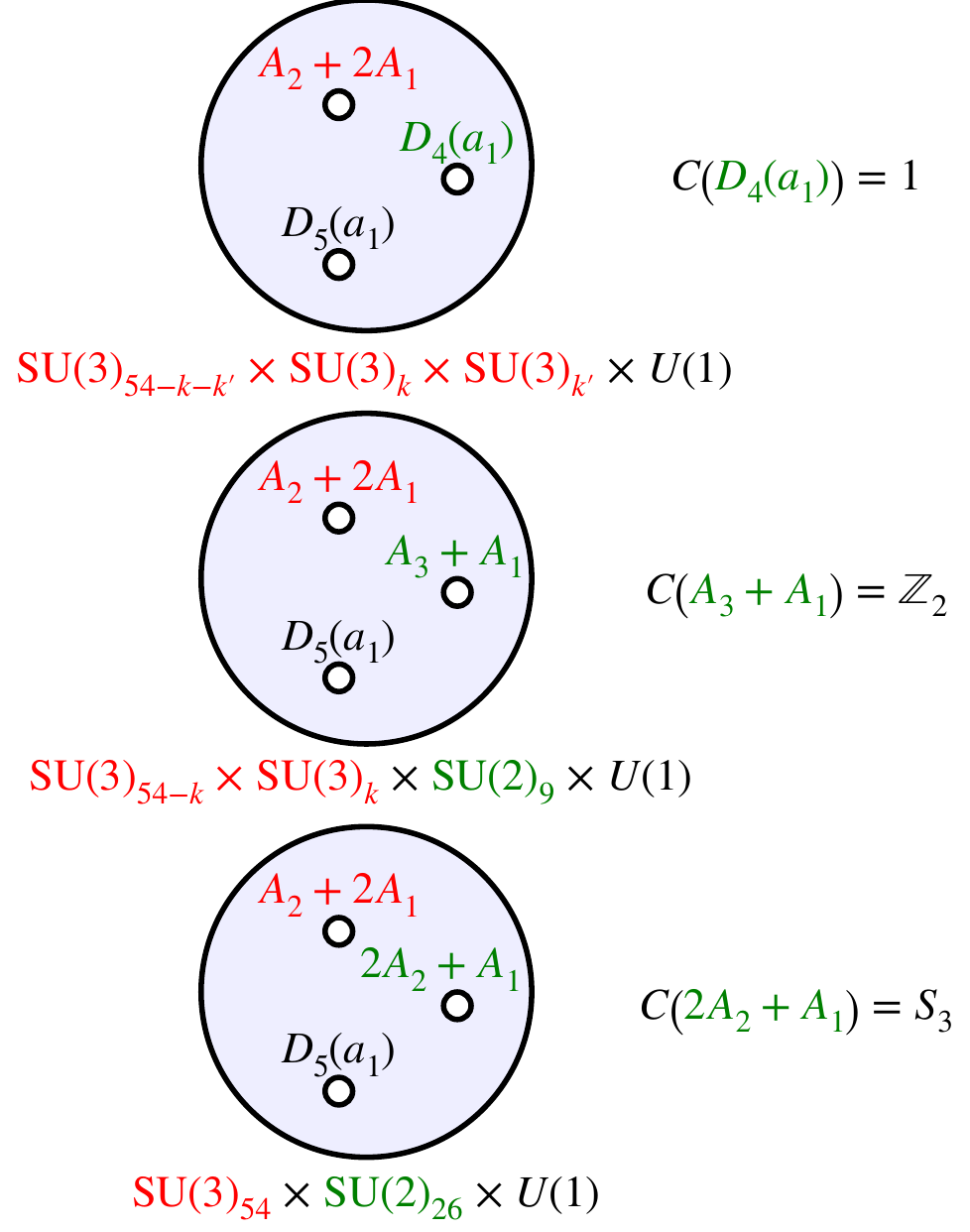}
\end{displaymath}
This example has an additional enhancement. As above, the manifest symmetry of the $A_2+2A_1$ puncture is enhanced

\begin{displaymath}
{SU(2)}_{54}\times U(1) \to {SU(2)}_{18}^3 \times U(1)
\end{displaymath}
with the further enhancement

\begin{displaymath}
{SU(2)}_{18}^3 \times {U(1)}^3 \to {SU(3)}_{18}^3
\end{displaymath}
Otherwise, this example fits the same pattern: the Sommers-Achar group, $C(O)$, acts by permutations on the ${SU(3)}_{18}^3$, and the global symmetry group of the fixture is the subgroup invariant under $C(O)$. That is, $k=k'=18$.

The twisted sector of the $E_6$ theory provides further examples of this phenomenon. Perhaps the most striking example is the fixture

\begin{displaymath}
 \includegraphics[width=89pt]{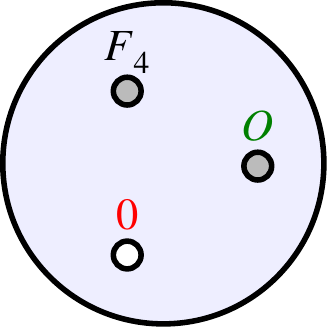}
\end{displaymath}
with an untwisted full puncture, a twisted simple puncture and a twisted puncture, ${\color{green}O}$. As we let the puncture, ${\color{green}O}$, vary over the special piece of $F_4(a_3)$, the ${\color{red}{(E_6)}_{24}}$ symmetry of the $0$ puncture is enhanced to (the $C({\color{green}O})$-invariant subgroup of) ${\color{red}{(E_6)}_{6}^4}$. The resulting SCFTs are products of the generalized $E_6$ Minahan-Nemeschansky SCFTs whose Higgs branches are the moduli space of $l$ $E_6$ instantons, $M(E_6,l)$\footnote{These SCFTs are realized in F-theory as the theory on $l$ D3-branes coincident with a $\text{IV}^*$ singularity.} .

{\footnotesize
\renewcommand{\arraystretch}{2}
\begin{longtable}{|c|c|c|c|c|c|c|}
\hline
\#&${\color{green}O}$&$C({\color{green}O})$&Theory&Higgs Branch&$\dim_{\mathbb{H}}\text{Higgs}$&$(n_h,n_v)$\\
\hline 
\endhead
1&$\underline{F_4(a_3)}$&1&$[(E_6)_6\, \text{ SCFT}]^4$&$M(E_6,1)^4$&$44$&$(64,20)$\\
\hline 
2&$\underline{C_3(a_1)}$&$\mathbb{Z}_2$&$[(E_6)_6\, \text{ SCFT}]^2 \times [(E_6)_{12}\times SU(2)_7\,\text{SCFT}]$&$M(E_6,1)^2 \times M(E_6,2)$&$45$&$(71,26)$\\
\hline 
3&$\underline{\widetilde{A_2}+A_1}$&$S_3$&$[(E_6)_6\,\text{ SCFT}]\times [(E_6)_{18}\times SU(2)_{20}\,\text{ SCFT}]$&$M(E_6,1)\times M(E_6,3)$&$46$&$(84,38)$\\
\hline 
4&$\underline{B_2}$&$\mathbb{Z}_2\times \mathbb{Z}_2$&$[(E_6)_{12} \times SU(2)_7\, \text{ SCFT}]^2$&$M(E_6,2)^2$&$46$&$(78,32)$\\
\hline 
&$\underline{A_2+\widetilde{A_1}}$&$S_4$&$[(E_6)_{24} \times SU(2)_{39}\,\text{ SCFT}]$&$M(E_6,4)$&$47$&$(103,56)$\\
\hline
\end{longtable}
}

Here the ${(E_6)}_{6l}$ global symmetry is realized as the $E_6$ global symmetry of $M(E_6,l)$. More subtle relations between instanton moduli spaces will be discussed below in \S\ref{InstantonModuliSpaces}.

In this example, the global symmetry groups and the levels were all determined by S-duality. In other examples, S-duality determines some, but not all of the levels of the enhanced global symmetries, and we can use the action of the Sommers-Achar group on the Higgs branch to fill in the missing levels\footnote{The action of $C(O)$ on the Higgs branch of mixed fixtures is not so transparent.}.

Two more sequences of fixtures, which have one puncture running over the special piece of $F_4(a_3)$, have global symmetry groups which are enhanced in this fashion, but levels we could not completely determine using S-duality\footnote{Interacting fixture 83 in the table above contains the puncture $\underline{A_2+\widetilde{A}_1}$, which is in the special piece of $F_4(a_3)$. However, three of the other four fixtures related by varying over the special piece are \emph{bad} (the other good fixture is mixed fixture 22). In particular, the fixture with the special puncture $F_4(a_3)$ is bad, so there is no enhancement of the form discussed above. Thus we don't know how to use this method to determine the levels for fixture 83.}:

In the case of
\begin{displaymath}
 \includegraphics[width=92pt]{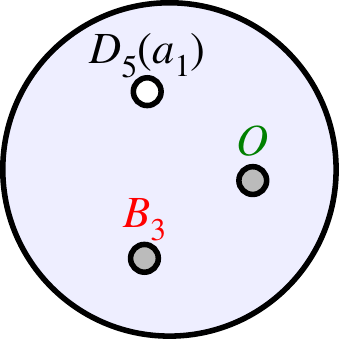}
\end{displaymath}
as ${\color{green}O}$ varies over the special piece of $F_4(a_3)$, the ${\color{red}{SU(2)}_{24}}$ global symmetry of $\underline{B_3}$ is enhanced.

{
\renewcommand{\arraystretch}{2}
\begin{longtable}{|c|c|c|c|}
\hline
\#&${\color{green} O}$&$C({\color{green} O})$&Global Symmetry\\
\hline 
\endhead
75&$\underline{F_4(a_3)}$&1&${\color{red}SU(2)_{24-k_1-k_2-k_3}\times SU(2)_{k_1} \times SU(2)_{k_2} \times SU(2)_{k_3}} \times {U(1)}^3$\\
\hline
77&$\underline{C_3(a_1)}$&$\mathbb{Z}_2$&${\color{red}SU(2)_{12}\times {SU(2)}^2_6} \times {\color{green}SU(2)_7} \times {U(1)}^2$\\
\hline
79&$\underline{\widetilde{A_2}+A_1}$&$S_3$&${\color{red}SU(2)_{24-k} \times SU(2)_k} \times {\color{green}SU(2)_{20}} \times U(1)$\\
\hline
81&$\underline{B_2}$&$\mathbb{Z}_2\times \mathbb{Z}_2$&${\color{red}{SU(2)}^2_{12}}\times {\color{green}{SU(2)}^2_{7}} \times U(1)$\\
\hline
&$\underline{A_2+\widetilde{A_1}}$&$S_4$&${\color{red}SU(2)_{24}}\times {\color{green}SU(2)_{39}}\times U(1)$\\
\hline
\end{longtable}
}
\noindent
Filling in the missing levels, we find $k_1=k_2=k_3=k=6$.
\bigskip

Similarly, as ${\color{green}O}$ in

\begin{displaymath}
 \includegraphics[width=92pt]{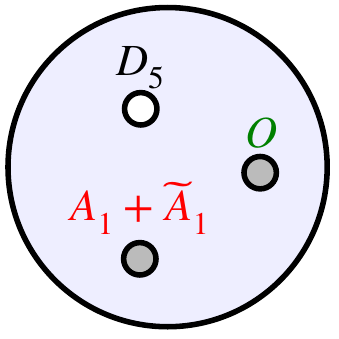}
\end{displaymath}
varies over the special piece of $F_4(a_3)$, the ${\color{red} SU(2)_{64}}\times {SU(2)}_{10}$ global symmetry of $\underline{A_1+\widetilde{A}_1}$ is enhanced.

{
\renewcommand{\arraystretch}{2}
\begin{longtable}{|c|c|c|c|}
\hline
\#&${\color{green} O}$&$C({\color{green} O})$&Global Symmetry\\
\hline
\endhead
97&$\underline{F_4(a_3)}$&1&${\color{red}SU(2)_{64-k_1-k_2-k_3} \times SU(2)_{k_1} \times SU(2)_{k_2} \times SU(2)_{k_3}} \times SU(2)_{10} \times U(1)$\\
\hline
101&$\underline{C_3(a_1)}$&$\mathbb{Z}_2$&${\color{red}SU(2)_{32} \times {SU(2)}^2_{16}} \times SU(2)_{10} \times {\color{green}{SU(2)}_7} \times U(1)$\\
\hline
105&$\underline{\widetilde{A_2}+A_1}$&$S_3$&${\color{red}SU(2)_{64-k} \times SU(2)_k} \times SU(2)_{10} \times {\color{green}{SU(2)}_{20}} \times U(1)$\\
\hline
109&$\underline{B_2}$&$\mathbb{Z}_2\times \mathbb{Z}_2$&${\color{red}{SU(2)}_{32}^2} \times {SU(2)}_{10} \times {\color{green}{SU(2)}_7^2} \times U(1)$\\
\hline
&$\underline{A_2+\widetilde{A_1}}$&$S_4$&${\color{red}SU(2)_{64}} \times SU(2)_{10} \times {\color{green}SU(2)_{39}} \times U(1)$\\
\hline
\end{longtable}
}
\noindent
Again, we can fill in the missing levels: $k_1=k_2=k_3=k=16$.
\bigskip

Additionally, we find the following three fixture by varying over special piece of the untwisted puncture $D_4(a_1)$:

\begin{displaymath}
 \includegraphics[width=80pt]{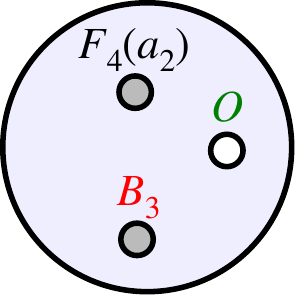}
\end{displaymath}
As ${\color{green} O}$ varies over the special piece of $D_4(a_1)$, the ${\color{red} SU(2)_{24}}$ global symmetry of $\underline{B_3}$ is enhanced.

{
\renewcommand{\arraystretch}{2}
\begin{longtable}{|c|c|c|c|}
\hline
\#&${\color{green} O}$&$C({\color{green} O})$&Global Symmetry\\
\hline
\endhead
25&$D_4(a_1)$&1&${\color{red}SU(2)_{24-k_1-k_2}\times SU(2)_{k_1}\times SU(2)_{k_2}}\times {\color{green}{U(1)}^2}$\\
\hline
26&$A_3+A_1$&$\mathbb{Z}_2$&${\color{red}SU(2)_{24-k}\times SU(2)_k} \times {\color{green}SU(2)_9 \times U(1)}$\\
\hline
&$2A_2+A_1$&$S_3$&${\color{red}SU(2)_{24}}\times {\color{green}SU(2)_{26}}$\\
\hline
\end{longtable}
}
\noindent
We fill in the missing levels $k_1=k_2=k=8$.

\section{$R_{2,5}$}\label{R25}

In \cite{Chacaltana:2010ks}, we introduced a series of $\mathcal{N}=2$ SCFTs, which we dubbed $R_{2,2n-1}$. $R_{2,2n-1}$ has a $\bigl({Spin(4n+2)}_{4n-2}\times U(1)\bigr)/\mathbb{Z}_2$ global symmetry (enhanced to ${(E_6)}_6$ for $n=2$), central charges $(n_h,n_v)=\bigl(4n^2,(n-1)(2n+1)\bigr)$ and graded Coulomb branch dimensions $n_{2k-1}=1$, for $k=2,\dots, n$.

These play an important role in the strong-coupling duals of various familiar gauge theories. Specifically
\begin{equation}
\begin{split}
SU(2n-1) +4\bigl( \includegraphics[width=9pt]{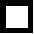}\bigr)+2\Bigl(\begin{matrix} \includegraphics[width=9pt]{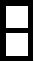}\end{matrix}\Bigr)
&\simeq Sp(n-1) + 1\bigl( \includegraphics[width=9pt]{fund}\bigr)
+ R_{2,2n-1}\\
SU(2n) +4\bigl( \includegraphics[width=9pt]{fund}\bigr)+2\Bigl(\begin{matrix} \includegraphics[width=9pt]{antisym}\end{matrix}\Bigr)
&\simeq Sp(n) + 3\bigl( \includegraphics[width=9pt]{fund}\bigr) + R_{2,2n-1}\\
SU(2n) +1\Bigl(\begin{matrix} \includegraphics[width=9pt]{antisym}\end{matrix}\Bigr)+1\bigl( \includegraphics[width=17pt]{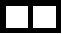}\bigr)&\simeq Spin(2n+1) + R_{2,2n-1}
\end{split}
\label{R22nm1dualities}\end{equation}
The realizations of the $R_{2,2n-1}$ are:

\begin{displaymath}
\begin{matrix} \includegraphics[width=107pt]{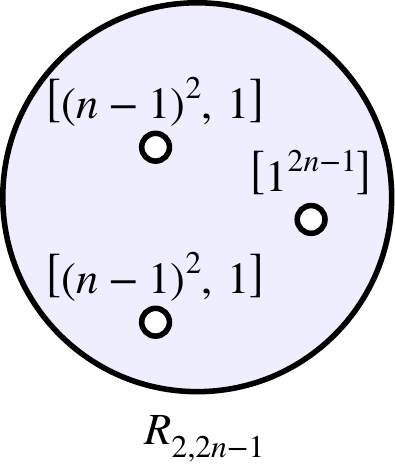}\end{matrix}
\quad
\begin{matrix} \includegraphics[width=107pt]{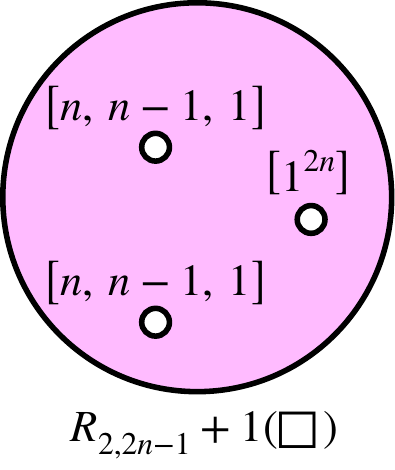}\end{matrix}
\quad
\begin{matrix} \includegraphics[width=107pt]{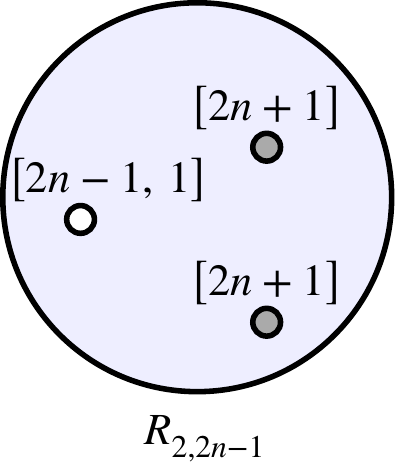}\end{matrix}
\end{displaymath}
in the $A_{2n-2}$, $A_{2n-1}$ and the twisted sector of the $A_{2n-1}$ theory, respectively. These different realizations expose different manifest subalgebras\footnote{As symmetry groups, they are, respectively, $S\bigl(U(2n-1)\times U(2)^2\bigr)$, $S\bigl(U(2n)\times {U(1)}^3\bigr)$ and $\bigl(Spin(2n+1)\times Spin(2n+1)\times U(1)\bigr)/\mathbb{Z}_2$.}  (respectively, ${\mathfrak{su}(2n-1)}_{4n-2}\times{\mathfrak{su}(2)}_{4n-2}^2\times {\mathfrak{u}(1)}^2$, ${\mathfrak{su}(2n)}_{4n-2}\times{\mathfrak{u}(1)}^4$ and ${\mathfrak{so}(2n+1)}_{4n-2}^2\times \mathfrak{u}(1)$) of the full global symmetry algebra of the $R_{2,2n-1}$ SCFT.

The twisted sector of the $E_6$ theory provides two new realizations of $R_{2,5}$:

\begin{displaymath}
\begin{matrix} \includegraphics[width=107pt]{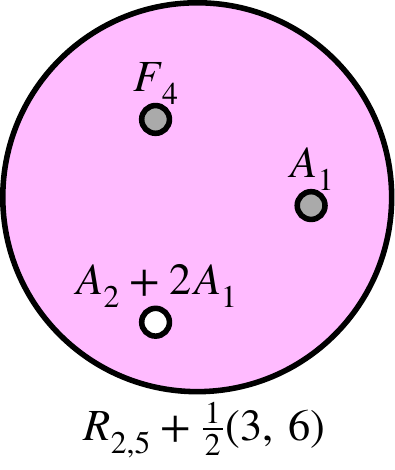}\end{matrix}
\quad
\begin{matrix} \includegraphics[width=107pt]{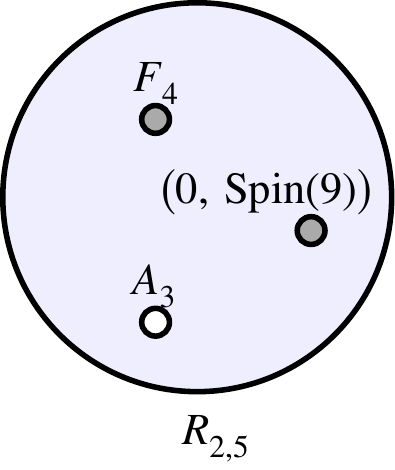}\end{matrix}
\end{displaymath}
which expose a manifest $\mathfrak{sp}(3)_{10}\times \mathfrak{su}(2)_{30}\times \mathfrak{u}(1)$ or ${\mathfrak{so}(9)}_{10}\times \mathfrak{sp}(2)_{10}\times \mathfrak{u}(1)$ subalgebra, respectively, of the ${\mathfrak{so}(14)}_{10}\times \mathfrak{u}(1)$ global symmetry algebra of $R_{2,5}$.

The latter realization will be useful to us in \S \ref{more_isomorphisms_among_hyperKahler_quotients}. The former provides, among other things, another realization of the aforementioned duality

\begin{displaymath}
SU(6) +4(6)+2(15) \simeq Sp(3) + 3(6) + R_{2,5}\quad
\end{displaymath}
via the 4-punctured sphere

\begin{displaymath}
 \includegraphics[width=220pt]{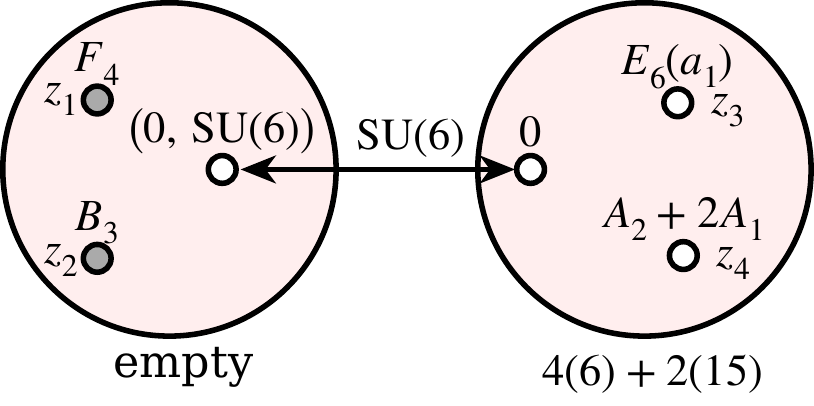}
\end{displaymath}
Here, the gauge coupling,

\begin{displaymath}
f(\tau)\equiv -\frac{\theta_2^4(0,\tau)}{\theta_4^4(0,\tau)} = \frac{w-1}{w+1}
\end{displaymath}
is a function on the double-cover of $\mathcal{M}_{0,4}$, where

\begin{displaymath}
w^2 = x\equiv \frac{z_{1 3} z_{2 4}}{z_{1 4} z_{2 3}}
\end{displaymath}
so that $f(\tau)=1$ at the degeneration

\begin{displaymath}
 \includegraphics[width=220pt]{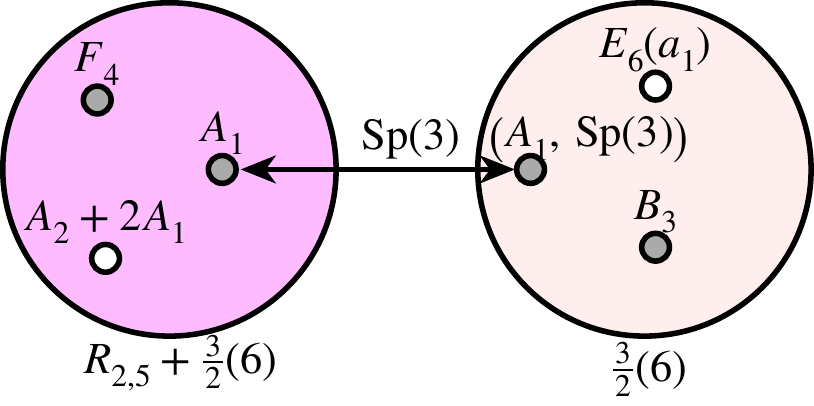}
\end{displaymath}
and the degeneration

\begin{displaymath}
 \includegraphics[width=220pt]{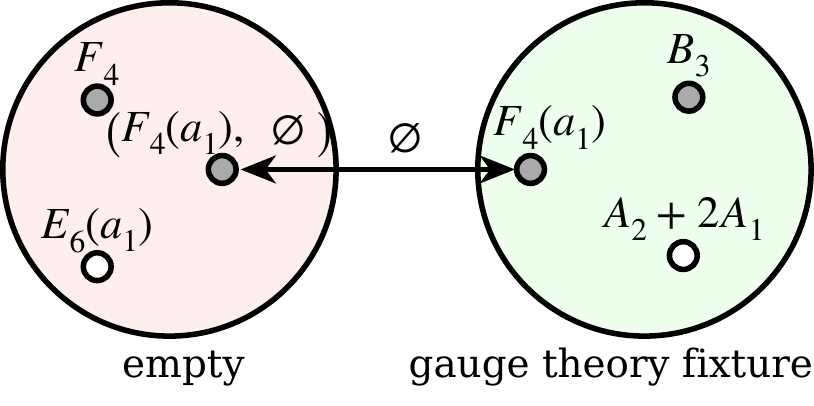}
\end{displaymath}
corresponds to the interior point of the gauge theory moduli space, $f(\tau)=-1$.

\section{Product SCFTs}\label{product_scfts}

The ${(E_6)}_{18}\times {(G_2)}_{10}$ SCFT occurs twice on our list of interacting fixtures: once, by itself, in

\begin{equation}
 \includegraphics[width=108pt]{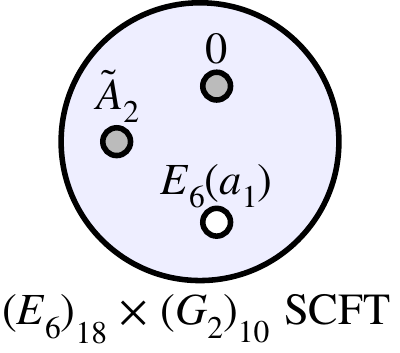}
\label{E6G2solo}\end{equation}
(interacting fixture 117) and once --- we claim --- as part of a product SCFT

\begin{equation}
 \includegraphics[width=192pt]{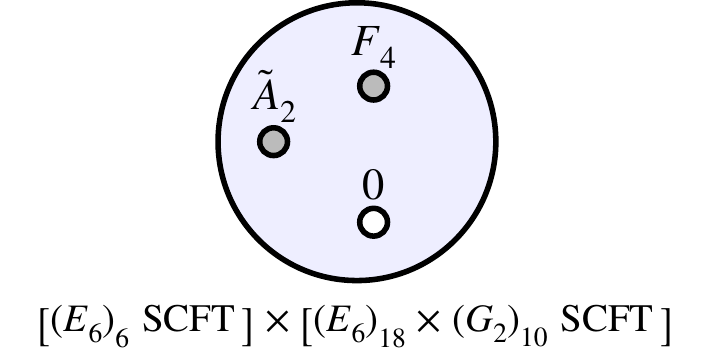}
\label{E6G2product}\end{equation}
(interacting fixture 5). We can check the latter claim, explicitly, by comparing the SCI for \eqref{E6G2product} with the SCI for \eqref{E6G2solo} and the (known) SCI for the ${(E_6)}_6$ SCFT.

Indeed, we find that, to second order in $q$, we have

\begin{displaymath}
\begin{split}
\mathcal{I}_{\#5}&=1+170q+14601q^2+\dots \\
&=(1+92q+4916q^2+\dots)(1+78q+2509q^2+\dots) \\
&=(\mathcal{I}_{\#117}\times\mathcal{I}_{(E_6)_6 SCFT})|_{q^2}
\end{split}
\end{displaymath}
Having established that \eqref{E6G2product} is a product SCFT, we can apply that knowledge to deduce that other fixtures are also product SCFTs. For instance, consider the 4-punctured sphere

\begin{displaymath}
 \includegraphics[width=269pt]{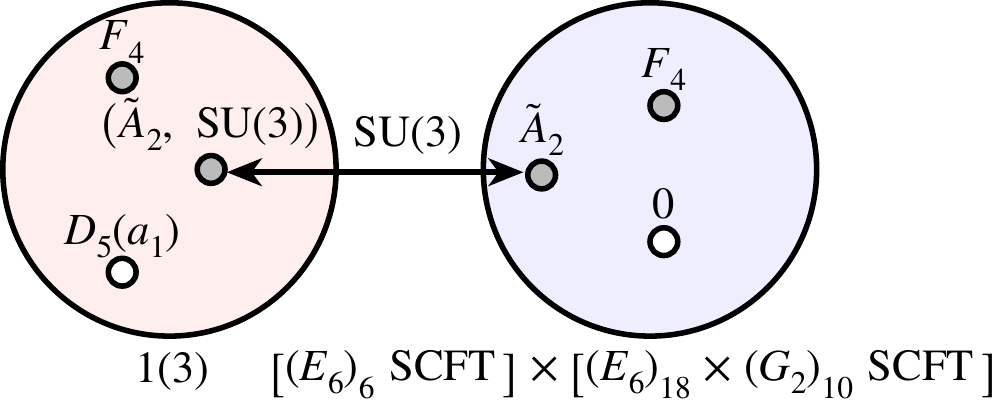}
\end{displaymath}
The $SU(3)$ gauges a subgroup of the ${(G_2)}_{10}$ symmetry of the ${(E_6)}_{18}\times {(G_2)}_{10}$ SCFT, leaving the ${(E_6)}_6$ SCFT decoupled. Taking the S-dual,

\begin{displaymath}
 \includegraphics[width=314pt]{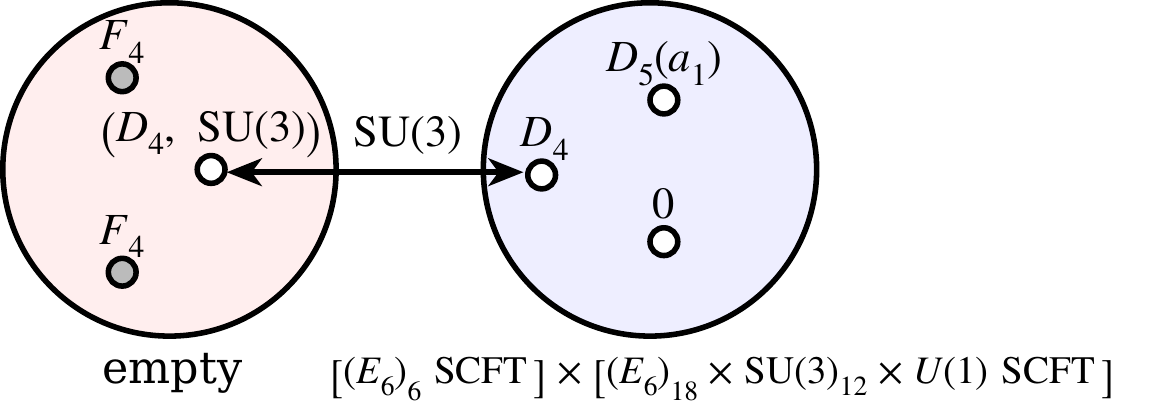}
\end{displaymath}
we conclude that fixture on the right also contains a decoupled ${(E_6)}_6$ SCFT and, hence, that

\begin{equation}
 \includegraphics[width=237pt]{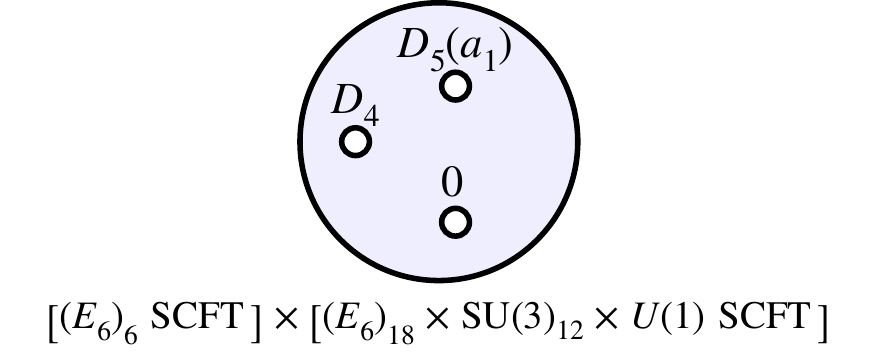}
\label{if61}\end{equation}
(interacting fixture 61 of \cite{Chacaltana:2014jba}) is also a product SCFT.

Similarly, in

\begin{displaymath}
 \includegraphics[width=269pt]{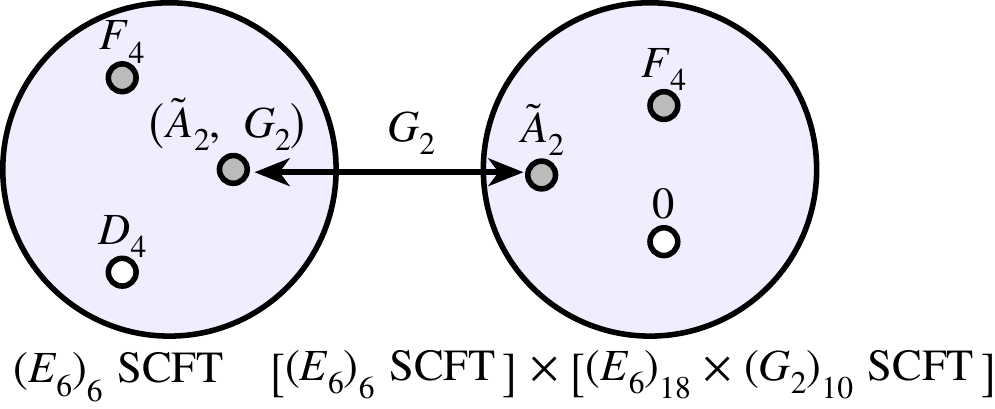}
\end{displaymath}
the gauging of the $G_2$ symmetry leaves the ${(E_6)}_6$ SCFT decoupled. Hence, in the S-dual,

\begin{displaymath}
 \includegraphics[width=276pt]{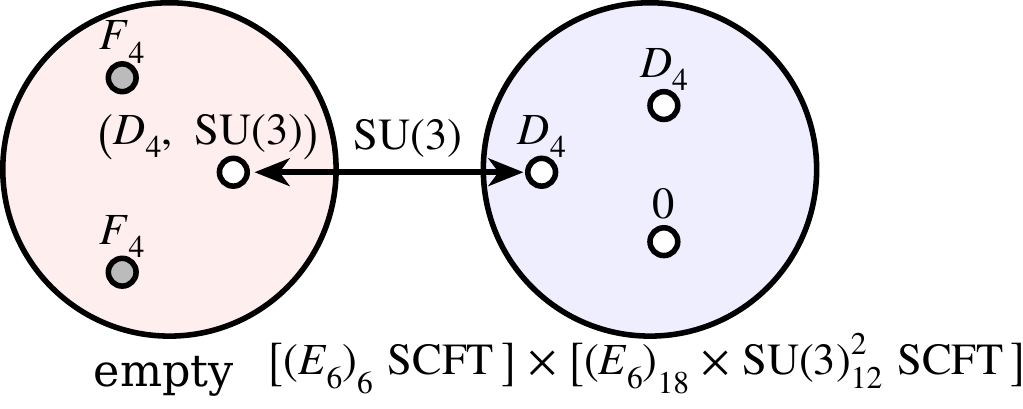}
\end{displaymath}
the fixture

\begin{equation}
 \includegraphics[width=198pt]{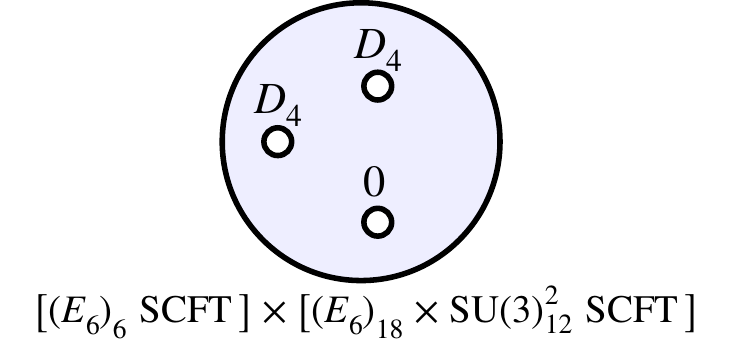}
\label{if99}\end{equation}
(interacting fixture 99 of \cite{Chacaltana:2014jba}) is, again, a product SCFT.

As a further check of these identifications, we can compare the SCIs for \eqref{if61} and \eqref{if99} with those of interacting fixtures 15 and 16 above, which directly realize, respectively, the ${(E_6)}_{18}\times{SU(3)}_{12}\times U(1)$ and ${(E_6)}_{18}\times{SU(3)}_{12}^2$ SCFTs.

Indeed, we find that
\begin{displaymath}
\begin{split}
\mathcal{I}_{\# 61}&=1+165q+164q^{\frac{3}{2}}+13451q^2+\dots\\
&=(1+87q+164q^{\frac{3}{2}}+4156q^2+\dots)(1+78q+2509q^2+\dots) \\
&=(\mathcal{I}_{\# 15} \times \mathcal{I}_{(E_6)_6 \text{ SCFT}})|_{q^2}
\end{split}
\end{displaymath}
and

\begin{displaymath}
\begin{split}
\mathcal{I}_{\# 99}&=1+172q+14886q^2+\dots\\
&=(1+94q+5045q^2+\dots)(1+78q+2509q^2+\dots) \\
&=(\mathcal{I}_{\# 16} \times \mathcal{I}_{(E_6)_6 \text{ SCFT}})|_{q^2}
\end{split}
\end{displaymath}
Similarly, we can check that interacting fixture 59 of \cite{Chacaltana:2014jba}

\begin{displaymath}
 \includegraphics[width=198pt]{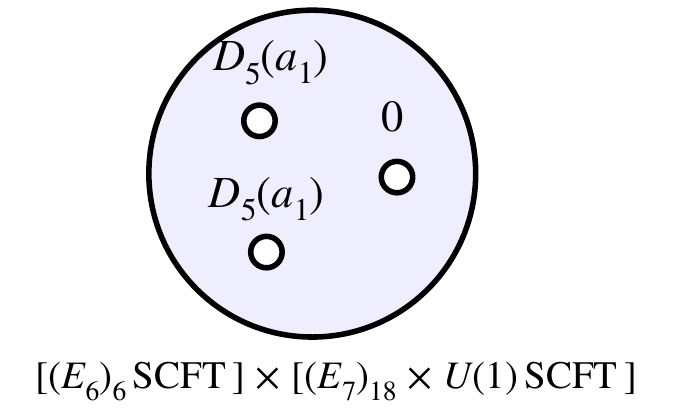}
\end{displaymath}
is a product SCFT by comparing the expansion of its SCI with that of interacting fixture 13 above, which directly realizes the $(E_7)_{18} \times U(1) \text{ \ SCFT}$. Indeed, one finds that

\begin{displaymath}
\begin{split}
\mathcal{I}_{\#59}&=1+212q+112q^{\frac{3}{2}}+22273q^2+\dots \\
&=(1+78q+2509q^2+\dots)(1+134q+112q^{\frac{3}{2}}+9312q^2+\dots)\\
&=(\mathcal{I}_{(E_6)_6 \text{ SCFT}} \times \mathcal{I}_{\#13})|_{q^2}.
\end{split}
\end{displaymath}
Finally, we claim that interacting fixture 111 above is the product of the $(E_6)_6$ SCFT and the $(F_4)_{12} \times {SU(2)}^2_7$ SCFT. The latter previously appeared in our list of interacting fixtures for the $D_4$ theory \cite{Chacaltana:2011ze} and appears in mixed fixture 30 above.

We find the expansion of the SCI for fixture 111 is given by

\begin{displaymath}
\mathcal{I}_{\# 111}=1+136q+104q^{\frac{3}{2}}+9036q^2+\dots
\end{displaymath}
That of mixed fixture 30 reads

\begin{displaymath}
\begin{split}
\mathcal{I}_{\# 30}&=1+2q^{\frac{1}{2}}+61q+226q^{\frac{3}{2}}+2394q^2+\dots \\
&=(1+2q^{\frac{1}{2}}+3q+6q^{\frac{3}{2}}+9q^2+\dots)(1+58q+104q^{\frac{3}{2}}+2003q^2+\dots) \\
&=(\mathcal{I}_{\text{free}} \times \mathcal{I}_{(F_4)_{12} \times {SU(2)}_7^2 SCFT})|_{q^2}
\end{split}
\end{displaymath}
Extracting the order $q^2$ expansion of the index of the $(F_4)_{12} \times {SU(2)}_7^2$ SCFT from the above, we see that

\begin{displaymath}
\begin{split}
\mathcal{I}_{(E_6)_6 SCFT} \times \mathcal{I}_{(F_4)_{12} \times {SU(2)}_7^2 SCFT}&=(1+78q+2509q^2+\dots)(1+58q+104q^{\frac{3}{2}}+2003q^2+\dots) \\
&=1+136q+104q^{\frac{3}{2}}+9036q^2+\dots \\
&=\mathcal{I}_{\# 111}|_{q^2}
\end{split}
\end{displaymath}

\section{Instanton moduli spaces}\label{InstantonModuliSpaces}

Let $M(G,k)$ denote the moduli space of $k$ instantons on $\mathbb{R}^4$, for gauge group $G$\footnote{Equivalently, the moduli space of \emph{framed} instantons on $S^4$.} . $M(G,k)$ is a hyperK\"ahler space of dimension

\begin{displaymath}
\dim_{\mathbb{H}}\bigl(M(G,k)\bigr)= \kappa_G k -1
\end{displaymath}
where $\kappa_G$ is the dual Coxeter number and the ``$-1$'' is present because we have removed the overall translational degree of freedom.

For $k=1$, $M(G,k)$ has hyperK\"ahler isometry group $G$. In fact, $M(G,1)$ is the minimal nilpotent orbit in $\mathfrak{g}_\mathbb{C}$ \cite{MR1072915}. For $k\gt 1$, the hyperK\"ahler isometry group of $M(G,k)$ is $G\times SU(2)$. The origin of the additional $SU(2)$ is as follows\footnote{We thank Andrew Neitzke for a discussion of this point.} . While we've removed the translational symmetry of $\mathbb{R}^4$, the $SO(4)=\bigl(SU(2)\times SU(2)\bigr)/\mathbb{Z}_2$ rotational symmetry still acts on the space of instanton solutions. One of the $SU(2)$s acts by rotating the complex structures of $M(G,k)$ among themselves. The other $SU(2)$ preserves the quaternionic structure. For $k=1$, it is easy to see that it acts trivially, whereas for $k\gt 1$ it acts nontrivially.

For the classical groups $G$, the ADHM construction \cite{Atiyah:1978ri} provides a realization of $M(G,k)$ as a hyperK\"ahler quotient of a quaternionic \emph{vector space}. When $G$ is exceptional, no such construction exists. But (at least for low $k$) something almost as nice exists. Namely: the hyperK\"ahler quotient $M(G,k)/\!/\!/ H$, for $H$ some subgroup of the isometry group of $M(G,k)$, has an alternative realization as a hyperK\"ahler quotient either of a quaternionic vector space or of some other well-known hyperK\"ahler space.

The first examples of this phenomenon come from the classic paper of Argyres-Seiberg \cite{Argyres:2007cn}

\begin{equation}
\begin{split}
\Bigl(M(E_6,1)\times \mathbb{H}^{2}\Bigr)/\!/\!/SU(2) &\simeq \mathbb{H}^{18}/\!/\!/ SU(3) \\
M(E_7,1)/\!/\!/SU(2) &\simeq \mathbb{H}^{24}/\!/\!/ Sp(2) \\
\end{split}
\label{AS}\end{equation}
They established something much stronger: the S-duality of a pair of $\mathcal{N}=2$ supersymmetric quantum field theories. The Higgs branch of one theory is the LHS; the Higgs branch of the other is the RHS. Because the Higgs branch geometry is independent of the gauge coupling, the S-duality of the two theories implies that the two Higgs branches are isomorphic. An independent, nontrivial check on the first of these isomorphisms was performed in \cite{Gaiotto:2008nz}. At the holomorphic-symplectic level, an axiomatization of this general construction is given in \cite{Moore:2011ee}.

Further examples of such isomorphisms of hyperK\"ahler quotients of instanton moduli spaces (implied, again, by the S-duality of the corresponding QFTs) appeared in our previous papers. In section 4.2.3 of \cite{Chacaltana:2013oka}, we found

\begin{equation}
M(E_8,1)/\!/\!/Sp(2) \simeq \mathbb{H}^{40}/\!/\!/ Sp(3)
\label{E81}\end{equation}
Here, the defining 6-dimensional representation and the 14-dimensional traceless 3-index antisymmetric tensor representation of $Sp(3)$ are both pseudo-real (have quaternionic structures) and hence induce, respectively, linear actions on $\mathbb{H}^{3}$ and $\mathbb{H}^{7}$. On the RHS of \eqref{E81}, we decompose $\mathbb{H}^{40}$ as 11 copies of the former and 1 copy of the latter. In the usual physics notation, we denote this by $\mathbb{H}^{40} \simeq \tfrac{11}{2}(6)\oplus\tfrac{1}{2}(14')$ (``11 half-hypermultiplets in the fundamental and 1 half-hypermultiplet in the $14'$ representation of $Sp(3)$''). Similarly, on the RHS of \eqref{AS}, we have $\mathbb{H}^{24}\simeq 6(4)$ (``6 full hypermultiplets in the fundamental representation of $Sp(2)$'').

In section 4.1.3 of \cite{Chacaltana:2014ica}, we found

\begin{equation}
M(E_7,2)/\!/\!/G_2 \simeq \mathbb{H}^{57}/\!/\!/ Spin(9)
\label{E72}\end{equation}
where, on the RHS, $\mathbb{H}^{57}$ decomposes as the $3(16)+1(9)$ of $Spin(9)$.

In this section, we will demonstrate five new identities of this sort.

\begin{equation}
\Bigl(M(E_6,2)\times \mathbb{H}\Bigr)/\!/\!/SU(2) \simeq M(E_8,1)/\!/\!/SU(3)
\label{E62}\end{equation}
\begin{equation}
M(E_6,2)/\!/\!/SU(3) \simeq \Bigl(M(E_6,1)\times M(E_6,1)\times \mathbb{H}^{7}\bigr)/\!/\!/G_2
\label{E62bis}\end{equation}
\begin{equation}
M(E_7,3)/\!/\!/Spin(8) \simeq \Bigl(M(E_7,1)\times M(E_7,1)\times M(E_7,1)\times \mathbb{H}^{26}\bigr)/\!/\!/F_4
\label{E73}\end{equation}
\begin{equation}
\Bigl(M(E_7,2)\times M(E_7,1)\Bigr)/\!/\!/Spin(8) \simeq \Bigl(M(E_7,1)\times M(E_7,1)\times M(E_7,1)\times \mathbb{H}^{9}\bigr)/\!/\!/Spin(9)
\label{E72E71}\end{equation}
and

\begin{equation}
\left(M(E_8,2)\times\mathbb{H}^{32}\right)/\!/\!/Spin(12) \simeq \Bigl(M(E_8,1)\times M(E_8,1)\times \mathbb{H}^{45}\bigr)/\!/\!/Spin(13)
\label{E82}\end{equation}
where, on the LHS, the two irreducible spinor representations of $Spin(12)$ are pseudoreal ($\mathbb{H}^{32}\simeq \tfrac{1}{2}(32)\oplus \tfrac{1}{2}(32')$) and, on the RHS, we have $\mathbb{H}^{45}\simeq \tfrac{1}{2}(64)+ 1(13)$.

\subsection{$M(E_6,2)/\!/\!/SU(3) \simeq \Bigl(M(E_6,1)\times M(E_6,1)\times \mathbb{H}^{7}\bigr)/\!/\!/G_2$}\label{_2}

\eqref{E62bis} is realized in the untwisted $D_4$ theory by the 4-punctured sphere

\begin{displaymath}
 \includegraphics[width=255pt]{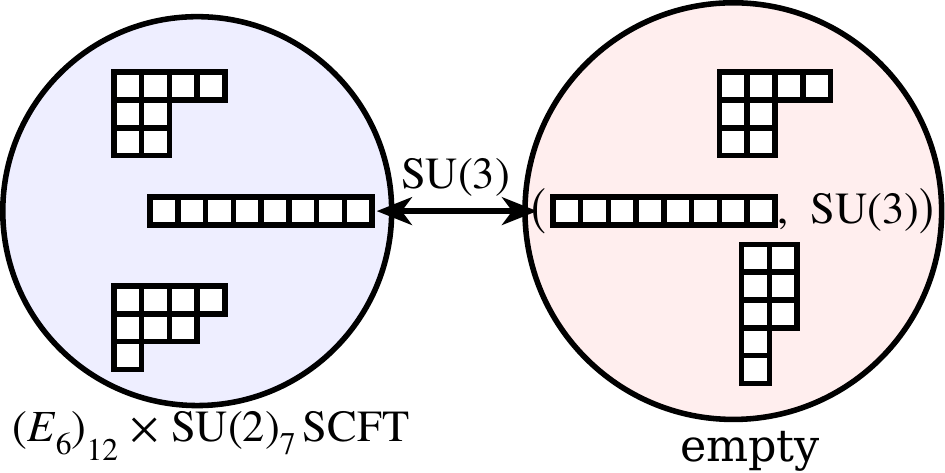}
\end{displaymath}
which is an $SU(3)$ gauging of the ${(E_6)}_{12}\times {SU(2)}_7$ SCFT (whose Higgs branch is $M(E_6,2)$). The S-dual theory

\begin{displaymath}
 \includegraphics[width=246pt]{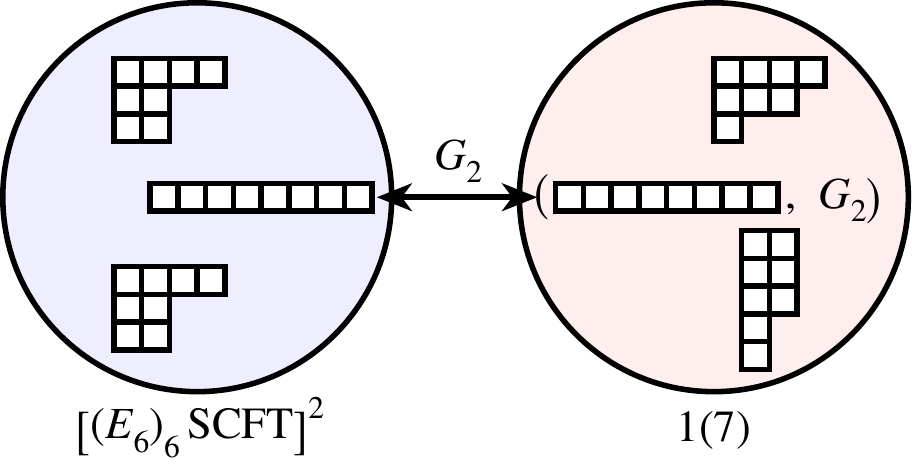}
\end{displaymath}
is a $G_2$ gauge theory coupled to two copies of the ${(E_6)}_6$ SCFT (whose Higgs branch is $M(E_6,1)$) and one hypermultiplet in the $7$.

\subsection{$\Bigl(M(E_6,2)\times \mathbb{H}\Bigr)/\!/\!/SU(2) \simeq M(E_8,1)/\!/\!/SU(3)$}\label{ME81SU3}

Recall that, for $k\gt 1$, $M(E_n,k)$ has an $E_n\times SU(2)$ isometry group. \eqref{E62} is unique among the examples listed here, in that on the LHS we use the $SU(2)$ action on $M(E_6,2)$, which commutes with $E_6$ action, to perform the hyperK\"ahler quotient.

A realization in the $D_4$ theory is

\begin{displaymath}
 \includegraphics[width=246pt]{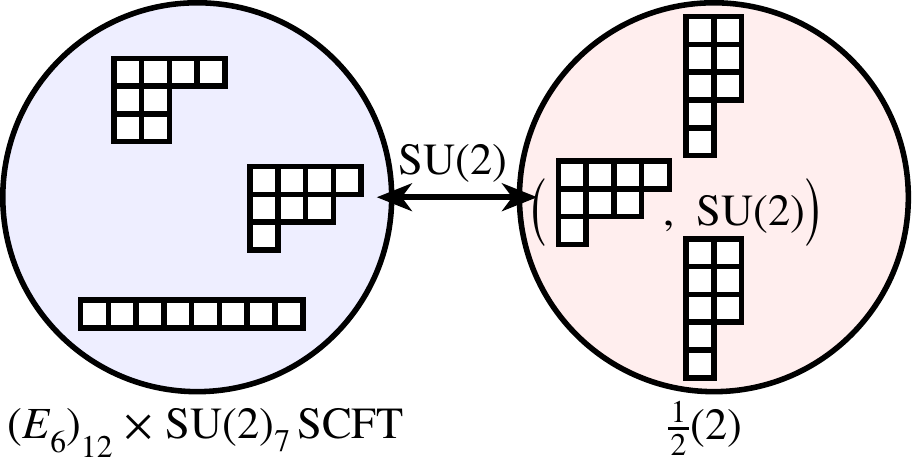}
\end{displaymath}
which is S-dual to

\begin{displaymath}
 \includegraphics[width=218pt]{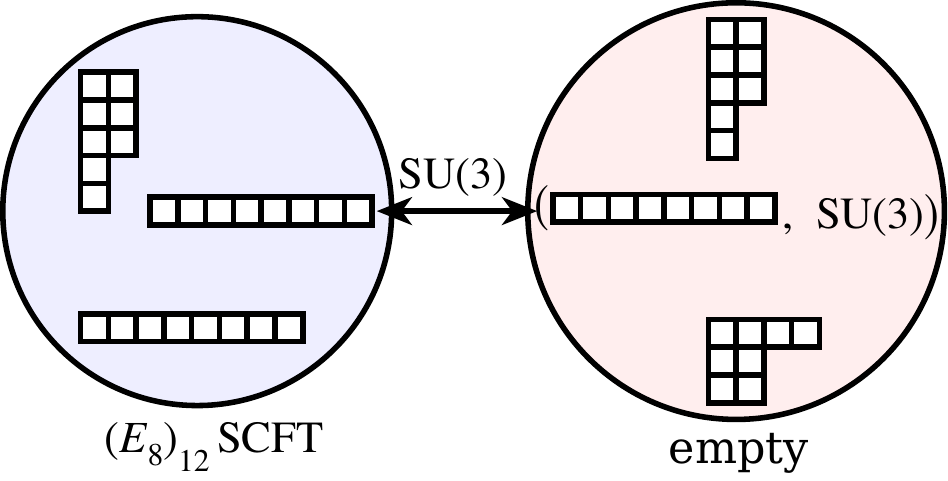}
\end{displaymath}
It is also realized in the untwisted $E_6$ theory as

\begin{displaymath}
 \includegraphics[width=218pt]{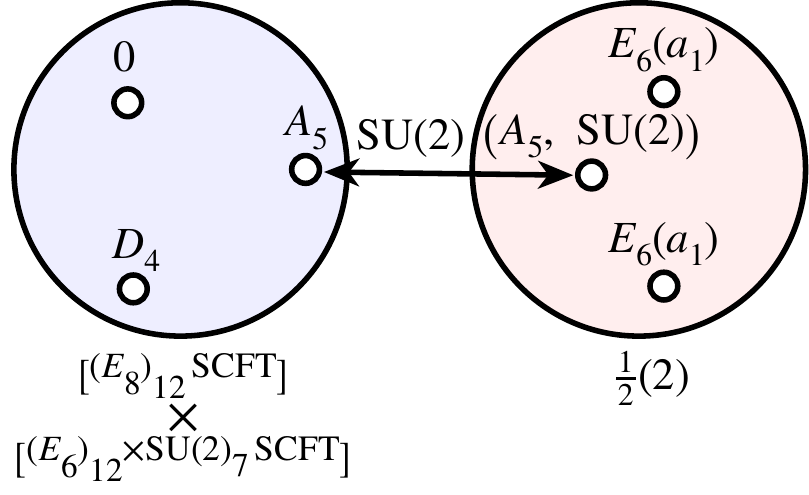}
\end{displaymath}
where the $SU(2)$ gauges the ${SU(2)}_7$ of the ${(E_6)}_{12}\times {SU(2)}_7$ SCFT and the ${(E_8)}_{12}$ SCFT is decoupled. The S-dual theory is

\begin{displaymath}
 \includegraphics[width=218pt]{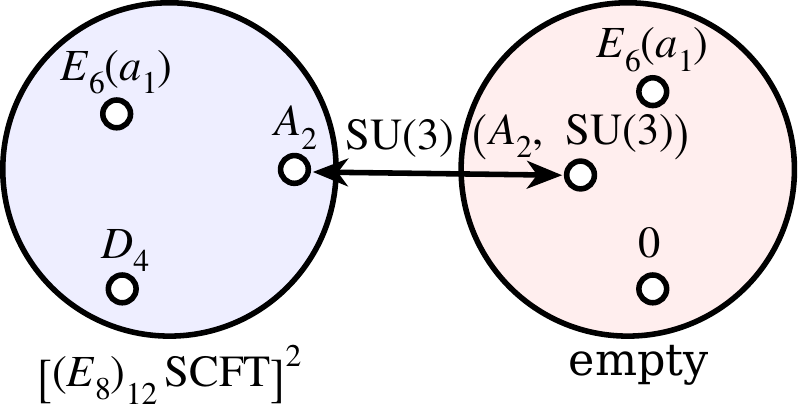}
\end{displaymath}
where we gauge an $SU(3)$ subgroup of one of the $E_8$s while the other ${(E_8)}_{12}$ SCFT is decoupled.

Another realization of \eqref{E62} appears in the twisted sector of the $E_6$ theory. In

\begin{displaymath}
 \includegraphics[width=218pt]{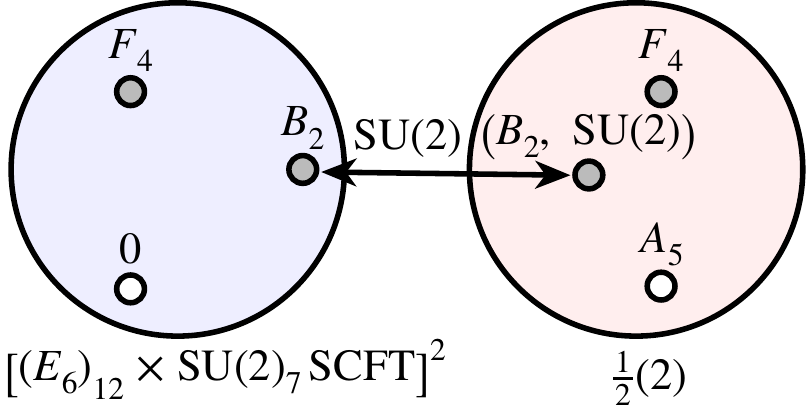}
\end{displaymath}
the $SU(2)$ gauges the ${SU(2)}_7$ of one of the ${(E_6)}_{12}\times {SU(2)}_7$ SCFTs, while the other ${(E_6)}_{12}\times {SU(2)}_7$ SCFT is decoupled. In the S-dual theory,

\begin{displaymath}
 \includegraphics[width=218pt]{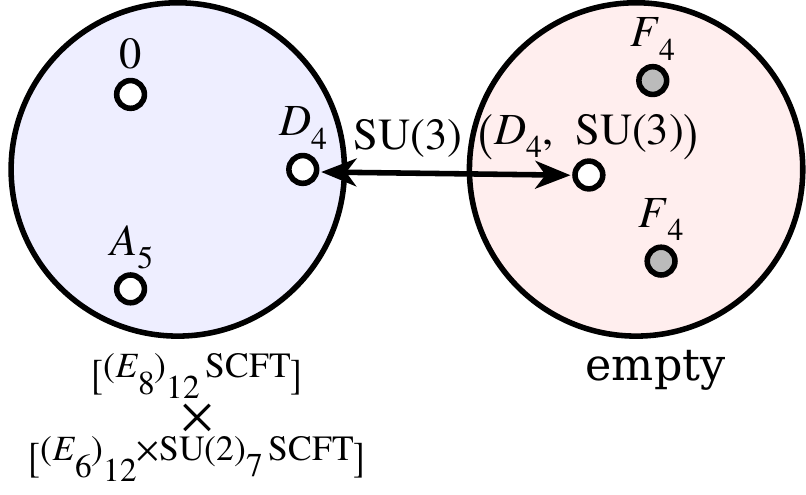}
\end{displaymath}
the $SU(3)$ gauges a subgroup of the $E_8$, while the ${(E_6)}_{12}\times {SU(2)}_7$ SCFT is decoupled.

A third realization, in which an ${(E_6)}_6$ SCFT is decoupled throughout, is given by

\begin{displaymath}
 \includegraphics[width=218pt]{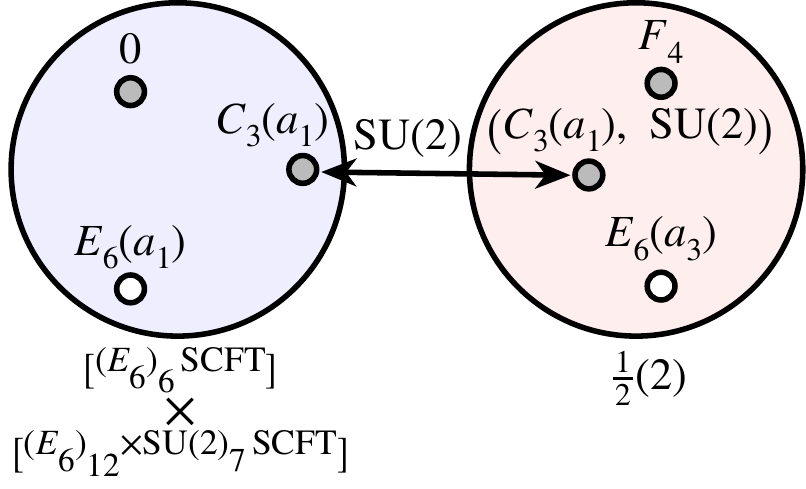}
\end{displaymath}
and has S-duals given by 

\begin{displaymath}
 \includegraphics[width=218pt]{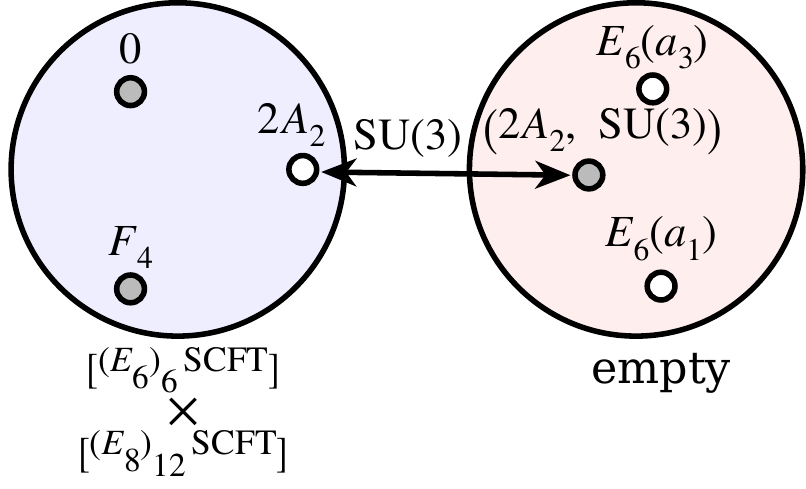}
\end{displaymath}
and the gauge-theory fixture

\begin{displaymath}
 \includegraphics[width=218pt]{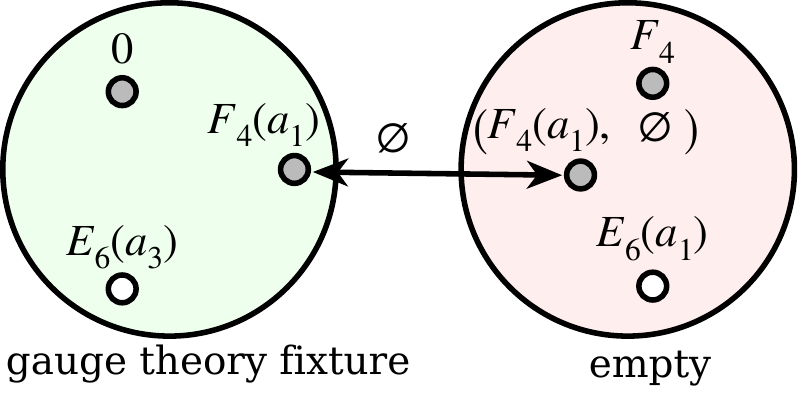}
\end{displaymath}

Finally, the 5-punctured sphere

\begin{displaymath}
 \includegraphics[width=122pt]{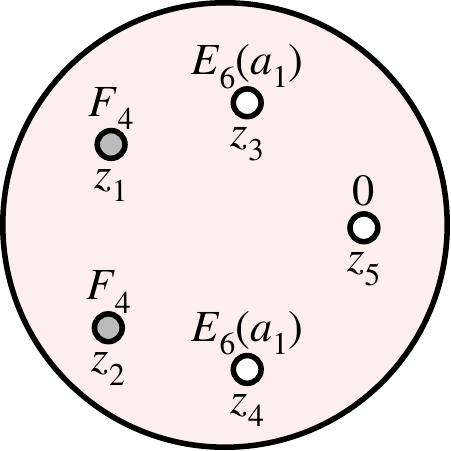}
\end{displaymath}
gives a realization of \emph{two decoupled copies} of this theory. The gauge theory moduli space is a 4-fold branched cover of $\mathcal{M}_{0,4}$, with coordinates $(y,w)$ given in terms of the cross-ratios as

\begin{displaymath}
y^2 = s_1 = \frac{z_{1 3}z_{2 5}}{z_{1 5}z_{2 3}},\qquad
w^2 = s_2 = \frac{z_{1 4}z_{2 5}}{z_{1 5}z_{2 4}}
\end{displaymath}
The gauge couplings are

\begin{displaymath}
f(\tau_1) = \frac{y-1}{y+1}\frac{w+1}{w-1},\qquad
f(\tau_2) = \frac{y-1}{y+1}\frac{w-1}{w+1}
\end{displaymath}
where

\begin{equation}
f(\tau)\equiv - \frac{\theta_2^4(0,\tau)}{\theta_4^4(0,\tau)}
\label{gaugecoupling}\end{equation}
and $\tau = \frac{\theta}{\pi}+\frac{8\pi i}{g^2}$. In the limits $f(\tau)\to 0,\infty$, the $SU(3) + \bigl[{(E_8)}_{12}\bigr]$ description is weakly-coupled. For $f(\tau)\to 1$, the $SU(2)+\tfrac{1}{2}(2) +\bigl[{(E_6)}_{12}\times {SU(2)}_7\bigr]$ description is weakly-coupled. Over the degeneration

\begin{displaymath}
 \includegraphics[width=342pt]{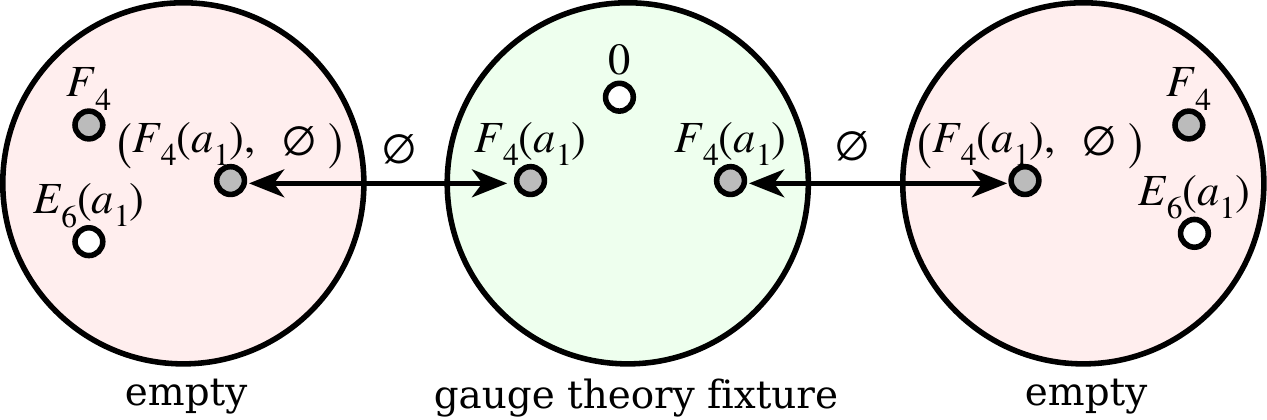}
\end{displaymath}
we have $(f(\tau_1),f(\tau_2))\to (-1,-1)$ and both descriptions are strongly-coupled.

\subsection{$M(E_7,3)/\!/\!/Spin(8) \simeq \Bigl({M(E_7,1)}^3\times \mathbb{H}^{26}\bigr)/\!/\!/F_4$ and $\Bigl(M(E_7,2)\times M(E_7,1)\Bigr)/\!/\!/Spin(8) \simeq \Bigl({M(E_7,1)}^3\times \mathbb{H}^{9}\bigr)/\!/\!/Spin(9)$}\label{ME73_and_ME72}

\eqref{E73} and \eqref{E72E71} both have realizations in the untwisted $E_6$ theory. The former is given by the duality between

\begin{displaymath}
 \includegraphics[width=242pt]{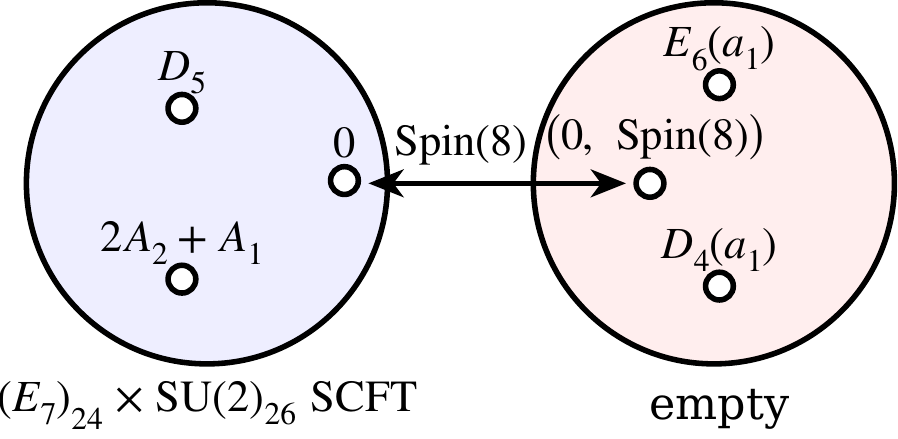}
\end{displaymath}
and

\begin{displaymath}
 \includegraphics[width=236pt]{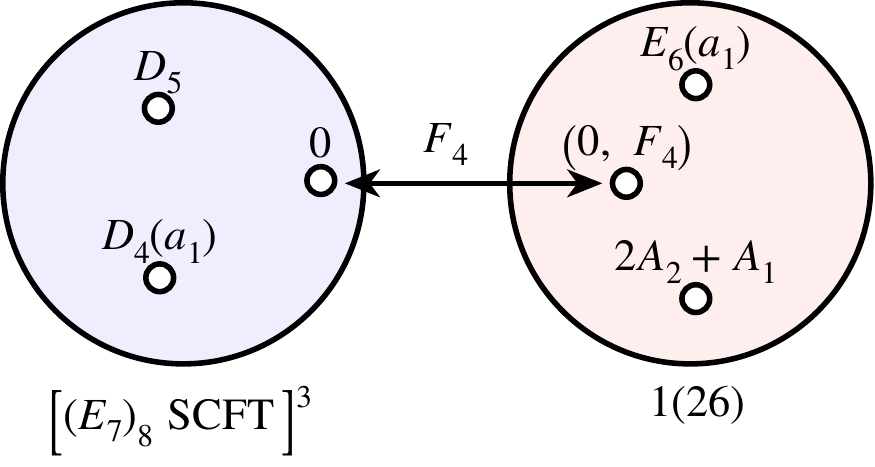}
\end{displaymath}
The latter is given by the duality between

\begin{displaymath}
 \includegraphics[width=236pt]{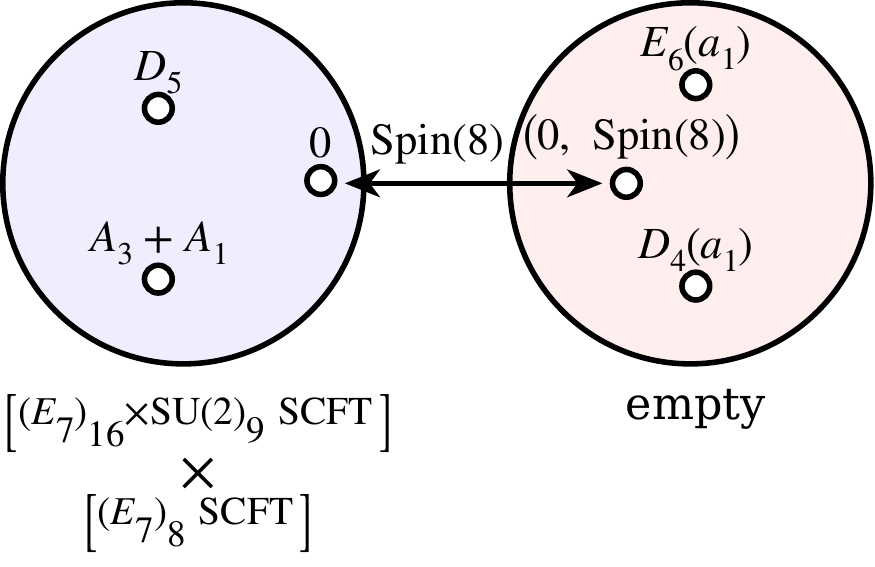}
\end{displaymath}
and

\begin{displaymath}
 \includegraphics[width=236pt]{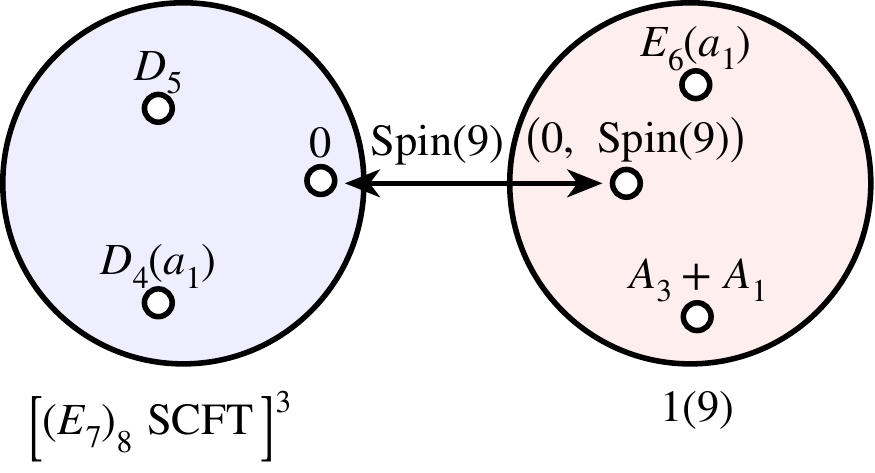}
\end{displaymath}
In both cases, unlike our previous examples, there is a third S-duality frame, respectively

\begin{displaymath}
 \includegraphics[width=268pt]{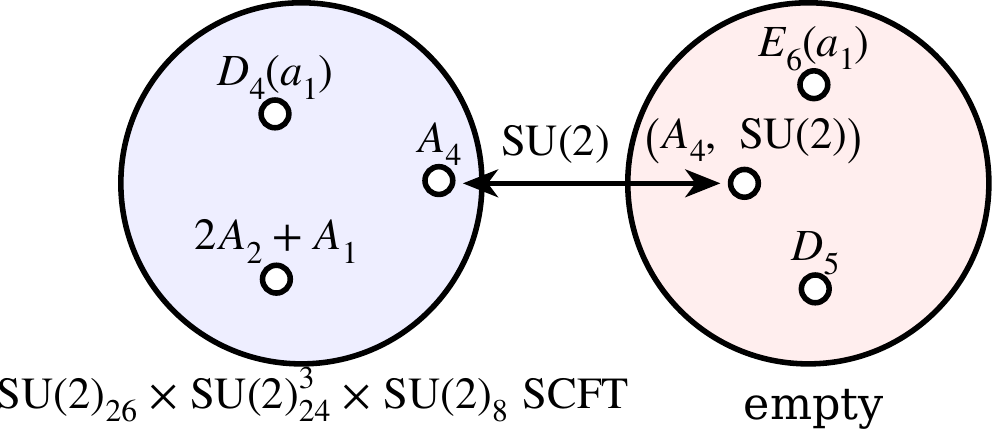}
\end{displaymath}
and

\begin{displaymath}
 \includegraphics[width=260pt]{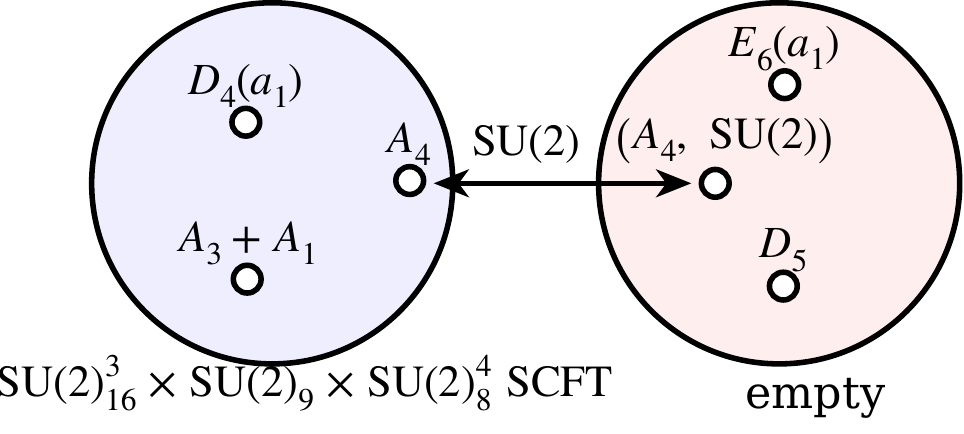}
\end{displaymath}
which are $SU(2)$ gaugings of some new non-Lagrangian SCFTs. Alas, since we don't have an independent construction of the Higgs branches of the latter theories, these isomorphisms don't shed much additional light on these instanton moduli spaces.

\subsection{$\left(M(E_8,2)\times\mathbb{H}^{32}\right)/\!/\!/Spin(12) \simeq \Bigl(M(E_8,1)\times M(E_8,1)\times \mathbb{H}^{45}\bigr)/\!/\!/Spin(13)$}\label{_4}

Turning to \eqref{E82}, there is a realization in the $D_7$ theory

\begin{displaymath}
 \includegraphics[width=326pt]{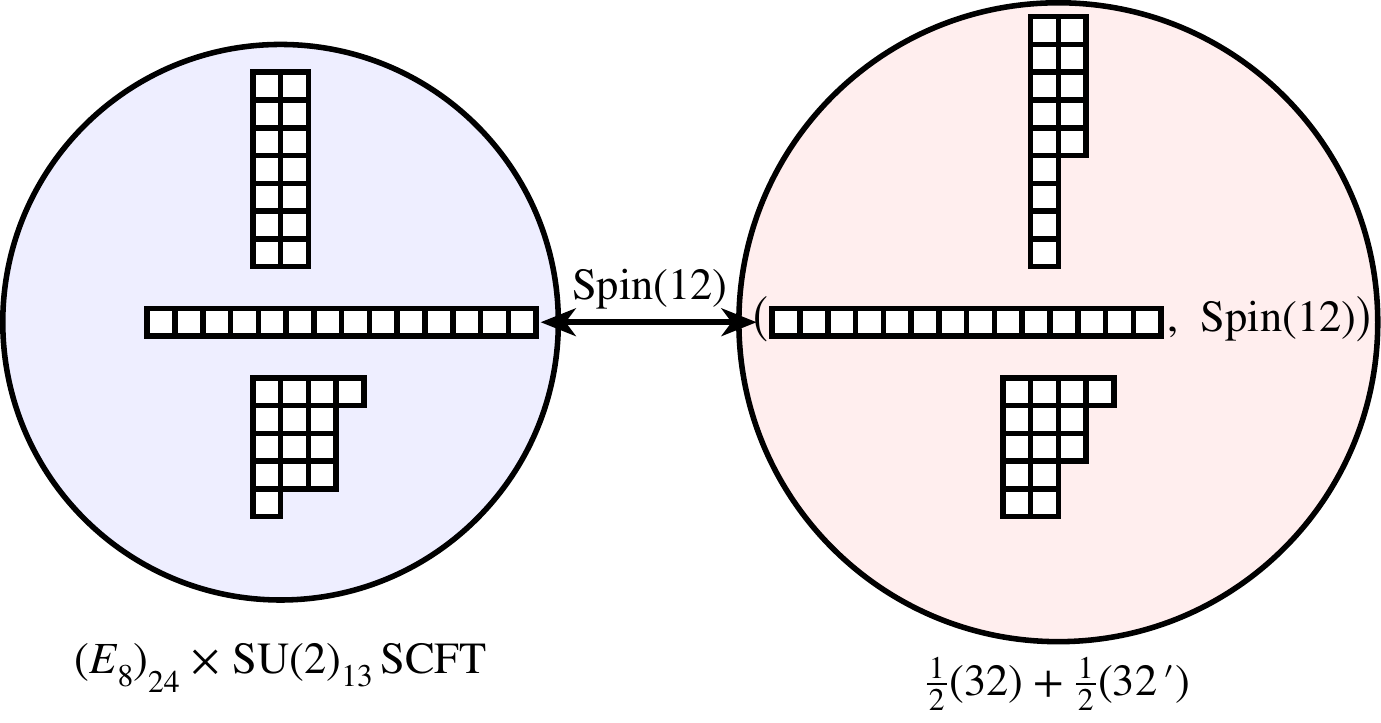}
\end{displaymath}
which has one S-dual presentation as

\begin{displaymath}
 \includegraphics[width=326pt]{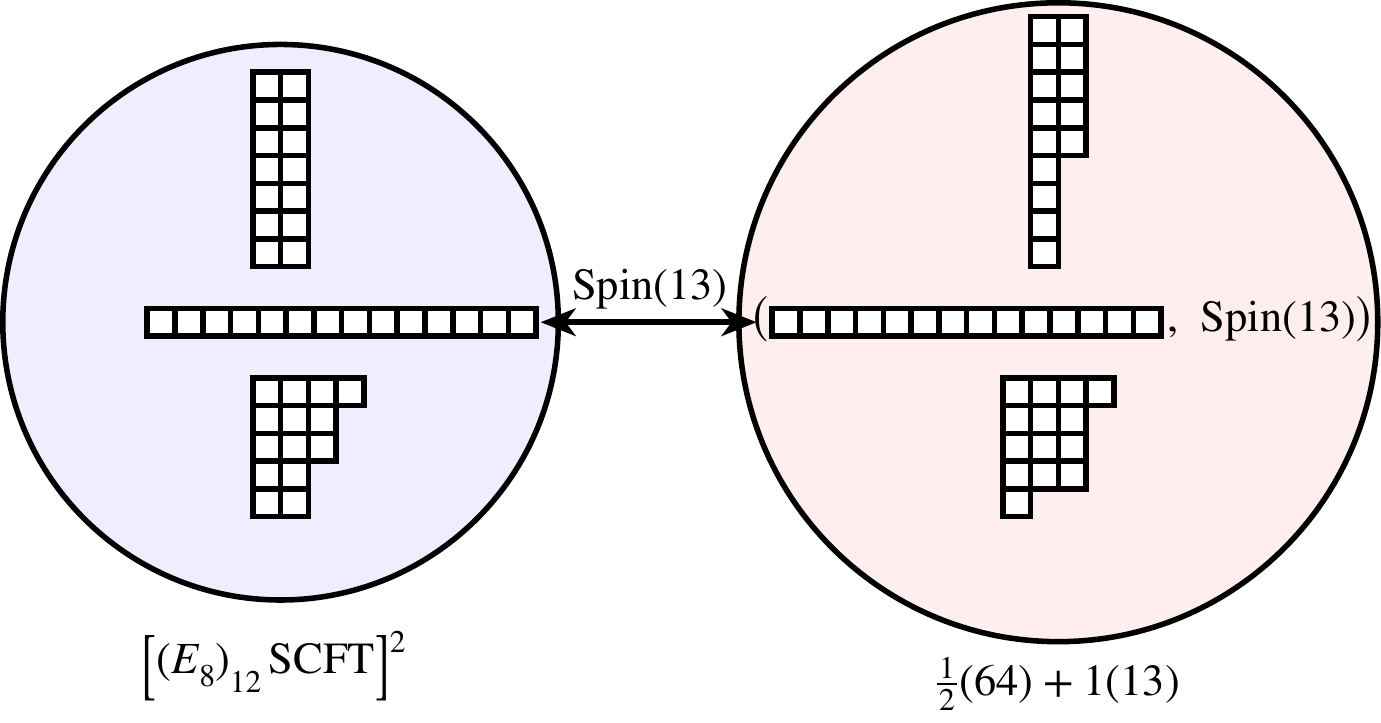}
\end{displaymath}
realizing the isomorphism of Higgs branches stated in \eqref{E82}.

This theory also has a third S-duality frame,
\begin{displaymath}
 \includegraphics[width=326pt]{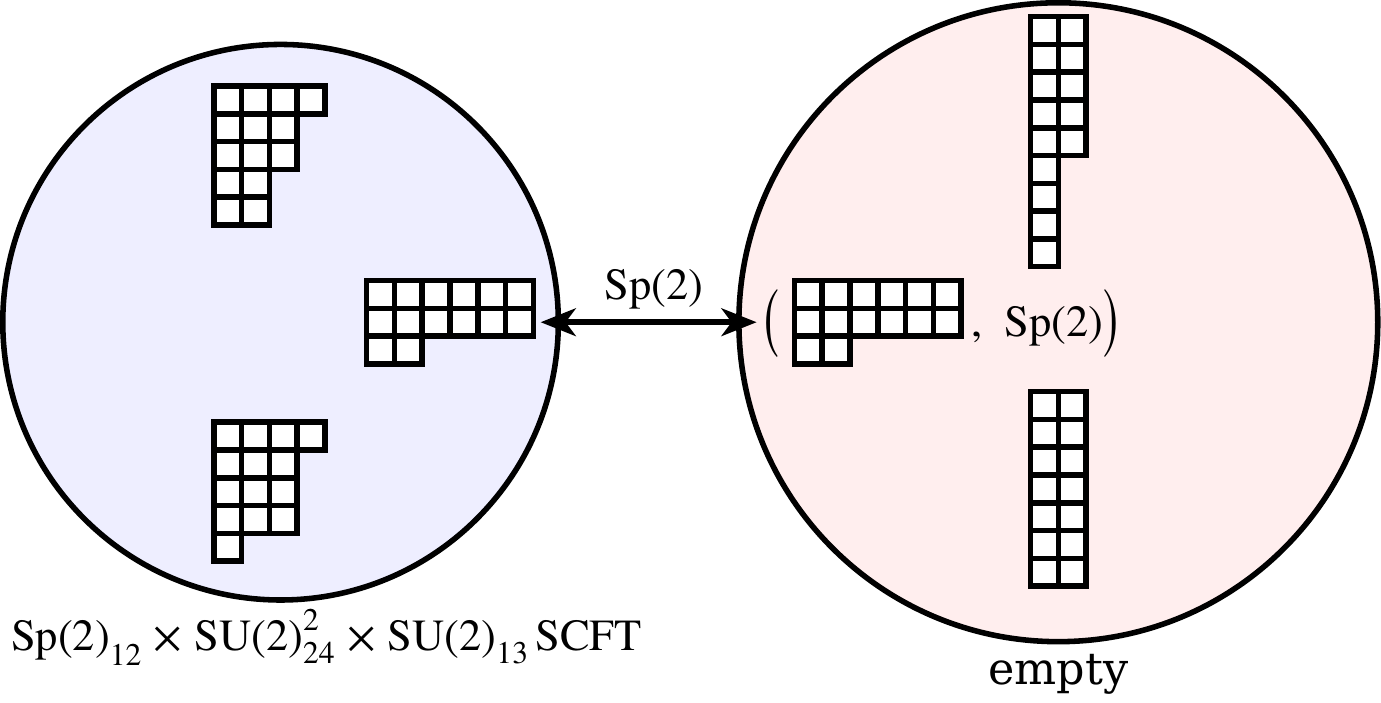}
\end{displaymath}

Alas, as in \S\ref{ME73_and_ME72}, we have no alternative construction of the Higgs branch of the $Sp(2)_{12}\times SU(2)_{24}^2\times SU(2)_{13}$ SCFT, so we don't learn anything new from this duality.

\subsection{Semi-simple quotients}\label{semisimple_quotients}

In \S\ref{InstantonModuliSpaces} we considered isomorphisms of hyperK\"ahler quotients of the form

\begin{equation}
X_1/\!/\!/G_1\simeq X_2/\!/\!/G_2
\label{simple}\end{equation}
where $G_i$ is a \emph{simple} subgroup of the group of hyperK\"ahler isometries of $X_i$. Let $H$ be the residual group of hyperK\"ahler isometries of the quotient. Of course, we can further quotient both sides of \eqref{simple} by a subgroup of $H$, but this would typically not yield anything new; all it would do is lose some of the information contained in \eqref{simple}.

There are, however, exceptions. For instance, we can combine \eqref{E81} with the first isomorphism in \eqref{AS} to obtain

\begin{displaymath}
M(E_8,1)/\!/\!/SU(3)\times Sp(2) \simeq \mathbb{H}^{40}/\!/\!/SU(3)\times Sp(3)\simeq \left(M(E_6,1)\times \mathbb{H}^{24}\right)/\!/\!/SU(2)\times Sp(3)
\end{displaymath}
where

\begin{displaymath}
\begin{split}
\mathbb{H}^{40} &= \tfrac{5}{2}(1,6)+\tfrac{1}{2}(1,14')+ (3,6)\quad\text{of}\, SU(3)\times Sp(3)\\
\mathbb{H}^{24} &= \tfrac{5}{2}(1,6)+\tfrac{1}{2}(1,14')+ (2,1)\quad\text{of}\, SU(2)\times Sp(3)
\end{split}
\end{displaymath}
These isomorphisms are realized in the twisted $D_4$ theory, as the 5-punctured sphere

\begin{displaymath}
 \includegraphics[width=463pt]{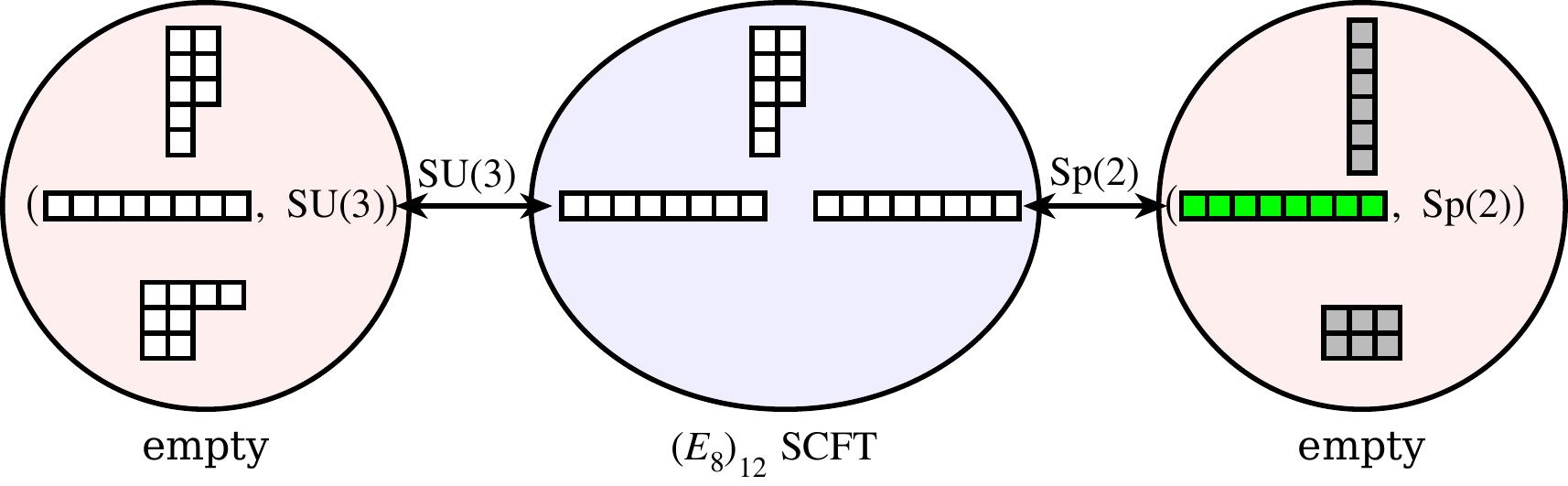}
\end{displaymath}
has, among its various other S-duality frames,

\begin{displaymath}
 \includegraphics[width=463pt]{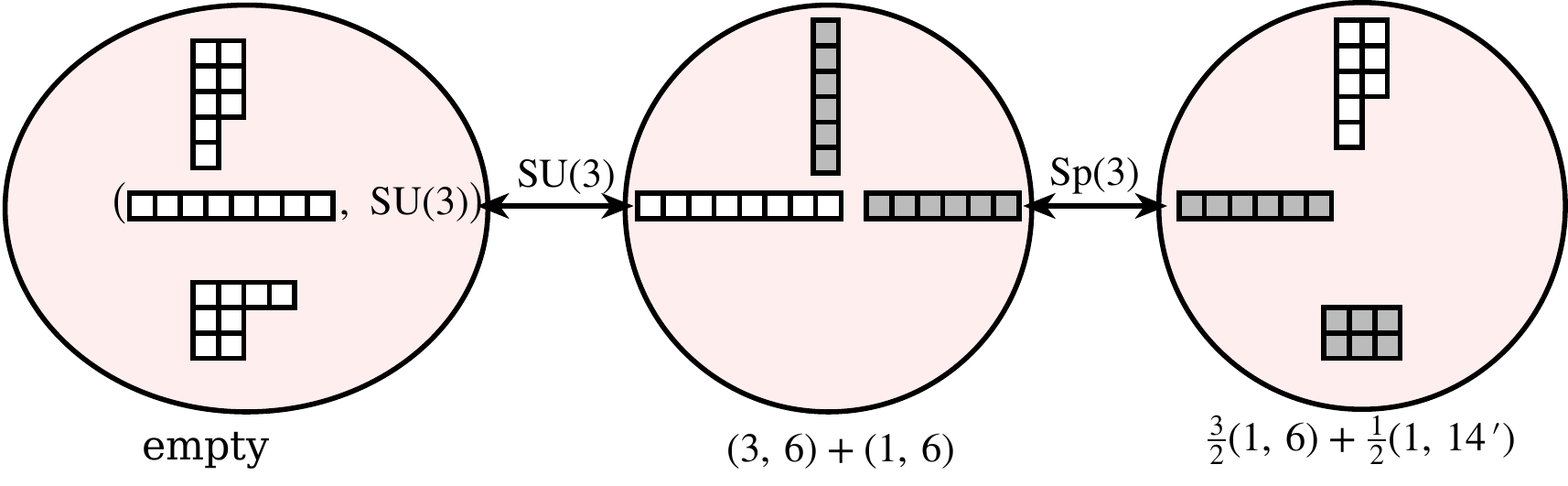}
\end{displaymath}
and

\begin{displaymath}
 \includegraphics[width=463pt]{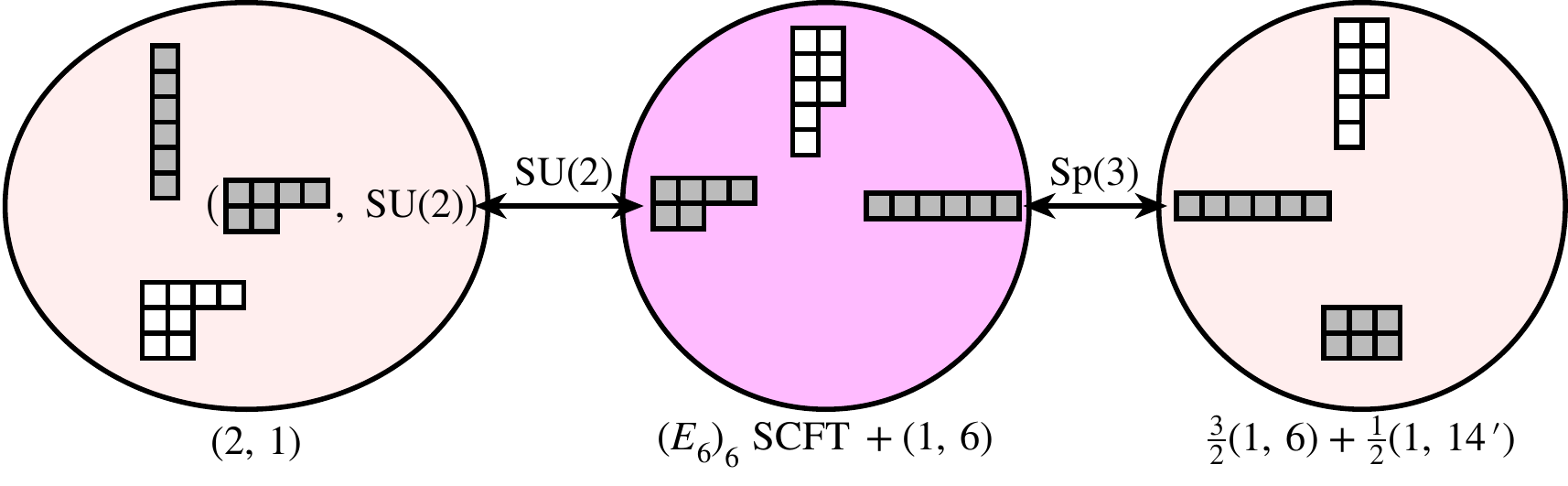}
\end{displaymath}
\subsection{More isomorphisms among hyperK\"ahler quotients}\label{more_isomorphisms_among_hyperKahler_quotients}

If we are are willing to venture a little further afield, we can find additional hyperK\"ahler quotient identities satisfied by the $M(G,k)$. In \S\ref{R25}, we recalled the $R_{2,2n-1}$ series of SCFTs. Let us denote the Higgs branch of $R_{2,2n-1}$ as $M_{2,2n-1}$. $M_{2,2n-1}$ has hyperK\"ahler isometry group $Spin(4n+2)\times U(1)$ and dimension

\begin{displaymath}
\dim_{\mathbb{H}}(M_{2,2n-1}) = 2n^2+n+1
\end{displaymath}
From the S-dualities in \eqref{R22nm1dualities}, certain hyperK\"ahler quotients of $M_{2,2n-1}$ are isomorphic to hyperK\"ahler quotients of quaternionic vector spaces

\begin{displaymath}
\begin{split}
\left(M_{2,2n-1}\times \mathbb{H}^{2(n-1)}\right)/\!/\!/Sp(n-1)&\simeq \mathbb{H}^{(2n-1)(2n+2)}/\!/\!/SU(2n-1)\\
\left(M_{2,2n-1}\times \mathbb{H}^{6n}\right)/\!/\!/Sp(n)&\simeq \mathbb{H}^{2n(2n+3)}/\!/\!/SU(2n)\\
M_{2,2n-1}/\!/\!/Spin(2n+1)&\simeq \mathbb{H}^{4n^2}/\!/\!/SU(2n)
\end{split}
\end{displaymath}
where, in the first two, the quaternionic vector space on the RHS transforms as $4\bigl( \includegraphics[width=9pt]{fund}\bigr)+2\Bigl(\begin{matrix} \includegraphics[width=9pt]{antisym}\end{matrix}\Bigr)$ and, in the third, it transforms as $1\Bigl(\begin{matrix} \includegraphics[width=9pt]{antisym}\end{matrix}\Bigr)+
1\bigl( \includegraphics[width=17pt]{sym}\bigr)$.

This isn't quite enough information to reconstruct $M_{2,2n-1}$. But, with a certain poetic license, we can proceed as if we understand that hyperK\"ahler space.

Using the realization of $M_{2,5}$ given in \S\ref{R25}, we have the new isomorphisms

\begin{equation}
\begin{split}
\left({M(E_6,1)}^4\times \mathbb{H}^{20}\right)/\!/\!/Spin(10)&\simeq\left({M(E_6,1)}^3\times M_{2,5}\right)/\!/\!/Spin(9)\\
\left(M(E_6,2)\times{M(E_6,1)}^2\times \mathbb{H}^{20}\right)/\!/\!/Spin(10)&\simeq\left(M(E_6,2)\times M(E_6,1)\times M_{2,5}\right)/\!/\!/Spin(9)\\
\left(M(E_6,3)\times M(E_6,1)\times \mathbb{H}^{20}\right)/\!/\!/Spin(10)&\simeq\left(M(E_6,3)\times M_{2,5}\right)/\!/\!/Spin(9)
\end{split}
\label{R25isoms}\end{equation}
from studying the 4-punctured spheres

\begin{displaymath}
 \includegraphics[width=261pt]{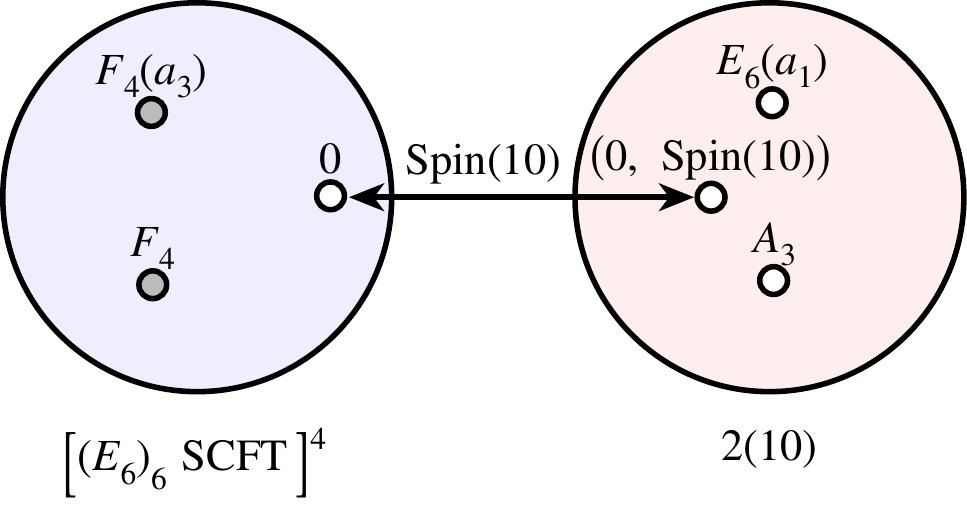}
\end{displaymath}
\begin{displaymath}
 \includegraphics[width=261pt]{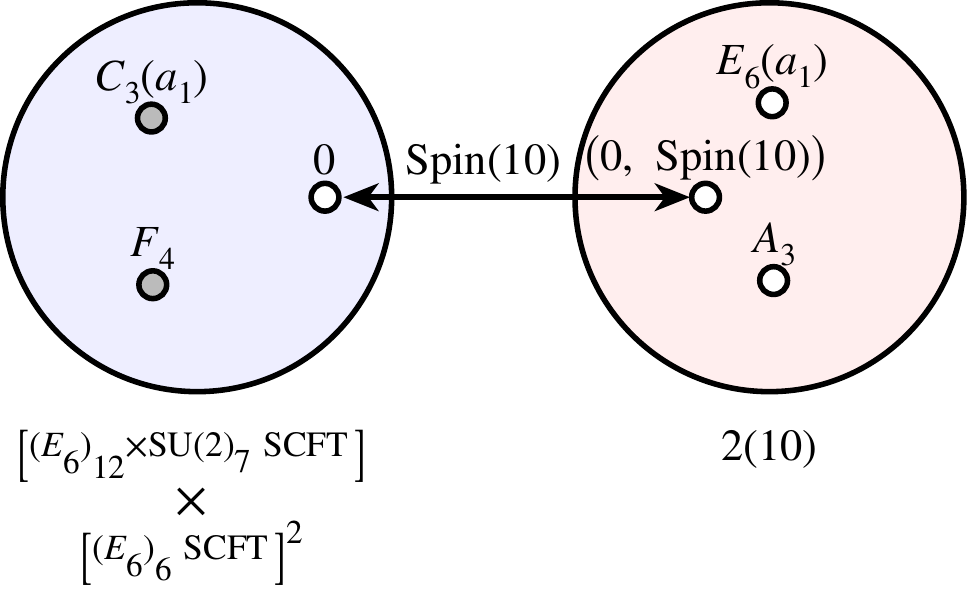}
\end{displaymath}
and

\begin{displaymath}
 \includegraphics[width=261pt]{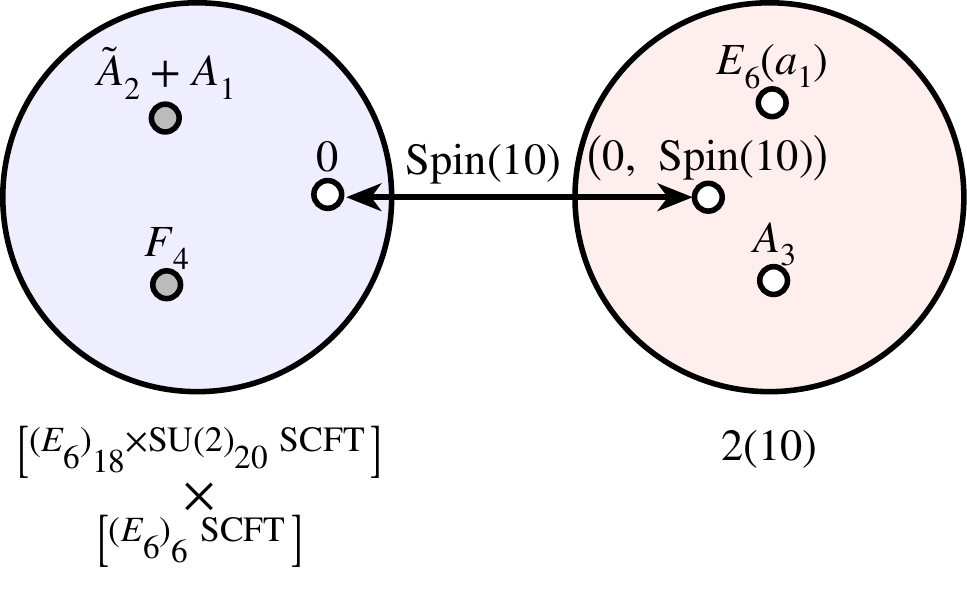}
\end{displaymath}
which are, respectively, S-dual to

\begin{displaymath}
 \includegraphics[width=261pt]{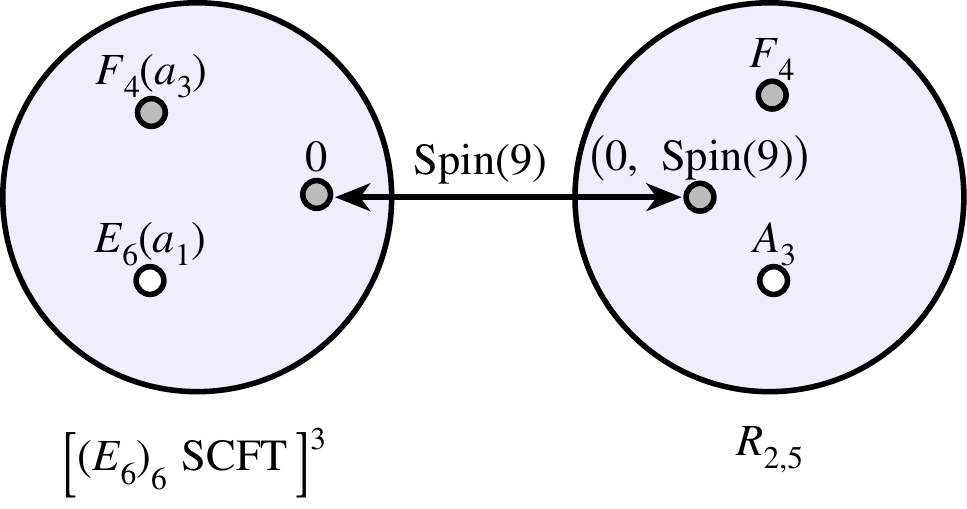}
\end{displaymath}
\begin{displaymath}
 \includegraphics[width=261pt]{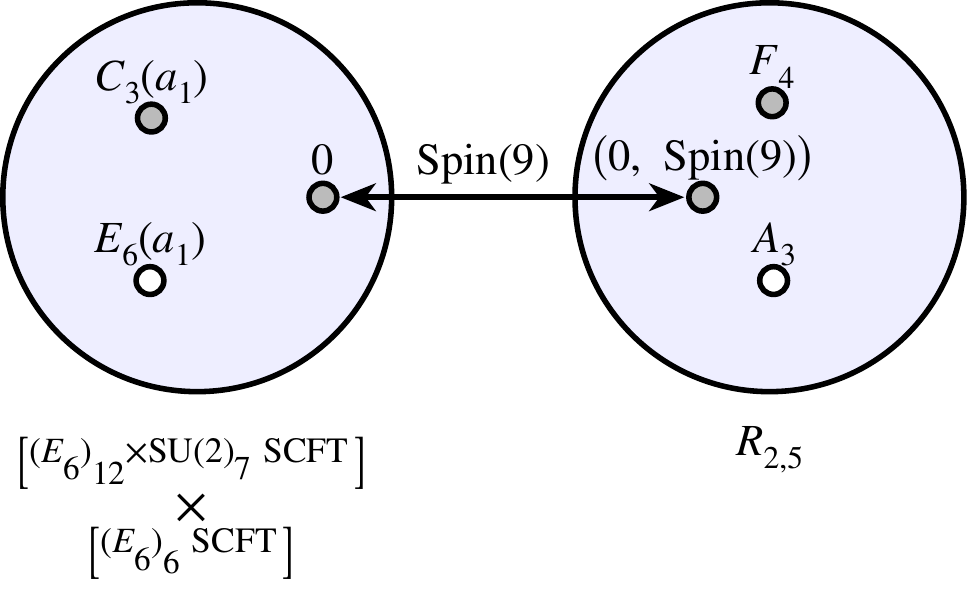}
\end{displaymath}
and

\begin{displaymath}
 \includegraphics[width=261pt]{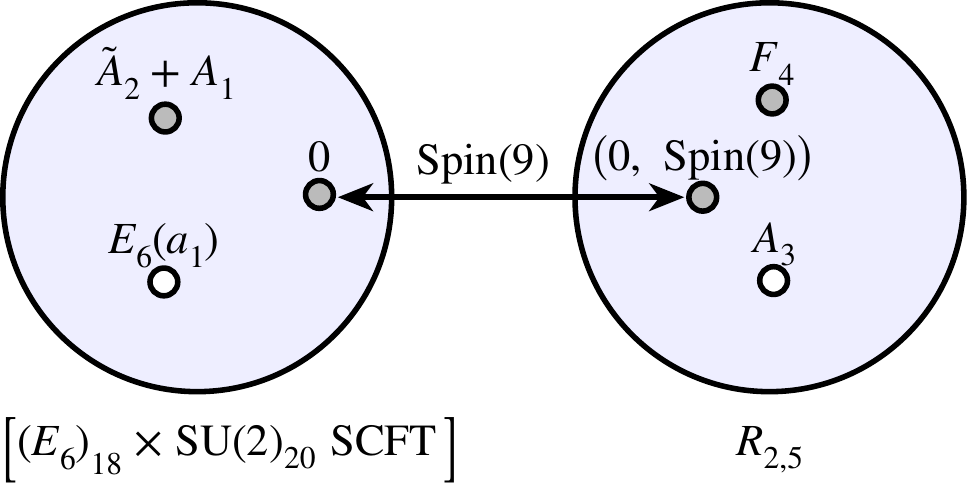}
\end{displaymath}

Note that:

\begin{itemize}%
\item The three examples are related by allowing the twisted puncture in the upper left corner to vary over the special piece of $F_4(a_3)$. (As discussed above, this special piece consists of \emph{five} nilpotent orbits. The other two involve theories whose Higgs branches are ``new'' hyperK\"ahler spaces.)
\item In each case, there's a third S-duality frame, which we won't write down, which is a gauge theory fixture.

\end{itemize}
\section{Instanton moduli spaces as affine algebraic varieties}\label{instanton_moduli_spaces_as_affine_algebraic_varieties}

As mentioned above, $M(G,1)$ admits a uniform description as the minimal nilpotent orbit in $\mathfrak{g}_\mathbb{C}$. For classical groups, $G$, the ADHM construction \cite{Atiyah:1978ri} gives a description of $M(G,k)$, for higher $k$, as a hyperK\"ahler quotient. For exceptional $G$, a concrete description of the $M(G,k)$ for higher $k$ is not known. However, in a series of papers \cite{Benvenuti:2010pq,Hanany:2012dm,Cremonesi:2014xha}, it was shown that the Hilbert series of $M(G,k)$ for $k \gt 1$ and classical $G$ can be written in terms of the root data of $G$ alone. This provides a natural conjecture for the Hilbert series of $M(G,k)$ for exceptional $G$, which has been shown to pass many tests.

The Hilbert series contains all information about the ring of holomorphic functions on $M(G,k)$. From this information, in \cite{Cremonesi:2014xha} the authors extracted the representations of the generators of $M(G,k)$ at each scaling dimension, and their lowest order chiral ring relations.

They conjectured that $M(G,k)$, as a complex variety, can be realized as an affine algebraic variety whose ring of functions has generators

\begin{displaymath}
\begin{split}
M &\in (1;\text{Adj}) \\
M_p &\in (p;\text{Adj}) \\
P_p &\in (p+1;1) \\
\end{split}
\end{displaymath}
transforming in the indicated representations\footnote{We label irreducible representations of $SU(2)$ by their dimension. In what follows, it is convenient to realize the $n$-dimensional irrep as a rank-$(n-1)$ symmetric tensor, $\Phi_{(\alpha_1\alpha_2\dots\alpha_{n-1})}$.}  of $SU(2) \times G$, for $p=2,\dots,k$.

These generators are subject to a set of polynomial relations. For $k=1$, the $M_p$ and $P_p$ are absent, and the only non-trivial relations are the celebrated Joseph relations \cite{MR0404366}

\begin{equation}
(M \otimes M) \vert_{\mathcal{I}_2}=0
\label{Joseph}\end{equation}
where the reducible representation, $\mathcal{I}_2$, is defined through

\begin{displaymath}
Sym^2(\text{Adj})=V(2\alpha)\oplus\mathcal{I}_2
\end{displaymath}
Here, $V(2\alpha)$ is the representation whose highest weight is twice the highest root. \eqref{Joseph} gives a realization of $M(G,1)$ as an affine algebraic variety.

For $k=2,3$, the lowest-order relations are given in \cite{Cremonesi:2014xha}.

The isomorphisms discussed in \S\ref{InstantonModuliSpaces} provide a strong test of this conjectured description of $M(G,k)$. Following \cite{Gaiotto:2008nz}, one can explicitly take the hyperK\"ahler quotient on each side, and compare the gauge-invariant generators and relations.

As an illustrative example, we consider $(M(E_7,1)^3 \times \mathbb{H}^{26})/\!/\!/F_4 \simeq M(E_7,3)/\!/\!/Spin(8)$. We will not give a precise mapping of the generators, as the methods of \cite{Cremonesi:2014xha} do not determine the constants appearing in the relations defining $M(E_7,3)$ but, up to a few unknown constants, we will be able to determine the form of the correspondence. The generators transform in representations of the $SU(2)_s \times SU(2)^3$ global symmetry. Moreover, there is an action of $S_3$ permuting the $SU(2)^3$. On the LHS, it acts by permuting the three $M(E_7,1)$s; on the RHS, it is the $S_3$ subgroup of $E_7$ which acts as triality on the $Spin(8)\subset E_7$. The generators of the ring of functions arrange themselves into representations of this $S_3$ action.

We first consider the proposed description of $M(E_7,3)$ above. Decomposing the 133 of $E_7$ under $SU(2)^3 \times Spin(8)$:

\begin{equation*}
\begin{split}
E_7 &\supset SU(2)^3 \times Spin(8) \\
133 &= (3,1,1;1)+(1,3,1;1)+(1,1,3;1)+(1,1,1;28)\\
&\quad +(2,2,1;8_c)+(2,1,2;8_s)+(1,2,2;8_v)
\end{split}
\end{equation*}
we have operators

{
\renewcommand{\arraystretch}{1.5}
\begin{longtable}{|c|c|c|}
\hline
Order&Operator&Representation of $SU(2)_s \times SU(2)^3 \times Spin(8)$\\
\hline 
\endhead
2&$\Psi_{(\alpha\beta)}$&$(3;1,1,1;1)$\\
&$J,K,L$&$(1;3,1,1;1),(1;1,3,1;1),(1;1,1,3;1)$\\
&$M,N,O$&$(1;2,2,1;8_c), (1;2,1,2;8_s), (1;1,2,2;8_v)$\\
&$P$&$(1;1,1,1;28)$\\
\hline
3&$\Phi_{(\alpha\beta\gamma)}$&$(4;1,1,1;1)$\\
&$Q_\alpha, R_\alpha, S_\alpha$&$(2;3,1,1;1), (2;1,3,1;1), (2;1,1,3;1)$\\
&$T_\alpha, U_\alpha, V_\alpha$&$(2;2,2,1;8_c), (2;2,1,2;8_s), (2;1,2,2;8_v)$\\
&$W_\alpha$&$(2;1,1,1;28)$\\
\hline
4&$X_{(\alpha\beta)}, Y_{(\alpha\beta)}, Z_{(\alpha\beta)}$&$(3;3,1,1;1),(3;1,3,1;1),(3;1,1,3;1)$\\
&$\widetilde{A}_{(\alpha\beta)}, \widetilde{B}_{(\alpha\beta)}, \widetilde{C}_{(\alpha\beta)}$&$(3;2,2,1;8_c), (3;2,1,2;8_s), (3;1,2,2;8_v)$\\
\hline
&$\widetilde{D}_{(\alpha\beta)}$&$(3;1,1,1;28)$\\
\hline
\end{longtable}
}

The lowest-order relation is at order 5, given by \cite{Cremonesi:2014xha}

\begin{displaymath}
(JQ_\alpha+a_1KR_\alpha+a_2LS_\alpha+a_3MT_\alpha+a_4NU_\alpha+a_5OV_\alpha+a_6PW_\alpha)|_{(2;1,1,1;1)}=0,
\end{displaymath}
where the $a_i$ are constants.

Let us now take the hyperK\"ahler quotient by Spin(8). The F-term constraint is simply

\begin{displaymath}
P=0.
\end{displaymath}
So, the gauge-invariant operators are given by

{
\renewcommand{\arraystretch}{1.5}
\begin{longtable}{|c|c|c|}
\hline
Order&Operator&Representation of $SU(2)_s \times SU(2)^3$\\
\hline
\endhead
2&$\Psi_{(\alpha\beta)}$&$(3;1,1,1)$\\
&$J,K,L$&$(1;3,1,1), (1;1,3,1), (1;1,1,3)$\\
\hline
3&$\Phi_{(\alpha\beta\gamma)}$&$(4;1,1,1)$\\
&$Q_\alpha, R_\alpha, S_\alpha$&$(2;3,1,1),(2;1,3,1),(2;1,1,3)$\\
\hline
4&$X_{(\alpha\beta)}, Y_{(\alpha\beta)}, Z_{(\alpha\beta)}$&$(3;3,1,1),(3;1,3,1),(3;1,1,3)$\\
&$M^2, N^2, O^2$&$(1;1,1,1)+(1;3,3,1), (1;1,1,1)+(1;3,1,3), (1;1,1,1)+(1;1,3,3)$\\
\hline
\end{longtable}
}
\noindent
subject to

\begin{equation}
(JQ_\alpha+a_1KR_\alpha+a_2LS_\alpha+a_3MT_\alpha+a_4NU_\alpha+a_5OV_\alpha)|_{(2;1,1,1;1)}=0.
\label{relation}\end{equation}
Let's see how this structure is reproduced on the $M(E_7,1)^3$ side. We first decompose the $(E_7)^3$ global symmetry under $SU(2)^3 \times (F_4)_\text{diag}$. Using the description of $M(E_7,1)$ as the minimal nilpotent orbit in $\mathfrak{e}_7$, we have operators at order 2 in the 133, subject to the Joseph relations at order 4 in the $\mathcal{I}_2 = 1+1539$. These representations decompose under $SU(2) \times F_4$ as

\begin{displaymath}
\begin{split}
E_7 &\supset SU(2) \times F_4 \\
133&=(3,1)+(3,26)+(1,52) \\
1539&=(1,1)+(1,26)+(1,324)+(3,26)+(3,273)+(3,52)+(5,1)+(5,26)
\end{split}
\end{displaymath}
Additionally, we have the generator of $\mathbb{H}^{26}$ at order 1. In total, we have the following generators of $M(E_7,1)^3 \times \mathbb{H}^{26}$:

{
\renewcommand{\arraystretch}{1.5}
\begin{longtable}{|c|c|c|}
\hline
Order&Operator&Representation of $SU(2)_s \times SU(2)^3 \times (F_4)_{diag}$\\
\hline
\endhead
1&$v_{\alpha}$&$(2;1,1,1;26)$\\
\hline
2&$A,B,C$&$(1;3,1,1;1), (1;1,3,1;1), (1;1,1,3;1)$\\
&$D,E,F$&$(1;3,1,1;26), (1;1,3,1;26), (1;1,1,3;26)$\\
&$G,H,I$&$(1;1,1,1;52), (1;1,1,1;52), (1;1,1,1;52)$\\
\hline
\end{longtable}
}
\noindent
subject to the Joseph relations at order 4.

To describe the hyperK\"ahler quotient by $(F_4)_\text{diag}$, we impose the F-term constraints
\begin{displaymath}
G+H+I+(v_{\alpha} v_{\beta})_{(1;1,1,1;52)}=0
\end{displaymath}
and form gauge-invariant generators. To order 4, these are given by:

{
\renewcommand{\arraystretch}{1.5}
\begin{longtable}{|c|c|c|}
\hline
Order&Operator&Representation of $SU(2)_s \times SU(2)^3$\\
\hline
\endhead
2&$(v_{\alpha} v_{\beta})_{(3;1,1,1;1)}$&$(3;1,1,1)$\\
&$A,B,C$&$(1;3,1,1;1), (1;1,3,1;1), (1;1,1,3;1)$\\
\hline
3&$(v_\alpha v_\beta v_\gamma)_{(4;1,1,1;1)}$&$(4;1,1,1)$\\
&$(Dv_\alpha)_{(2;3,1,1;1)}, (Ev_\alpha)_{(2;1,2,1;1)}, (Fv_\alpha)_{(2;1,1,3;1)}$&$(2;3,1,1), (2;1,3,1), (2;1,1,3)$\\
\hline
4&$\begin{gathered}((v_\alpha v_\beta)_{(3;1,1,1;26)}D)_{(3;3,1,1;1)},\\ ((v_\alpha v_\beta)_{(3;1,1,1;26)}E)_{(3;1,3,1;1)},\\ ((v_\alpha v_\beta)_{(3;1,1,1;26)}F)_{(3;1,1,3;1)}\end{gathered}$&$(3;3,1,1), (3;1,3,1), (3;1,1,3)$\\
&$(D^2)_{(1;1+5,1,1;1)}, (E^2)_{(1;1,1+5,1;1)}, (F^2)_{(1;1,1,1+5;1)}$&$\begin{gathered}(1;1+5,1,1),\\ (1;1,1+5,1),\\ (1;1,1,1+5)\end{gathered}$\\
&$(G^2)_{(1;1,1,1;1)}, (H^2)_{(1;1,1,1;1)}, (I^2)_{(1;1,1,1;1)}$&$(1;1,1,1), (1;1,1,1), (1;1,1,1)$\\
&$(DE)_{(1;3,3,1;1)}, (DF)_{(1;3,1,3;1)}, (EF)_{(1;1,3,3;1)}$&$(1;3,3,1), (1;3,1,3), (1;1,3,3)$\\
&$(GH)_{(1;1,1,1;1)}, (GI)_{(1;1,1,1;1)}, (HI)_{(1;1,1,1;1)}$&$(1;1,1,1), (1;1,1,1), (1;1,1,1)$\\
\hline
\end{longtable}
}

The gauge-invariant relations at order 4 are given by

\begin{equation}
(A^2+c_1D^2+c_2G^2)|_{(1;1,1,1)}=0
\label{r1}\end{equation}
\begin{equation}
(B^2+c_1E^2+c_2H^2)|_{(1;1,1,1)}=0
\label{r2}\end{equation}
\begin{equation}
(C^2+c_1F^2+c_2I^2)|_{(1;1,1,1)}=0
\label{r3}\end{equation}
\begin{equation}
(A^2+c_3D^2)|_{(1;5,1,1)}=0
\label{r4}\end{equation}
\begin{equation}
(B^2+c_3E^2)|_{(1;1,5,1)}=0
\label{r5}\end{equation}
\begin{equation}
(C^2+c_3F^2)|_{(1;1,1,5)}=0
\label{r6}\end{equation}
where the $c_i$ are constants which can be fixed by evaluating a few points on the nilpotent orbit \cite{Gaiotto:2008nz}.

We see that the correspondence between the generators is given by\footnote{We have multiple generators with the same quantum numbers, so, without knowing the constants $a_i$, the correspondence between these generators is only up to a permutation (or linear combination).}

{
\renewcommand{\arraystretch}{2}
\begin{longtable}{|c|c|}
\hline
$(M(E_7,1)^3\times \mathbb{H}^{26})/\!/\!/F_4$&$M(E_7,3)/\!/\!/Spin(8)$\\
\hline
\hline
\endhead
$(v_{\alpha} v_{\beta})_{(3;1,1,1;1)}$&$\Psi_{(\alpha\beta)}$\\
\hline
$A,B,C$&$J,K,L$\\
\hline
$(v_{\alpha} v_{\beta}v_{\gamma})_{(4;1,1,1;1)}$&$\Phi_{(\alpha\beta\gamma)}$\\
\hline
$((v_\alpha D )_{(2;3,1,1;1)},(v_\alpha E)_{(2;1,3,1;1)},(v_\alpha F)_{(2;1,1,3;1)}$&$Q_\alpha,R_\alpha,S_\alpha$\\
\hline
$\begin{gathered}((v_\alpha v_\beta)_{(3;1,1,1;26)}D)_{(3;3,1,1;1)},\\ ((v_\alpha v_\beta)_{(3;1,1,1;26)}E)_{(3;1,3,1;1)},\\ ((v_\alpha v_\beta)_{(3;1,1,1;26)}F)_{(3;1,1,3;1)}\end{gathered}$&$X_{(\alpha\beta)},Y_{(\alpha\beta)},Z_{(\alpha\beta)}$\\
\hline
$\begin{gathered}(GH)_{(1;1,1,1;1)},\\(DE)_{(1;3,3,1;1)}\end{gathered}$&$\begin{gathered}M^2_{(1;1,1,1;1)},\\M^2_{(1;3,3,1;1)}\end{gathered}$\\
\hline
$\begin{gathered}(GI)_{(1;1,1,1;1)},\\(DF)_{(1;3,1,3;1)}\end{gathered}$&$\begin{gathered}N^2_{(1;1,1,1;1)},\\N^2_{(1;3,1,3;1)}\end{gathered}$\\
\hline
$\begin{gathered}(HI)_{(1;1,1,1;1)},\\(EF)_{(1;1,3,3;1)}\end{gathered}$&$\begin{gathered}O^2_{(1;1,1,1;1)},\\O^2_{(1;1,3,3;1)}\end{gathered}$\\
\hline
\end{longtable}
}

The ``extra'' generators at order 4, $(D^2)_{(1;1+5,1,1;1)}$, $(E^2)_{(1;1,1+5,1;1)}$, $(F^2)_{(1;1,1,1+5;1)}$ and $(G^2)_{(1;1,1,1;1)}$, $(H^2)_{(1;1,1,1;1)}$, $(I^2)_{(1;1,1,1;1)}$, are removed from the chiral ring by the Joseph relations \eqref{r1}-\eqref{r6}.

We find the order 5 relation \eqref{relation} on the $M(E_7,1)^3$ side by adding the order 4 Joseph relations

\begin{equation}
(AD)_{(1;3,1,1;26)}+c_{4}(DG)_{(1;3,1,1;26)}=0
\label{r7}\end{equation}
\begin{equation}
(BE)_{(1;1,3,1;26)}+c_{4}(EH)_{(1;1,3,1;26)}=0
\label{r8}\end{equation}
\begin{equation}
(CF)_{(1;1,1,3;26)}+c_{4}(FI)_{(1;1,1,3;26)}=0
\label{r9}\end{equation}
and contracting with $v_\alpha$:

\begin{displaymath}
(vAD+vBE+vCF+c_{4}(vDG+vEH+vFI))_{(2;1,1,1;1)}=0.
\end{displaymath}
Following \cite{Cremonesi:2014xha}, one can extract the higher-order relations for $M(E_7,3)$ and compare them with those on the $M(E_7,1)^3$ side obtained from the remaining Joseph relations. It would be interesting to carry out this analysis for the other examples in \S\ref{InstantonModuliSpaces} as well.

\section*{Acknowledgements}\label{Acknowledgements}
\addcontentsline{toc}{section}{Acknowledgements}
We would like to thank Tom Mainiero, Noppadol Mekareeya, Andy Neitzke, Yuji Tachikawa, and Fei Yan for helpful discussions. The work of J.D. and A.T. was supported in part by the National Science Foundation under Grant No. PHY-1316033. The work of O.C. was supported in part by the INCT-Matem\'atica and the ICTP-SAIFR in Brazil through a Capes postdoctoral fellowship. O.C. would like to thank the Johns Hopkins University, and especially Jared Kaplan, for hospitality while this work was being completed. A.T. thanks the hospitality of the Kavli IPMU at the University of Tokyo where part of this work was completed under NSF EAPSI award number IIA-1413868 and Japan Society for the Promotion of Science (JSPS) Summer Program 2014, and the Yukawa Institute for Theoretical Physics at Kyoto University for hospitality during the workshop YITP-W-14-4 ``Strings and Fields". He would especially like to thank Yuji Tachikawa and Yu Nakayama for their kind hospitality during his stay in Japan.

\begin{appendices}
\section{Constraints}\label{appendix_constraints}

{
\renewcommand{\arraystretch}{2}
\begin{longtable}{|c|c|c|}
\hline
Bala-Carter&New parameters&Constraints\\
\hline
\endhead
$\widetilde{A}_1$&$h_6 \equiv \frac{a^{(6)}_{11/2}}{z^{11/2}}$&$\phi_{12} - h_6^2 \sim \frac{1}{z^{10}}$\\
\hline
$\widetilde{A}_2$&$h_3 \equiv \frac{a^{(3)}_{5/2}}{z^{5/2}}$&${\begin{gathered}\phi_8 - \phi_5 h_3  \sim \frac{1}{z^6}\\ \phi_{12} + \phi_6^2 + \frac{1}{16}\phi_6 h_3^2 +3\phi_9h_3+\frac{1}{1024}h_3^4 \sim \frac{1}{z^{9}}\end{gathered}}$\\
\hline
$B_2$&${\begin{gathered}h_3 \equiv \frac{a^{(3)}_{5/2}}{z^{5/2}}\\h_6 \equiv \frac{a^{(6)}_{5}}{z^{5}}\end{gathered}}$&${
\begin{gathered}\phi_9 - \frac{1}{12}h_3(h_3^2 +h_6+2\phi_6)  \sim \frac{1}{z^{13/2}}\\
\phi_{12} - h_3^2(h_3^2-h_6)-\frac{1}{4}\left(h_3^2-h_6-2\phi_6\right)^2  \sim \frac{1}{z^9}\end{gathered}
}$\\
\hline
$\widetilde{A}_2+A_1$&$h_3 \equiv \frac{a^{(3)}_{5/2}}{z^{5/2}}$&$\phi_{12} -24\phi_9 h_3 + \left(\phi_6 + 2h_3^2\right)^2 \sim \frac{1}{z^{9}}$\\
\hline
$C_3(a_1)$&$\begin{gathered}h_3  \equiv \frac{a^{(3)}_{5/2}}{z^{5/2}}\\ h'_3  \equiv \frac{a'^{(3)}_{5/2}}{z^{5/2}}\end{gathered}$&$\begin{gathered}\phi_9 - \frac{1}{12}h_3(h_3^2 +{h'}_3^2+2\phi_6)  \sim \frac{1}{z^{13/2}}\\ \phi_{12} - h_3^2(h_3^2-{h'}_3^2)-\frac{1}{4}\left(h_3^2-{h'}_3^2-2\phi_6\right)^2  \sim \frac{1}{z^9}\end{gathered}$\\
\hline
$F_4(a_3)$&$\begin{gathered}h_3  \equiv \frac{a^{(3)}_{5/2}}{z^{5/2}}\\ h'_3  \equiv \frac{{a'}^{(3)}_{5/2}}{z^{5/2}}\\ h''_3  \equiv \frac{{a''}^{(3)}_{5/2}}{z^{5/2}}\end{gathered}$&$\begin{gathered}\phi_6 + 2h_3^2 +{h'}_3^2-3{h''}_3^2 \sim \frac{1}{z^{4}}\\ \phi_9 - \frac{1}{3}h_3({h'}_3^2 + 3{h''}_3^2) \sim \frac{1}{z^{13/2}}\\ \phi_{12} + 8h_3^2 ({h'}_3^2 - 3{h''}_3^2) + \left({h'}_3^2 + 3{h''}_3^2\right)^2 \sim \frac{1}{z^9}\end{gathered}$\\
\hline
$B_3$&$h_4 \equiv \frac{a^{(4)}_{3}}{z^{3}}$&$\begin{gathered}\phi_8 -48h_4^2  \sim \frac{1}{z^{5}}\\ \phi_9 -\phi_5 h_4  \sim \frac{1}{z^{11/2}}\\ \phi_{12} + 96h_4^3 \sim \frac{1}{z^8} \end{gathered}$\\
\hline
$C_3$&$\begin{gathered}h_3  \equiv \frac{a^{(3)}_{5/2}}{z^{5/2}}\\ h_4  \equiv \frac{a^{(4)}_{3}}{z^{3}}\end{gathered}$&$\begin{gathered}\phi_6 +6h_3^2  \sim \frac{1}{z^{4}}\\ \phi_8 -16\phi_2 h_3^2-8\phi_5 h_3-48h_4^2 \sim \frac{1}{z^{5}}\\ \phi_9 + \frac{1}{3}\phi_6 h_3+\phi_5 h_4-2\phi_2 h_4 h_3 + \frac{2}{3}h_3^3 \sim \frac{1}{z^{11/2}}\\ \phi_{12} + \phi_6^2 + 24\phi_9 h_3 - 3\phi_8 h_4 - 12\phi_6\phi_2 h_4 + 4\phi_6 h_3^2 \\ + 24\phi_5h_4h_3 +36 \phi_2^2h_4^2-24\phi_2 h_4 h_3^2+4h_3^4+48h_4^3 \sim \frac{1}{z^7} \end{gathered}$\\
\hline
$F_4(a_2)$&$h_4 \equiv \frac{a^{(4)}_{3}}{z^{3}}$&$\begin{gathered}\phi_8 - 48 h_4^2  \sim \frac{1}{z^{5}}\\ \phi_9 + \phi_5 h_4 \sim \frac{1}{z^{11/2}}\\ \phi_{12} + \phi_6^2-12h_4(\frac{1}{4}\phi_8- 3\phi_2^2 h_4+\phi_6 \phi_2 -4h_4^2) \sim \frac{1}{z^7} \end{gathered}$\\
\hline
$F_4(a_1)$&$\begin{gathered} h_2  \equiv \frac{a^{(2)}_{3/2}}{z^{3/2}}\\ h_3 \equiv \frac{a^{(3)}_2}{z^2}\\ h_6  \equiv \frac{a^{(6)}_{7/2}}{z^{7/2}}\\ \end{gathered}$&$\begin{gathered} \phi_5 - 2h_3h_2 \sim \frac{1}{z^{5/2}}\\ \phi_6 -6\phi_2 h_2^2 - h_3^2 \sim \frac{1}{z^3}\\ \phi_8 +4 \phi_2 h_3^2 + 48h_2(h_6-h_2^3) \sim \frac{1}{z^4}\\ \phi_9 + \phi_5 h_2^2 - h_6h_3 \sim \frac{1}{z^{9/2}}\\ \phi_{12} -3\phi_8 h_2^2 +2\phi_6 h_3^2 - 12\phi_2 h_3^2 h_2^2 - 36h_6^2 - h_3^4 \sim \frac{1}{z^6}\end{gathered}$\\
\hline
$F_4$&$h_3 \equiv \frac{a^{(3)}_{3/2}}{z^{3/2}}$&$\begin{gathered}\phi_5 - \phi_2 h_3  \sim \frac{1}{z^{3/2}}\\ \phi_6 + \frac{3}{2}\phi_2^3 + \frac{3}{2} h_3^2 \sim \frac{1}{z^2}\\ \phi_8 + 4\phi_6\phi_2 + 3\phi_2^4 - 4\phi_5 h_3 + 2\phi_2 h_3^2 \sim \frac{1}{z^2}\\ \phi_9 +\frac{1}{6}\phi_6 h_3 - \frac{1}{4}\phi_5\phi_2^2+\frac{1}{4}\phi_2^3h_3 +\frac{1}{12}h_3^3 \sim \frac{1}{z^{5/2}}\\ \phi_{12} + \phi_6^2 + 12\phi_9 h_3 + \phi_6 h_3^2 + \frac{3}{2}\phi_2^6+3\phi_6\phi_2^3 \\ +\frac{3}{4}\phi_8\phi_2^2-3\phi_5\phi_2^2 h_3+ \frac{3}{2}\phi_2^3 h_3^2 + \frac{1}{4}h_3^4 \sim \frac{1}{z^3}\end{gathered}$\\
\hline
\end{longtable}
}

\section{Appendix: Embeddings of $SU(2)$ in $F_4$}\label{appendix_embeddings_of__in_}

{
\renewcommand{\arraystretch}{1.5}
\begin{longtable}{|c|c|c|l|l|}
\hline
Bala-Carter&$\mathfrak{f}$&\mbox{\shortstack{\\Embedding\\indices}}&$26$&$52$\\
\hline 
\endhead
$A_1$&$\mathfrak{sp}(3)$&$(1,1)$&$(2,6) + (1,14)$&$(3,1)+(2,14')+(1,21)$\\
\hline
$\widetilde{A}_1$&$\mathfrak{su}(4)$&$(2,1)$&$\begin{gathered}(2,4)+(2,\overline{4})+(1,6)\\+(3,1)+(1,1)\end{gathered}$&$\begin{gathered}(3,1)+(3,6)\\+(2,4)+(2,\overline{4})+(1,15)\end{gathered}$\\
\hline
$A_1+\widetilde{A}_1$&$\mathfrak{su}(2)\times \mathfrak{su}(2)$&$(3,1,2)$&$(1;5,1)+(2;3,2)+(3;3,1)$&${\begin{gathered}(1;3,1)+(1;1,3)+\\(2;5,2)+(3;1,1)+\\(3;5,1)+(4;1,2)\end{gathered}}$\\
\hline
$A_2$&$\mathfrak{su}(3)$&$(4,2)$&$(3,3)+(3,\overline{3})+(1,8)$&$\begin{gathered}(5,1)+(3,6)+(3,\overline{6})\\+(3,1)+(1,8)\end{gathered}$\\
\hline
$\widetilde{A}_2$&$\mathfrak{g}_2$&$(8,1)$&$(3,7)+(5,1)$&$(5,7)+(3,1)+(1,14)$\\
\hline
$A_2+\widetilde{A}_1$&$\mathfrak{su}(2)$&$(6,6)$&$\begin{gathered}(4,2)+(3,3)+\\(2,4)+(1,1)\end{gathered}$&$\begin{gathered}(5,3)+(4,2)+(3,5)\\+(3,1)+(2,4)+(1,3)\end{gathered}$\\
\hline
$B_2$&$\mathfrak{su}(2)\times\mathfrak{su}(2)$&$(10,1,1)$&${\begin{aligned}&(5,1,1)+(4,2,1)+(4,1,2)\\&+(1,2,2)+(1,1,1)\end{aligned}}$&${\begin{gathered}(7,1,1)+(5,2,2)+\\(4,2,1)+(4,1,2)\\+(3,1,1)+(1,3,1)+(1,1,3)\end{gathered}}$\\
\hline
$\widetilde{A}_2+A_1$&$\mathfrak{su}(2)$&$(9,3)$&$\begin{gathered}(5,1)+(4,2)+\\(3,3)+(2,2)\end{gathered}$&$\begin{gathered}(6,2)+(5,3)+(4,2)+\\2(3,1)+(2,4)+(1,3)\end{gathered}$\\
\hline
$C_3(a_1)$&$\mathfrak{su}(2)$&$(11,1)$&$\begin{gathered}2(5,1)+(4,2)+(3,1)\\+(2,2)+(1,1)\end{gathered}$&$\begin{gathered}(7,1)+(6,2)+(5,1)+\\2(4,2)+3(3,1)+(1,3)\end{gathered}$\\
\hline
$F_4(a_3)$&$-$&$12$&$3(5)+3(3)+2(1)$&$2(7)+4(5)+6(3)$\\
\hline
$B_3$&$\mathfrak{su}(2)$&$(28,8)$&$(1;5)+(7,3)$&$\begin{gathered}(1;3)+(3;1)+\\(7;5)+(11;1)\end{gathered}$\\
\hline
$C_3$&$\mathfrak{su}(2)$&$(35,1)$&$(9,1)+(6,2)+(5,1)$&$\begin{gathered}(11,1)+(10,2)+(7,1)\\+(4,2)+(3,1)+(1,3)\end{gathered}$\\
\hline
$F_4(a_2)$&$-$&$36$&$(9)+(7)+2(5)$&$\begin{gathered}2(11)+(9)+\\(7)+(5)+3(3)\end{gathered}$\\
\hline
$F_4(a_1)$&$-$&$60$&$(11)+(9)+(5)+(1)$&$\begin{gathered}(15)+2(11)+\\(7)+(5)+(3)\end{gathered}$\\
\hline
$F_4$&$-$&$156$&$(17)+(9)$&$(23)+(15)+(11)+(3)$\\
\hline
\end{longtable}
}

\section{Projection matrices}\label{projection_matrices}

{
\renewcommand{\arraystretch}{1.5}
\begin{longtable}{|c|c|c|}
\hline
Bala-Carter&$\mathfrak{f}$&Projection Matrix\\
\hline
\endhead
$\underline{A_1}$&$Sp(3)_{13}$&$\begin{pmatrix}1&0&0&0 \\ 0&0&1&0 \\ 0&0&0&1 \\ 1&2&1&0\end{pmatrix}$\\
\hline
$\underline{\widetilde{A}_1}$&$SU(4)_{12}$&$\begin{pmatrix}0&2&1&0\\1&1&1&0\\0&1&1&1\\1&1&0&0\end{pmatrix}$\\
\hline
$\underline{A_1 + \widetilde{A}_1}$&$SU(2)_{64}\times SU(2)_{10}$&$\begin{pmatrix}1&3&3&2\\4&8&4&2\\1&1&1&0\end{pmatrix}$\\
\hline
$\underline{A_2}$&$SU(3)_{16}$&$\begin{pmatrix}4&6&4&2\\0&0&1&1\\0&2&1&0\end{pmatrix}$\\
\hline
$\underline{\widetilde{A}_2}$&$(G_2)_{10}$&$\begin{pmatrix}4&8&6&4\\1&0&1&0\\0&1&0&0\end{pmatrix}$\\
\hline
$\underline{A_2 + \widetilde{A}_1}$&$SU(2)_{39}$&$\begin{pmatrix}2&6&5&3\\4&6&3&1\end{pmatrix}$\\
\hline
$\underline{B_2}$&$SU(2)_7^2$&$\begin{pmatrix}4&10&7&4\\1&1&0&0\\1&1&1&0\end{pmatrix}$\\
\hline
$\underline{\widetilde{A}_2 + A_1}$&$SU(2)_{20}$&$\begin{pmatrix}4&9&7&4\\2&3&1&0\end{pmatrix}$\\
\hline
$\underline{C_3(a_1)}$&$SU(2)_7$&$\begin{pmatrix}5&11&8&4\\1&1&0&0\end{pmatrix}$\\
\hline
$\underline{F_4(a_3)}$&$-$&$\begin{pmatrix}6&12&8&4\end{pmatrix}$\\
\hline
$\underline{B_3}$&$SU(2)_{24}$&$\begin{pmatrix}6&16&12&6\\4&4&2&2\end{pmatrix}$\\
\hline
$\underline{C_3}$&$SU(2)_6$&$\begin{pmatrix}9&19&14&8\\1&1&0&0\end{pmatrix}$\\
\hline
$\underline{F_4(a_2)}$&$-$&$\begin{pmatrix}10&20&14&8\end{pmatrix}$\\
\hline
$\underline{F_4(a_1)}$&$-$&$\begin{pmatrix}14&26&18&10\end{pmatrix}$\\
\hline
$\underline{F_4}$&$-$&$\begin{pmatrix}22&42&30&16\end{pmatrix}$\\
\hline
\end{longtable}
}

\end{appendices}

\bibliographystyle{utphys}
%\small\baselineskip=.93\baselineskip
%\let\bbb\bibitem\def\bibitem{\itemsep1pt\bbb}
\bibliography{ref}

\end{document}